\documentclass[useAMS,usenatbib]{mn2e}

\usepackage{epsfig}

\newcommand{\lsim}{\la}

\title[MUNICS IX: Galaxy Evolution to $z \sim 2$ From Optical
  Catalogues]{The Munich Near-Infrared Cluster Survey (MUNICS) --
  IX. Galaxy Evolution to $z \sim 2$ From Optically Selected
  Catalogues\footnotemark[2]\footnotemark[3]\footnotemark[4]}

\author[Georg Feulner et al.]
{Georg~Feulner$^{1,2,3}$\footnotemark[1], 
Yuliana~Goranova$^{1,2}$\footnotemark[5],  
Ulrich~Hopp$^{1,2}$,
Armin~Gabasch$^{1,2}$\footnotemark[6],\newauthor
Ralf~Bender$^{1,2}$,
Christine~S.~Botzler$^{1,4,5}$,
Niv~Drory$^{2,6}$\vspace*{.25em} 
\\
$^1$Universit\"ats-Sternwarte M\"unchen, Scheinerstra\ss e 1, D--81679
M\"unchen, Germany\\
$^2$Max-Planck-Institut f\"ur Extraterrestrische Physik,
Giessenbachstra\ss e, D--85748 Garching bei M\"unchen, Germany\\
$^3$Potsdam--Institut f\"ur Klimafolgenforschung, Postfach 60~12~03,
D--14412 Potsdam, Germany\\
$^4$University of Auckland, Private Bag 92019, Morrin Road, 
Glen Innes, Auckland, New Zealand\\
$^5$University of Canterbury, Private Bag 4800, Christchurch, 
New Zealand\\
$^6$University of Texas at Austin, Austin, Texas 78712
}

\begin{document}

\date{Accepted 2007 March 28. Received 2007 March 28; in original form 2006
  December 20}

\pagerange{\pageref{firstpage}--\pageref{lastpage}} \pubyear{2007}

\maketitle

\label{firstpage}

\begin{abstract}
We present $B$, $R$, and $I$-band selected galaxy catalogues based on
the Munich Near-Infrared Cluster Survey (MUNICS) which, together with
the previously used $K$-selected sample, serve as an important probe
of galaxy evolution in the redshift range $0 \lsim z \lsim
2$. Furthermore, used in comparison they are ideally suited to study
selection effects in extragalactic astronomy.  The construction of the
$B$, $R$, and $I$-selected photometric catalogues, containing $\sim
9000$, $\sim 9000$, and $\sim 6000$ galaxies, respectively, is
described in detail. The catalogues reach 50\% completeness limits for
point sources of $B \simeq 24.5$~mag, $R \simeq 23.5$~mag, and $I
\simeq 22.5$~mag and cover an area of about 0.3 square
degrees. Photometric redshifts are derived for all galaxies with an
accuracy of $\delta z / (1+z) \simeq 0.057$, very similar to the
$K$-selected sample. Galaxy number counts in the $B$, $V$, $R$, $I$,
$J$, and $K$ bands demonstrate the quality of the dataset.

The rest-frame colour distributions of galaxies at different selection
bands and redshifts suggest that the most massive galaxies have formed
the bulk of their stellar population at earlier times and are
essentially in place at redshift unity.  We investigate the influence
of selection band and environment on the specific star formation rate
(SSFR). We find that $K$-band selection indeed comes close to
selection in stellar mass, while $B$-band selection purely selects
galaxies in star formation rate. We use a galaxy group catalogue
constructed on the $K$-band selected MUNICS sample to study possible
differences of the SSFR between the field and the group environment,
finding a marginally lower average SSFR in groups as compared to
the field, especially at lower redshifts.

The field-galaxy luminosity function in the $B$ and $R$ band as
derived from the $R$-selected sample evolves out to $z \simeq 2$ in
the sense that the characteristic luminosity increases but the number
density decreases. This effect is smaller at longer rest-frame
wavelengths and gets more pronounced at shorter
wavelengths. Parametrising the redshift evolution of the Schechter
parameters as $M^* (z) = M^* (0) + a \, \ln ( 1 + z )$ and $\Phi^* (z)
= \Phi^* (0) ( 1 + \, z )^b$ we find evolutionary parameters $a \simeq
-2.1$ and $b \simeq -2.5$ for the $B$ band, and $a \simeq -1.4$ and $b
\simeq -1.8$ for the $R$ band.

\end{abstract}


\footnotetext[1]{E-mail: feulner@usm.lmu.de}

\footnotetext[2]{Based on observations collected at the Centro
  Astron\'omico Hispano Alem\'an (CAHA), operated by the
  Max-Planck-Institut f\"ur Astronomie, Heidelberg, jointly with the
  Spanish National Commission for Astronomy.}

\footnotetext[3]{Based on observations collected at the VLT (Chile)
  operated by the European Southern Observatory in the course of the
  observing proposals 66.A-0123 and 66.A-0129.}

\footnotetext[4]{Based on observations obtained with the Hobby-Eberly 
  Telescope, which is a joint project of the University of Texas at
  Austin, the Pennsylvania State University, Stanford University,
  Ludwig-Maximilians-Universit\"at M\"unchen, and 
  Georg-August-Universit\"at G\"ottingen.}

\footnotetext[5]{Current address: Leiden Observatory, P.O. Box 9513,
NL--2300 RA Leiden, The Netherlands}

\footnotetext[6]{Current address: European Southern Observatory,
Karl--Schwarzschild--Stra\ss e 2, D--85748 Garching bei M\"unchen,
Germany}

%
% INTRODUCTION
%
\section{Introduction}
\label{s:intro}

Investigating the evolution of galaxies with time requires the
analysis of statistical properties of galaxies at various cosmic
epochs. Studies of the luminosity function of galaxies at different
wavelengths, of the galaxies' mass function or star formation rates at
different redshifts are important methods to address the problem of
the formation and evolution of galaxies within observational
cosmology. These investigations rely on galaxy surveys large enough to
yield sufficient statistical accuracy and deep enough to trace the
properties of the galaxy population to high redshifts and faint
intrinsic luminosities.

Since different wavelengths sample light from different stellar
populations within the galaxies, the selection band of a galaxy survey
is of fundamental importance. At the reddest wavelengths, a galaxy's
light is dominated by old stellar populations and hence reflects its
total stellar mass \citep{RR93}. Moving to bluer wavelengths, light
from younger stellar populations becomes important, and samples
selected in bluer filters become biased toward on-going star formation
activity.

Historically, by the 1980s the increased number of 4-m class
telescopes and advances in electronic detector technology enabled deep
surveys of the universe out to redshifts $z \; \sim \; 1$, when the
universe had only about half of its present age (see
\citealt{Colless1997, Ellis1997} for reviews).

Traditionally, many surveys were conducted in the $B$ band using
photographic plates at Schmidt telescopes. When multi-object
spectrographs became available at 4-m class telescopes, many groups
started follow-up observations of these surveys. One noteworthy
redshift survey of that kind is the Autofib Redshift Survey probing
the evolution of the luminosity function out to redshifts $z \; \sim
\; 0.8$ with roughly 1700 galaxy redshifts \citep{ECBHG96, HCEB97}.

\begin{figure*}

\epsfig{figure=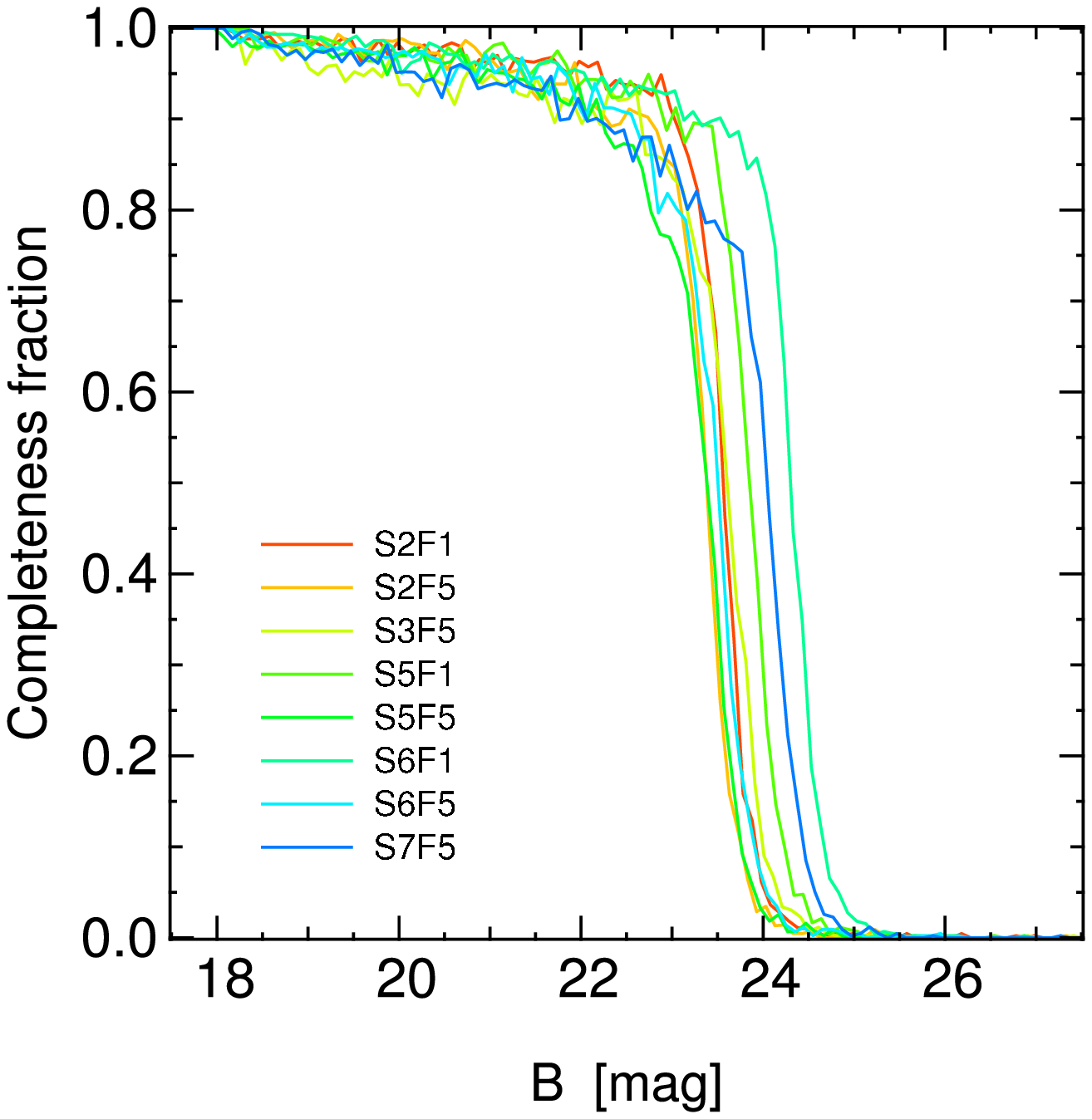,width=0.4\textwidth}
\hspace*{1cm}
\epsfig{figure=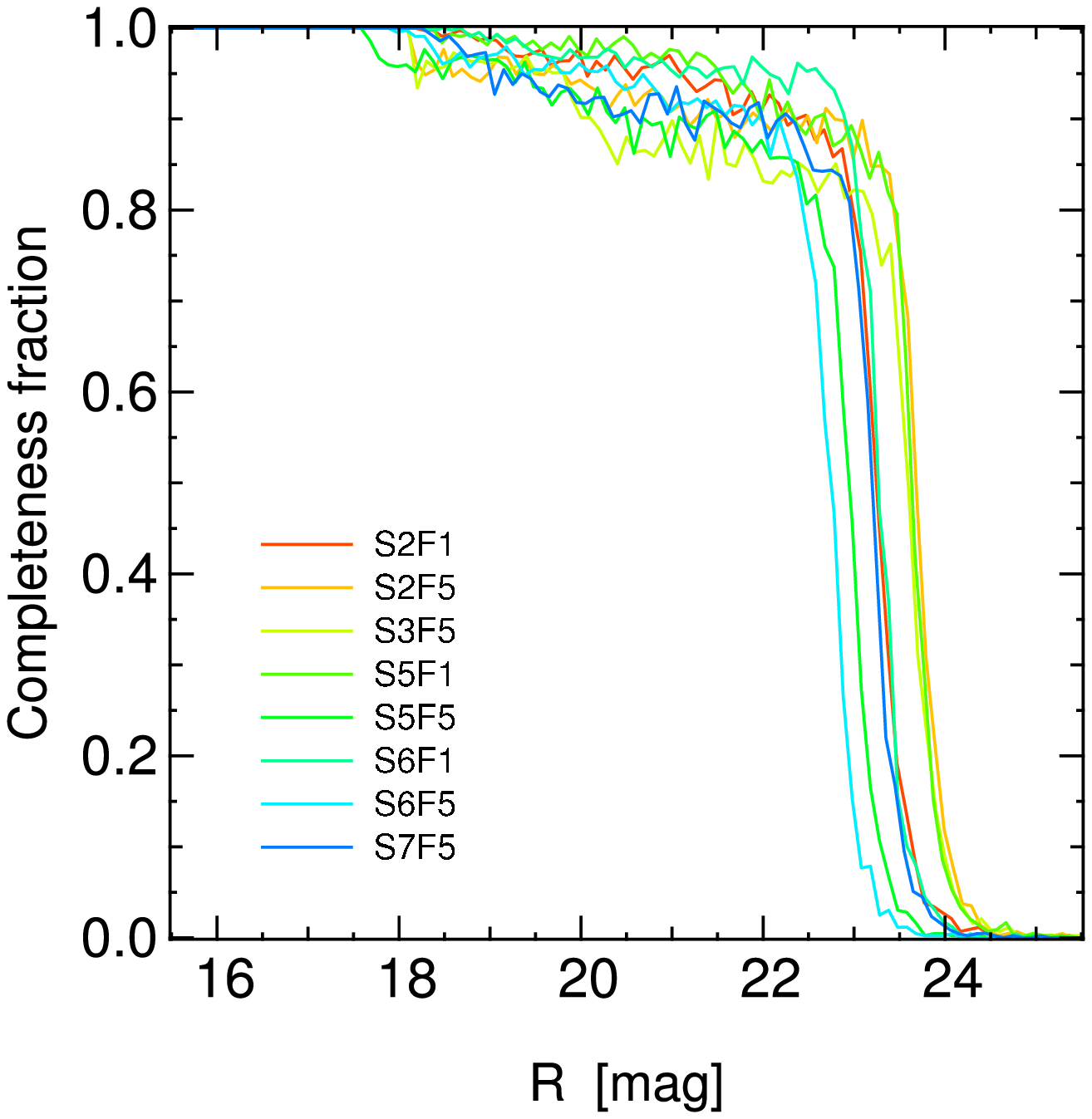,width=0.4\textwidth}

\vspace*{.5cm}

\epsfig{figure=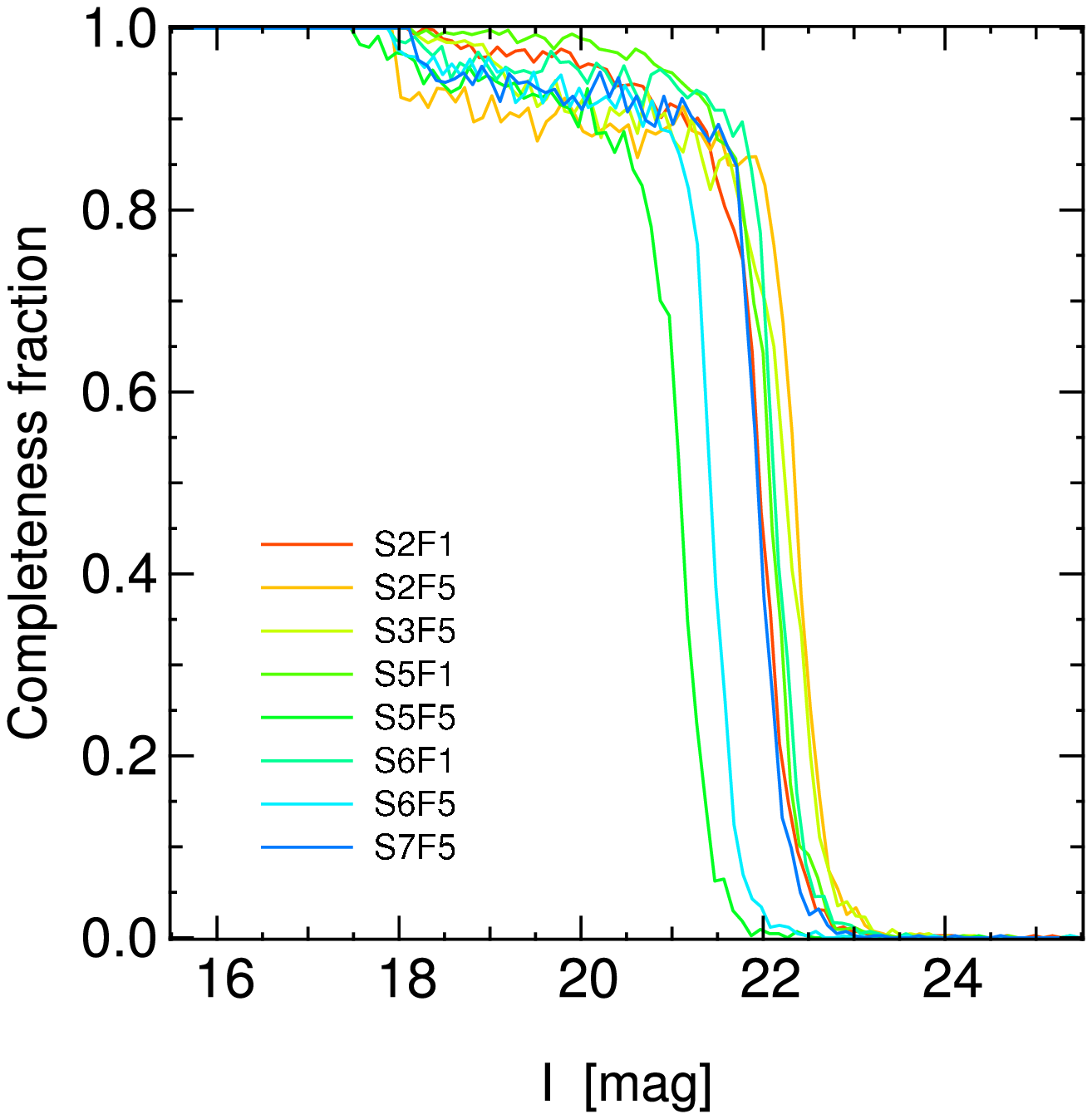,width=0.4\textwidth}
\hspace*{1cm}
\epsfig{figure=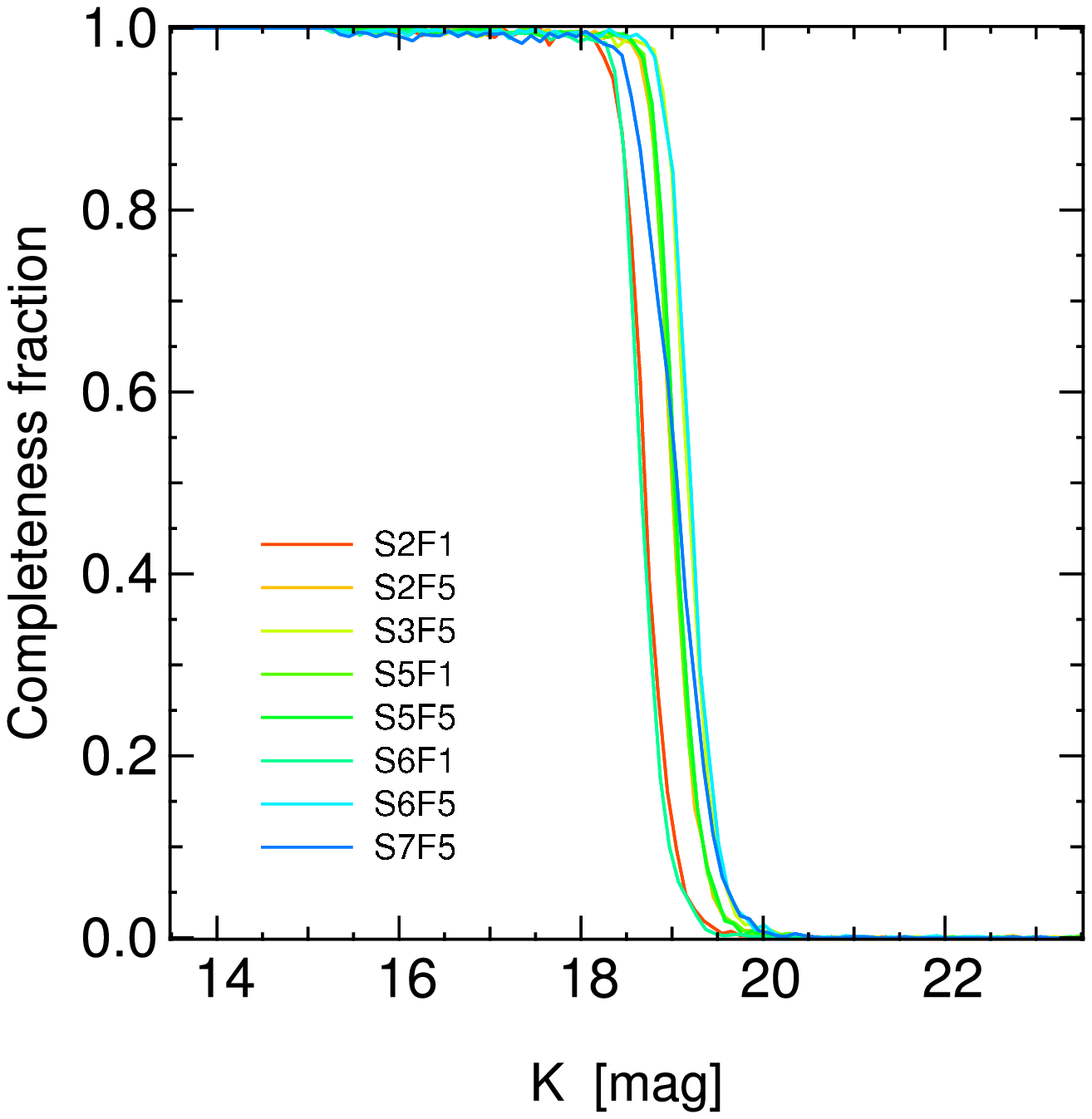,width=0.4\textwidth}

\caption{Result of Monte-Carlo completeness simulations with
  artificial point sources placed in the $B$ (\textit{upper left-hand
  panel}), $R$ (\textit{upper right-hand panel}), $I$ (\textit{lower
  left-hand panel}), and $K$-band images (\textit{lower right-hand
  panel}) of MUNICS.}
  \label{f:complbrik}

\end{figure*}

In the middle of the 1990s, the $I$-band selected Canada--France
Redshift Survey (CFRS) marked an important point in the study of
galaxy evolution with redshift surveys. The selection in the $I$ band
made it possible to study the evolution of the evolved, massive
galaxies rather than star forming galaxies picked up in the more
traditional $B$-band selected samples. The CFRS comprises spectra of
more than 1000 objects with $17.5 \le I_{AB} \le 22.5$, 591 of which
are galaxies with secure redshifts in the range $0 \le z \le 1.3$ and
enabled studies of the evolution of the luminosity function
\citep{CFRS6} and of the luminosity density and star formation rate
density of the universe \citep{CFRS96}, finding very little evolution
in the luminosity and number density of red galaxies in the redshift
range $0 < z < 1$.

Selection in near-infrared filters like the $K$-band comes even closer
to a selection in stellar mass. When suitable near-infrared detectors
became available at large telescopes, a number of $K$-band selected
surveys were undertaken which can probe the evolution of massive
galaxies out to redshift $z \; \sim \; 1$. Examples for such surveys
are the Hawaii Deep Fields \citep{CGHSHW94}, the K20 Survey
\citep{K20_3} and the Munich Near-Infrared Cluster Survey (MUNICS;
\citeauthor{munics1} 2001b).

In this paper we present $B$, $R$, and $I$ band selected MUNICS
catalogues which can be used for comparison with previous work and,
together with the $K$-band selected catalogue, to study selection
effects in extra-galactic surveys. They are an important tool to trace
different parts of the field galaxy population, and to discriminate
between selection biases and real evolutionary effects. Furthermore,
this paper also serves as a general update to \citeauthor{munics1}
(2001b). Since the publication of that paper, the survey's
high-quality multi-colour imaging part has not only grown in area, but
$B$-band imaging has become available, and the catalogue has been
improved in a number of ways.

This paper is organised as follows. Section~\ref{s:munics} gives a
brief overview of the Munich Near-Infrared Cluster Survey (MUNICS),
while Section~\ref{s:optical} describes the construction and content
of the optically selected catalogues presented in this paper. In
Section~\ref{s:nc} we present galaxy number counts in all six filters
and compare them to previous studies. Section~\ref{s:colours}
discusses results on the rest-frame colour distributions of galaxies
in the different catalogues, Sections~\ref{s:ssfrsel} and
\ref{s:ssfrenv} elaborate on the influence of selection effects and
environment on the SSFR, respectively, while Section~\ref{s:lf}
presents results on the evolution of luminosity functions, before we
summarise our findings in Section~\ref{s:summ}.  Throughout this work
we assume $\Omega_M = 0.3$, $\Omega_\Lambda = 0.7$ and $H_0 = 70 \,
\mathrm{km} \, \mathrm{s}^{-1} \, \mathrm{Mpc}^{-1}$. All magnitudes
are given in the Vega system.

\begin{table}
\caption[MUNICS fields]{The ten MUNICS fields. The table gives
the field name, the field coordinates (right ascension $\alpha$ and
declination $\delta$ as well as galactic coordinates $l$ and $b$) for
the equinox 2000.0 and the effective area of the field in square
arc-minutes. The effective area is the area of the sky covered by
observations in all six filters.}
\label{t:fields}
\begin{center}
\begin{tabular}{lccccc}
\hline
Field & $\alpha$ & $\delta$ & $l$ & $b$ & Area\\
& \multicolumn{2}{c}{(2000.0)} &
\multicolumn{2}{c}{(2000.0)} & [arcmin$^2$] \\
\hline
S2F1 & 03:06:41 & $+$00:01:12 & 178.66 & $-$47.67 &  119.1 \\
S2F5 & 03:06:41 & $-$00:13:30 & 178.93 & $-$47.84 &  123.7 \\
S3F1 & 09:04:38 & $+$30:02:56 & 195.37 & $+$40.64 &  116.6 \\
S3F5 & 09:03:44 & $+$30:02:56 & 195.32 & $+$40.45 &  107.6 \\
S4F1 & 03:15:00 & $+$00:07:41 & 180.58 & $-$46.05 &  105.6 \\
S5F1 & 10:24:01 & $+$39:46:37 & 180.85 & $+$57.04 &  115.7 \\
S5F5 & 10:25:14 & $+$39:46:37 & 180.76 & $+$57.27 &  105.3 \\
S6F1 & 11:55:58 & $+$65:35:55 & 131.90 & $+$50.55 &  117.1 \\
S6F5 & 11:57:56 & $+$65:35:55 & 131.60 & $+$50.62 &  116.5 \\
S7F5 & 13:34:44 & $+$16:51:44 & 349.45 & $+$75.66 &  119.1 \\
\hline
All  &          &             &        &          & 1146.3 \\
\hline
\end{tabular}
\end{center}
\end{table}

%
% MUNICS
%
\section{The Munich Near-Infrared Cluster Survey (MUNICS)}
\label{s:munics}

The Munich Near-Infrared Cluster Survey (or MUNICS for short) is a
wide-field medium-deep imaging survey in the near-infrared and optical
initially described in \citeauthor{munics1} (2001b). Dedicated follow-up
spectroscopy is available for $\sim 600$ galaxies \citep{munics5}. The
main part of the survey consists of 10 fields the details of which are
summarised in Table~\ref{t:fields}. For all these fields photometry in
$K'$, $J$, $I$, $R$, $V$, and $B$ is available, with limiting
magnitudes ranging from $K' \simeq 19.5$ to $B \simeq 24.5$ (50\%
completeness for point sources; \citealt{munics4}). The total area of
this part of the survey is 1146.3 square arcmin (or 0.32 square
degrees). However, the fields S3F1 and S4F1 have less than
satisfactory quality, especially concerning their photometric
calibration. Since an accurate calibration of photometric colours is
essential for high-quality photometric redshifts, we exclude these two
fields from any further analysis (as has been done with the
$K$-selected sample). The total area of the eight high-quality fields
is then 924.1 square arcmin (or 0.26 square degrees).

Most of the research on field galaxy evolution and galaxy clusters in
the MUNICS project has been carried out with the $K$-band selected
catalogue presented in \citeauthor{munics1} (2001b). This includes
studies of luminosity function evolution \citep{munics5, munics2}, the
stellar mass function of galaxies (\citeauthor{munics3} 2001a,
\citeyear{munics6}), the integrated specific star formation rate of
groups \citep{intssfr}, and the cluster catalogue described in
\citet{munics8}. \citeauthor{munics7} (2005b), however, a study of the
evolution of the specific star-formation rate with redshift, is based
on the $I$-band selected catalogue. In this paper, the $I$-, $R$-, and
$B$-selected MUNICS samples will be described for the first time in
detail.

\begin{figure*}
\epsfig{figure=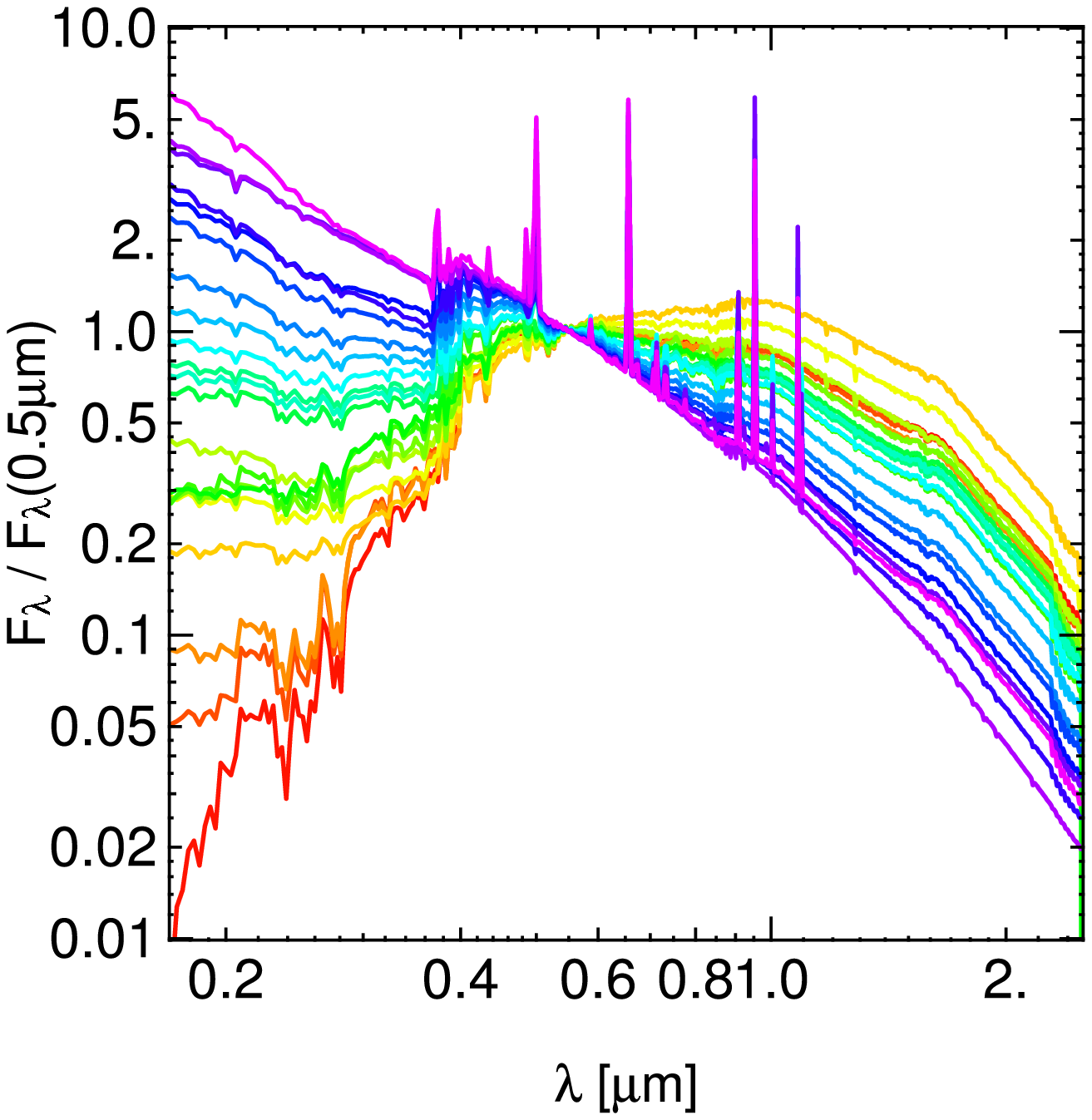,width=0.4\textwidth}
\hspace*{1cm}
\epsfig{figure=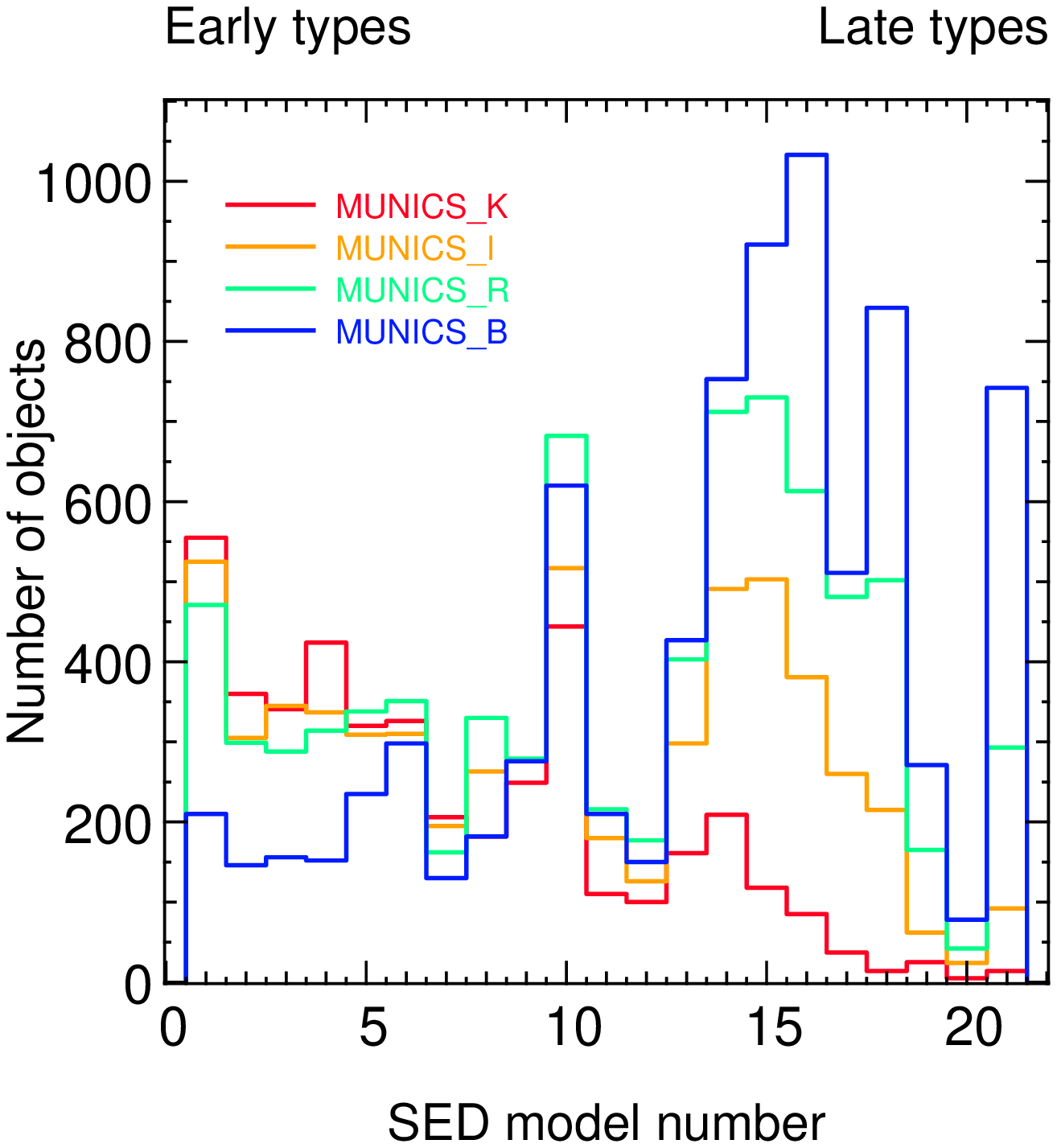,width=0.4\textwidth}

\caption{\textit{Left-hand panel:} Template spectral energy
  distributions (SEDs) for galaxies used in the computation of
  photometric redshifts for the $K$-, $I$-, $R$-, and $B$-selected
  samples. \textit{Right-hand panel:} The distribution of SED types in
  the four MUNICS catalogues. The SED number is an internal SED
  identification: Small numbers refer to redder (early-type) galaxies,
  higher numbers to bluer (late-type) galaxies.}
\label{f:model-seds}

\end{figure*}

%
% OPTICAL CATALOGUES
%
\section{The Construction of Optically Selected Catalogues}
\label{s:optical}

In this section we want to discuss the construction and the properties
of the $B$, $R$, and $I$-band selected photometric catalogues as well
as the measurement and reliability of photometric redshifts. We will
often refer to the different catalogues as MUNICS\_B, MUNICS\_R,
MUNICS\_I, and MUNICS\_K for short.

\begin{figure*}

\epsfig{figure=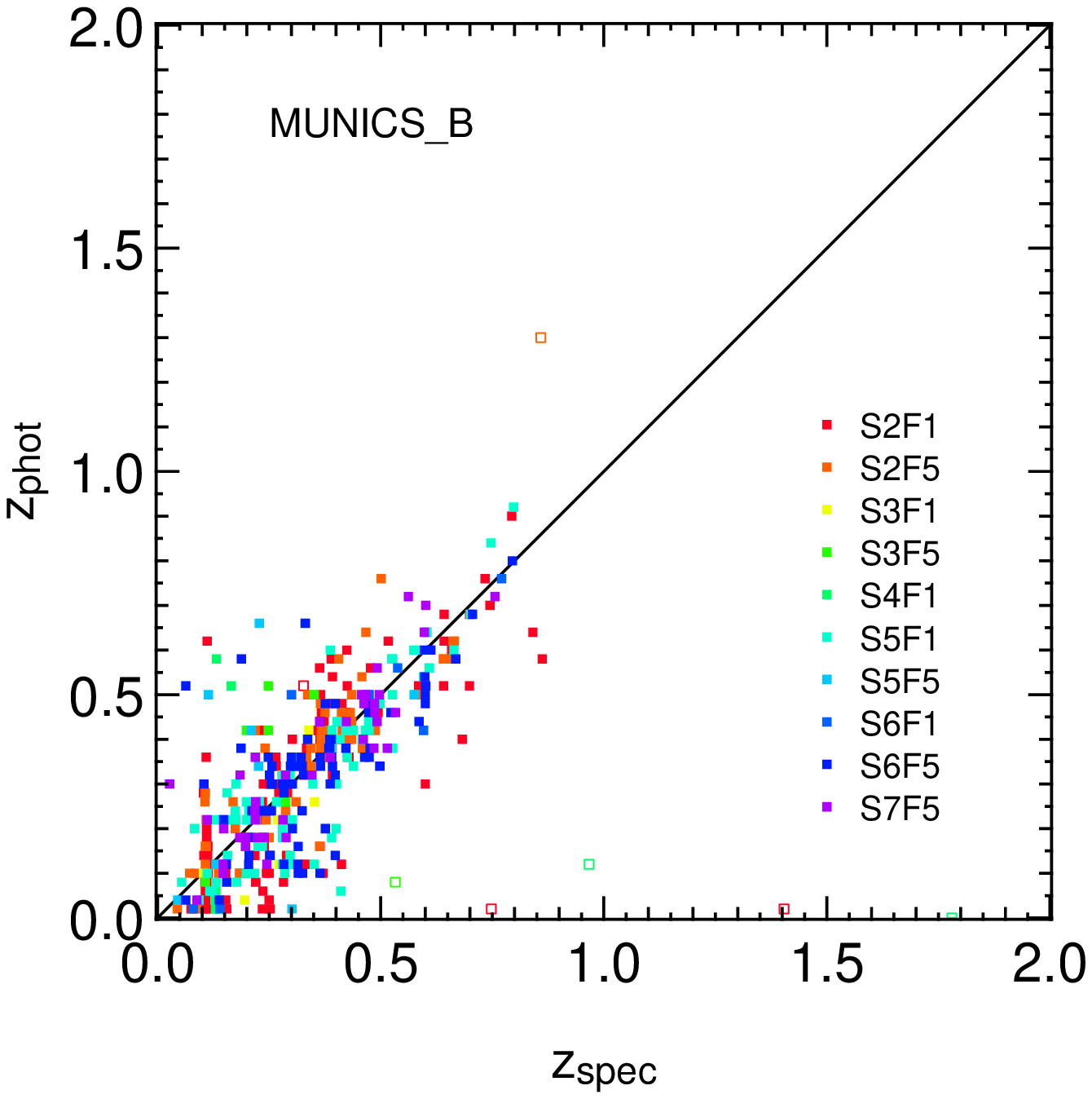,width=0.3\textwidth}
\hspace*{.5cm}
\epsfig{figure=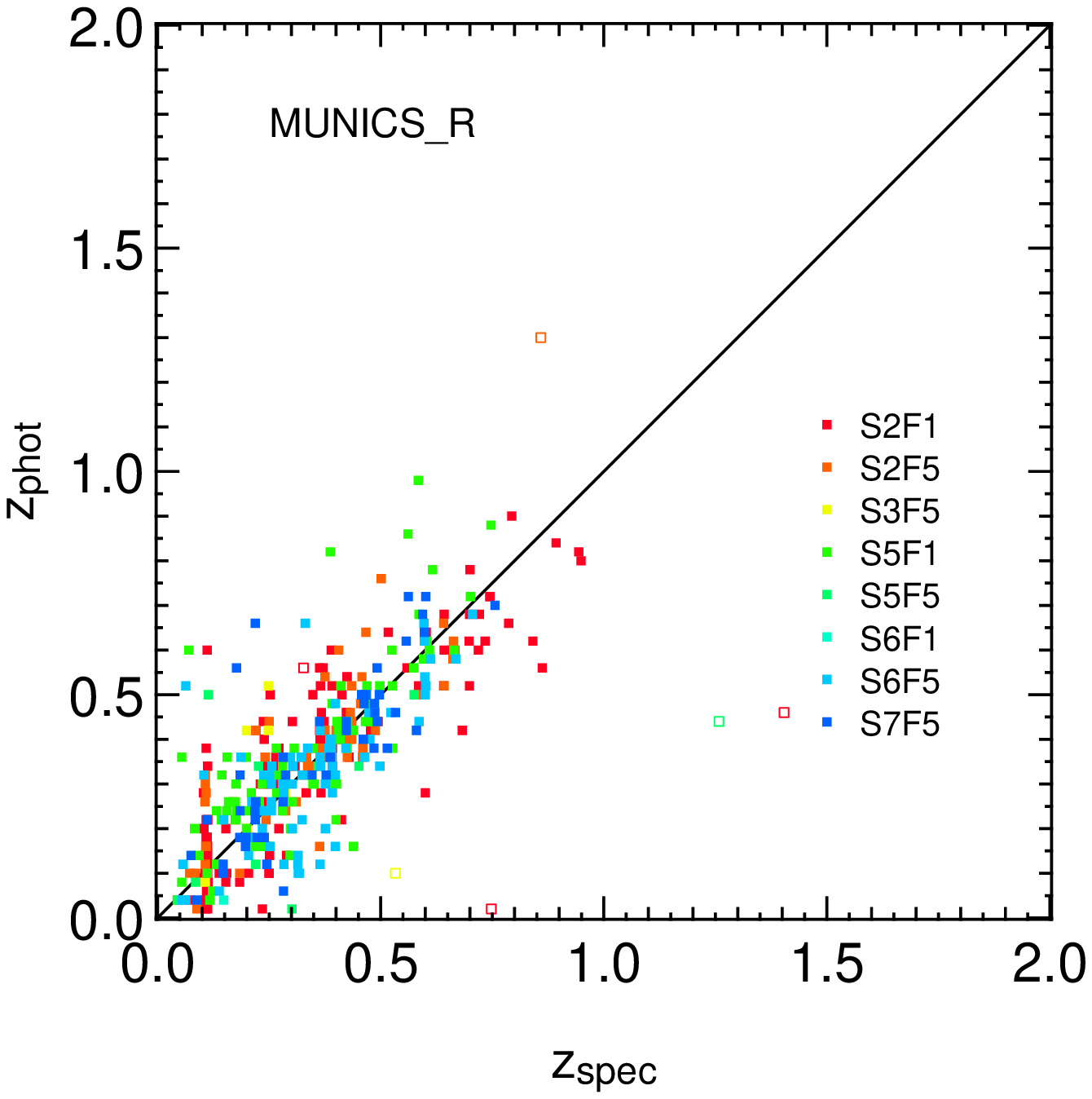,width=0.3\textwidth}
\hspace*{.5cm}
\epsfig{figure=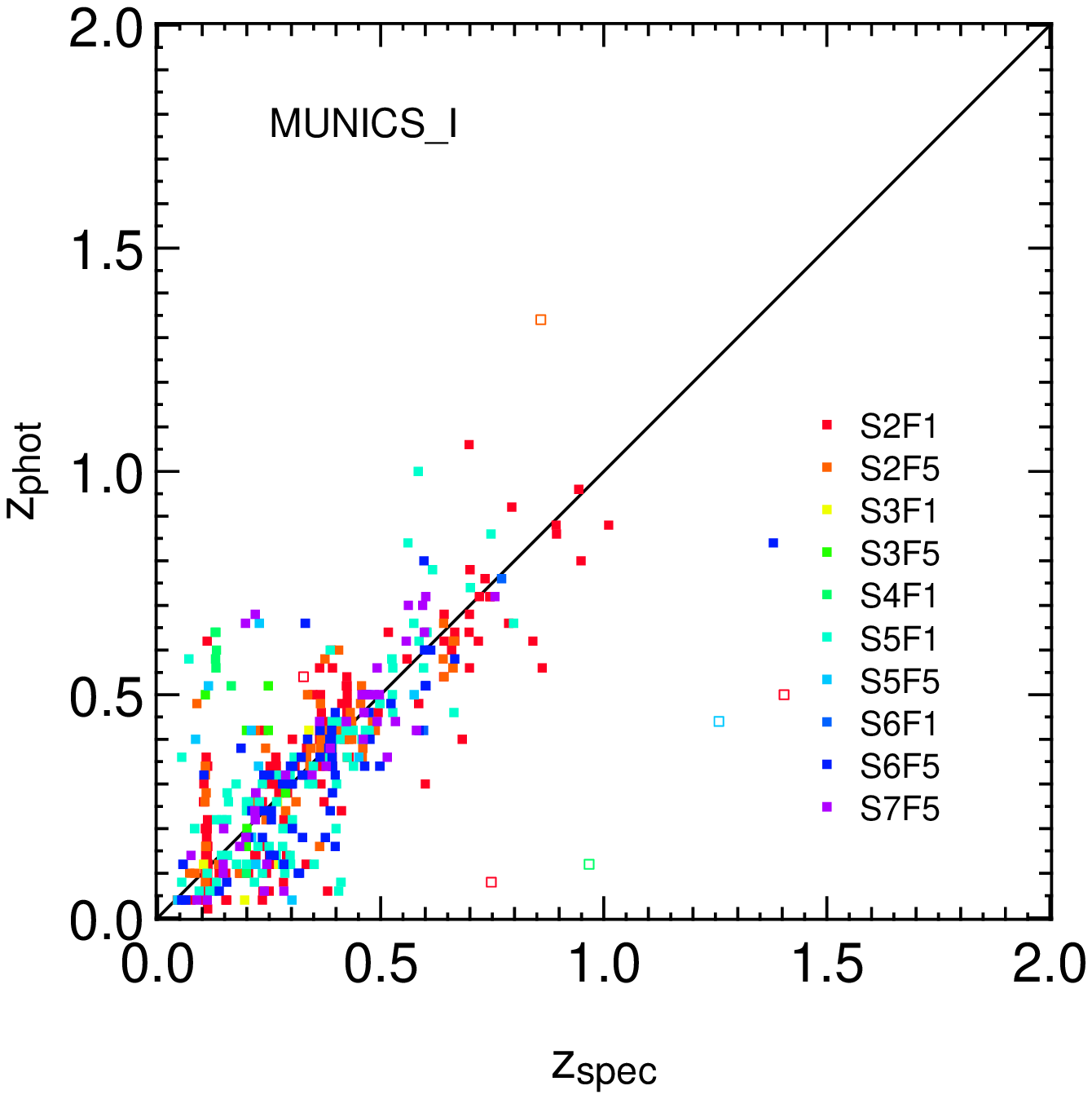,width=0.3\textwidth}

\vspace*{.5cm}

\epsfig{figure=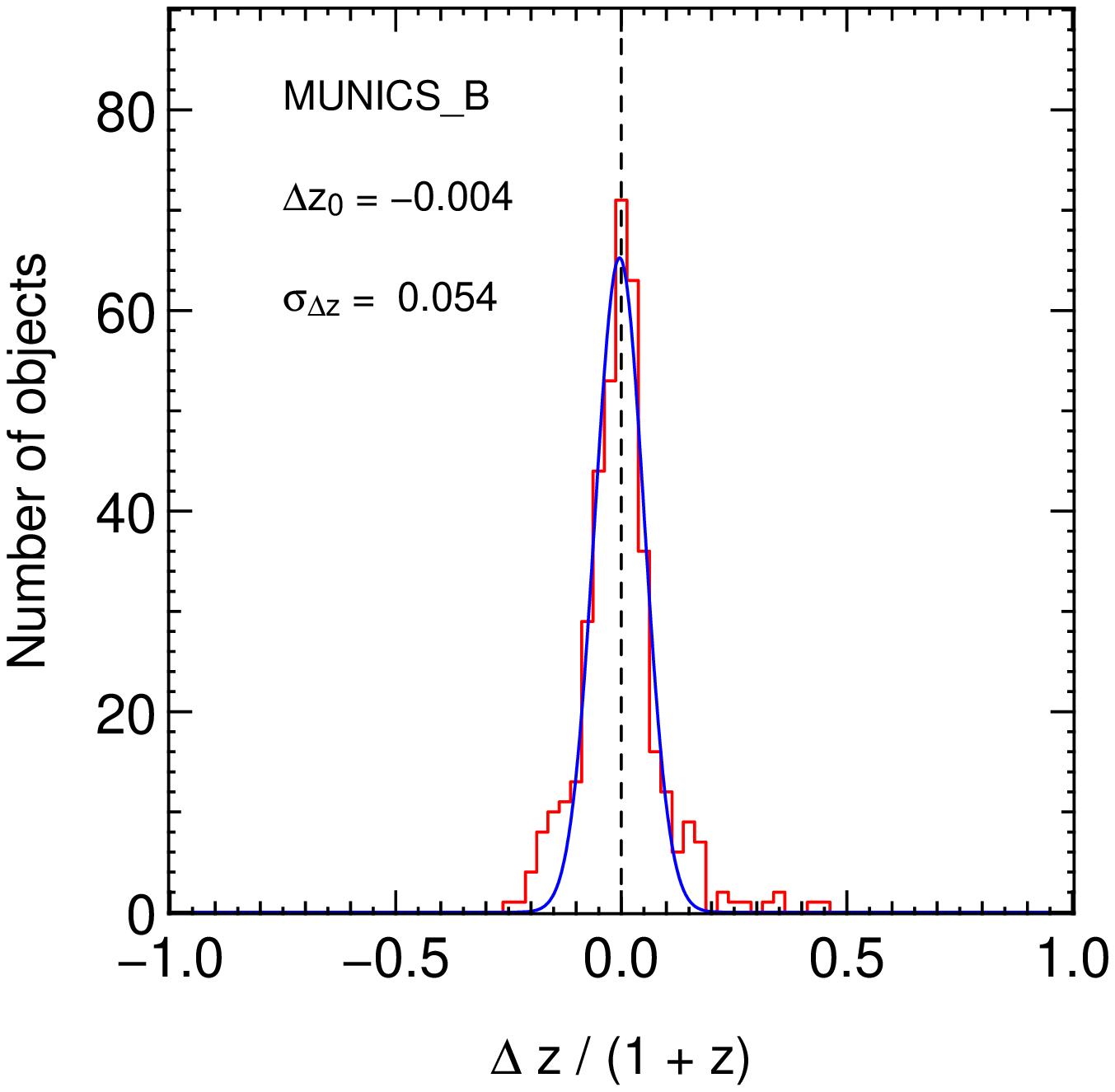,width=0.3\textwidth}
\hspace*{.5cm}
\epsfig{figure=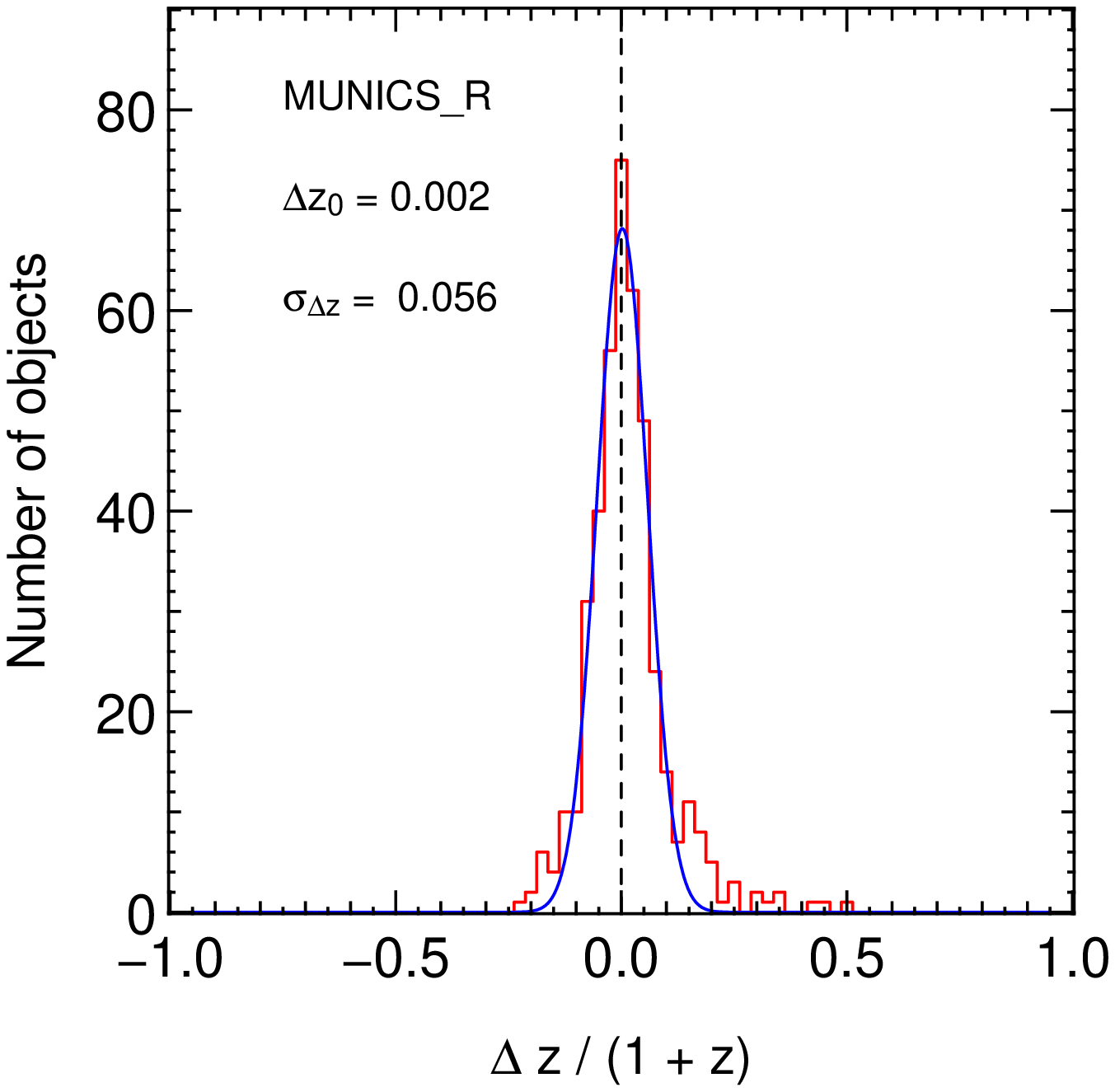,width=0.3\textwidth}
\hspace*{.5cm}
\epsfig{figure=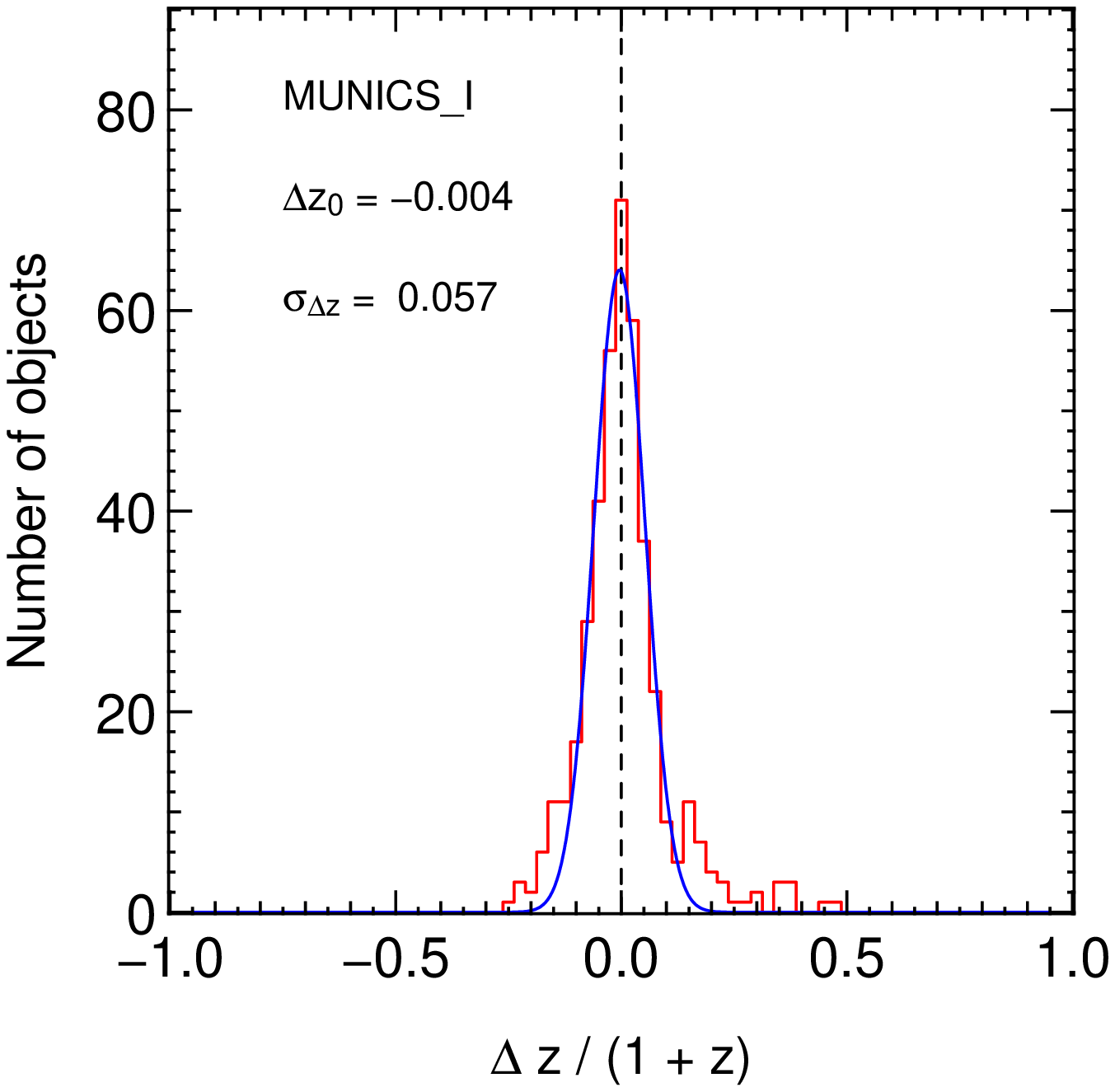,width=0.3\textwidth}

\caption[Comparison of photometric and spectroscopic
  redshifts]{\textit{Upper panels:} A comparison of photometric and
  spectroscopic redshifts for MUNICS\_B (\textit{left}), MUNICS\_R
  (\textit{middle}), and MUNICS\_I (\textit{right}). Filled symbols
  are galaxies, open symbols quasars. \textit{Lower panels:} The
  corresponding histograms of the redshift differences together with a
  Gaussian fit showing the quality of the photometric redshifts. The
  corresponding diagrams for MUNICS\_K are published in
  \citet{munics2}.}

\label{f:zz}

\end{figure*}

\subsection{Object Detection}
\label{s:det}

As for the $K$-selected catalogue, object detection on the $B$, $R$,
and $I$-band images was performed using the {\sc yoda} source
extraction software \citep{yoda}. Sources are detected by requiring a
minimum number $N_{pix}$ of consecutive pixels to lie above a certain
threshold $t$ expressed in units of the local RMS $\sigma$ of the
background noise.  To ensure secure detection of faint sources, the
images are convolved with a Gaussian of full width at half maximum
(FWHM) similar to the seeing in the image. The choice of the number of
consecutive pixels $N_{pix}$ and the threshold $t$ is a compromise
between limiting magnitude at some completeness fraction, say 50 per
cent, and the number of tolerable spurious detections per unit image
area (\citealt{Saha95}, see also the discussion in
\citeauthor{munics1} 2001b).

To find reasonable values for $N_{pix}$ and $t$ we performed
simulations with varying minimum number of consecutive pixels on the
optical images of one of our mosaic fields (S6F5). Instead of fixing
two parameters (the threshold and the convolution kernel) and keeping
only one free parameter (as with the $K$-band detection), we chose
only one fixed parameter: The size of the convolution kernel was set
to be equal to the (Gaussian) point-spread function (PSF) of the
images. Thus we kept both the threshold and the number of consecutive
pixels as free parameters in our simulations. Choosing a PSF-like
Gaussian kernel for the image convolution is necessary to ensure
secure detection of faint, compact objects in the images. The number
of false detections was measured by running the detection algorithm on
the same images multiplied by $-1$.

From these simulations, the best choice of parameters is $t=4\sigma$
for the detection threshold, and $N_{pix}=\pi ({\mathrm FWHM}/2)^2$
for the minimum number of consecutive pixels required for an
object. With these settings, the contamination rate (the rate of false
detections) is actually below 1 per cent.

Since one of the ideas of this paper is a comparison of properties
of the $K$-band and the optically selected galaxies, we have convinced
ourselves that this change of detection parameters does not influence
the results presented here. The reason is that different detection
parameter choices mostly affect the object statistics at the very
faintest apparent magnitudes. Since these objects also have large
photometric errors, they are usually excluded from any further
analysis and thus cannot influence the results of our work.

In Figure~\ref{f:complbrik} we show the completeness functions for the
$B$, $R$, and $I$-band images of the eight high-quality MUNICS mosaic
fields, based on Monte-Carlo simulations of artificial point-sources
placed at random in the images. In these simulations, $200$ artificial
objects with a Moffat-type PSF having the same FWHM as stars in the
frames and a constant magnitude distribution were added to the images
using the {\sc iraf} {\sc artdata} package. After running the object
detection algorithm on the images applying the same detection
parameters as for the main catalogue, the fraction of re-detected
artificial objects as a function of magnitude is computed. This
procedure was repeated $500$ times in order to decrease statistical
errors. An extensive discussion of completeness simulations
for extended objects in MUNICS $K$-band images can be found in
\citet{munics4}.

One can see from the diagram that the 50\% completeness limits for
point sources as derived from these simulations are $B \simeq
24.5$~mag, $R \simeq 23.5$~mag, $I \simeq 22.5$~mag, and $K \simeq
19.5$~mag.

\begin{figure*}

\epsfig{figure=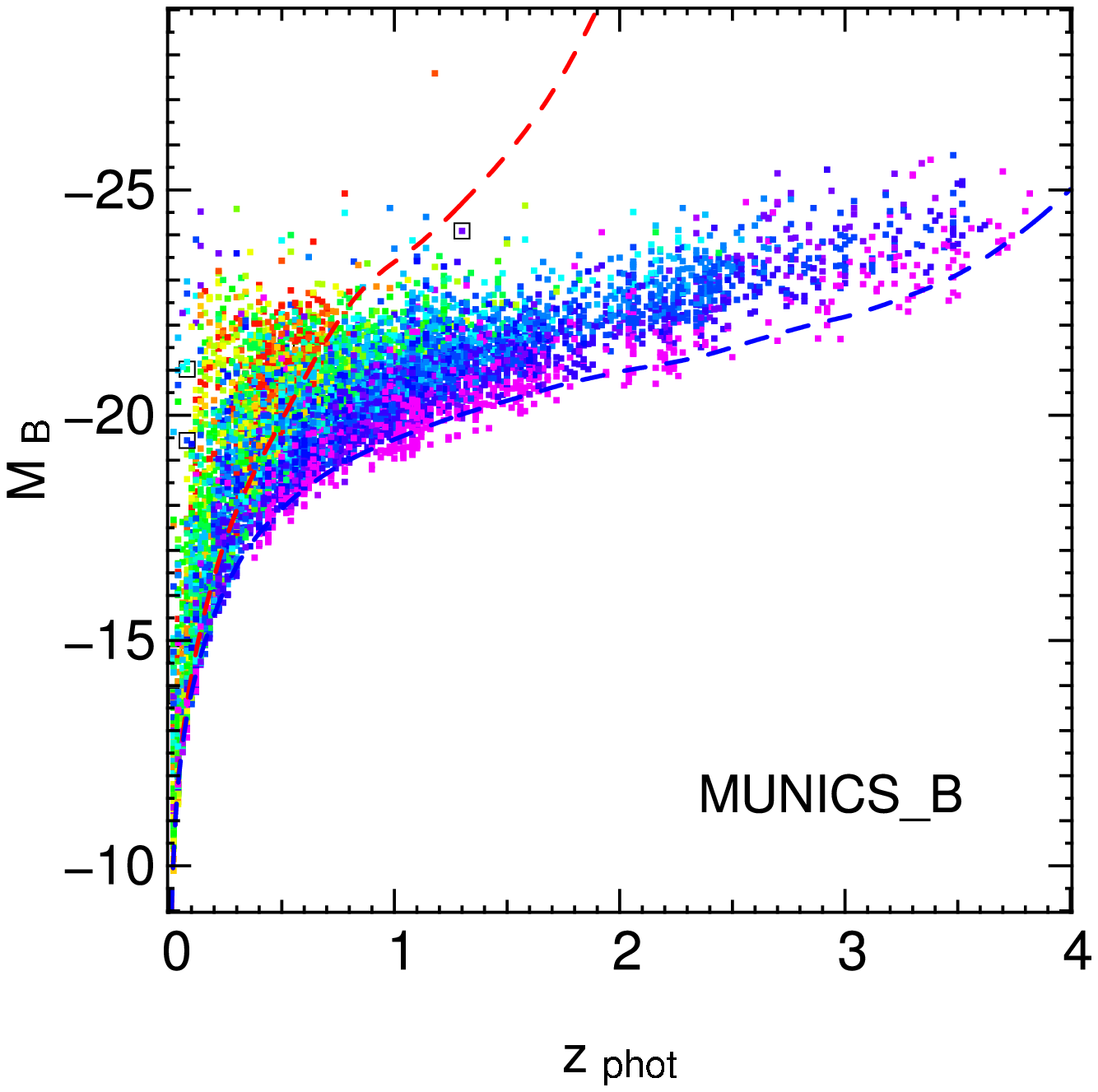,width=0.4\textwidth}
\hspace*{1cm}
\epsfig{figure=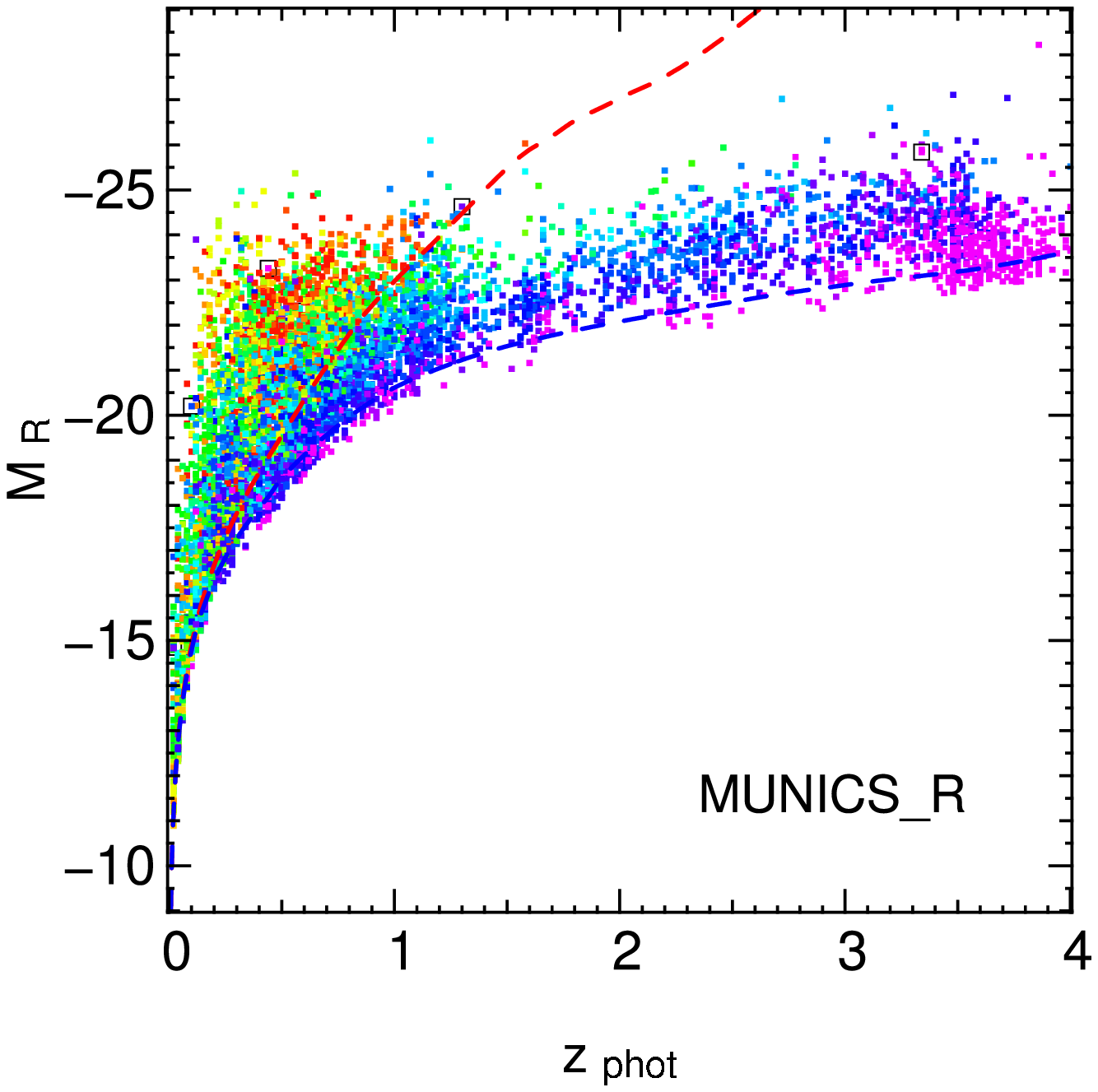,width=0.4\textwidth}

\epsfig{figure=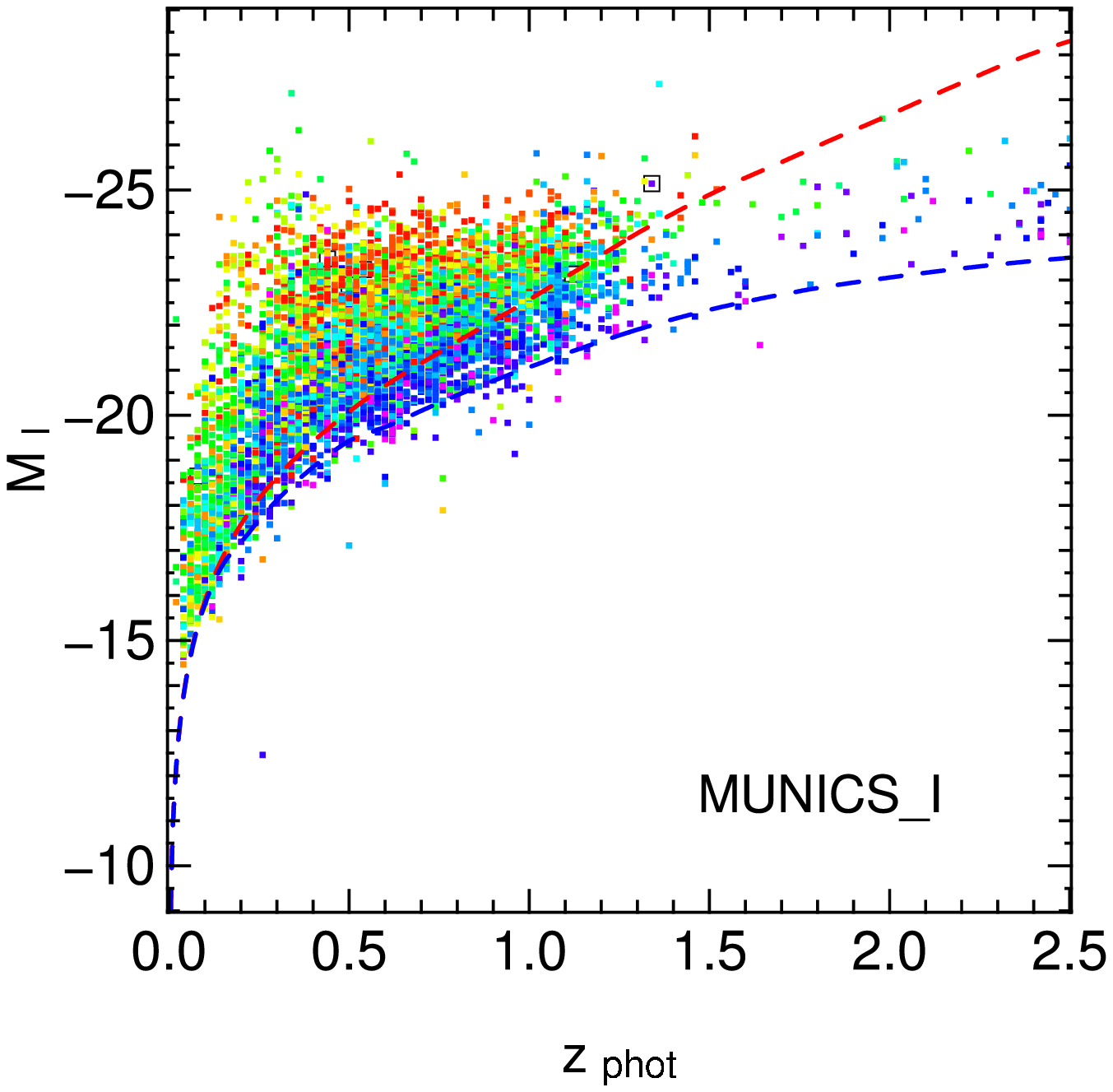,width=0.4\textwidth}
\hspace*{1cm}
\epsfig{figure=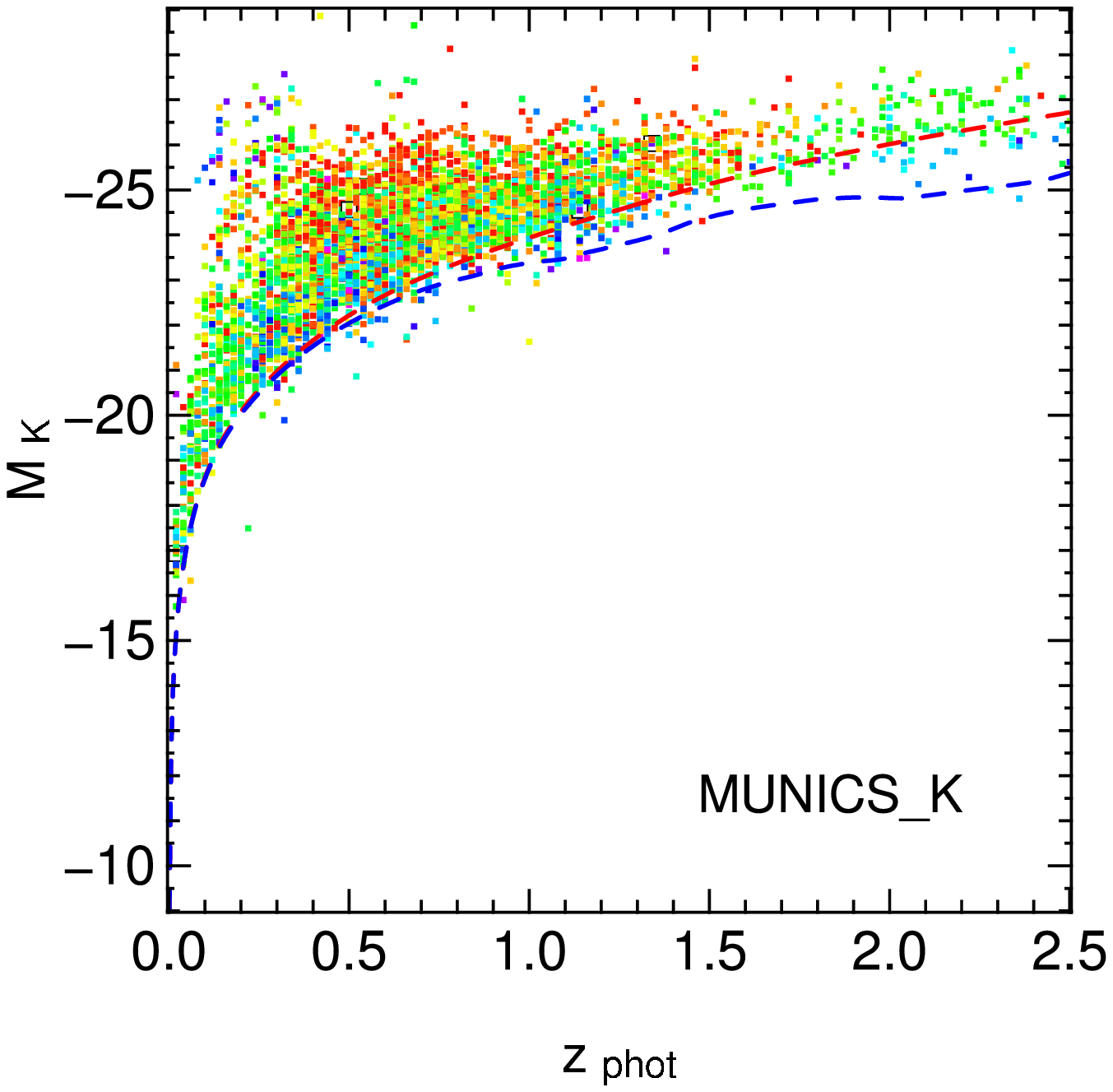,width=0.4\textwidth}

\caption{Plots of the absolute magnitudes $M_B$, $M_R$, $M_I$, and
  $M_K$ versus the photometric redshift $z_\mathrm{phot}$ for
  MUNICS\_B (\textit{upper left-hand panel}), MUNICS\_R (\textit{upper
  right-hand panel}), MUNICS\_I (\textit{lower left-hand panel}) and
  MUNICS\_K (\textit{lower right-hand panel}), respectively. The
  different colours denote the different model SEDs, where the colour
  corresponds roughly to the model colour, i.e.\ early-type galaxies
  are shown in red, late-type galaxies in blue. Objects
  spectroscopically classified as active galactic nuclei (AGN) are
  additionally marked as open squares. The dashed lines represent the
  expected absolute magnitudes as a function of redshift for an
  early-type SED (\textit{red dashed line}) and a very late type SED
  (\textit{blue dashed line}) at the limiting magnitudes of the
  samples, i.e.\ $B = 24.5$, $R = 23.5$, $I = 22.5$, and $K' = 19.5$.}
\label{f:BRIKz}

\end{figure*}

\begin{figure*}

\epsfig{figure=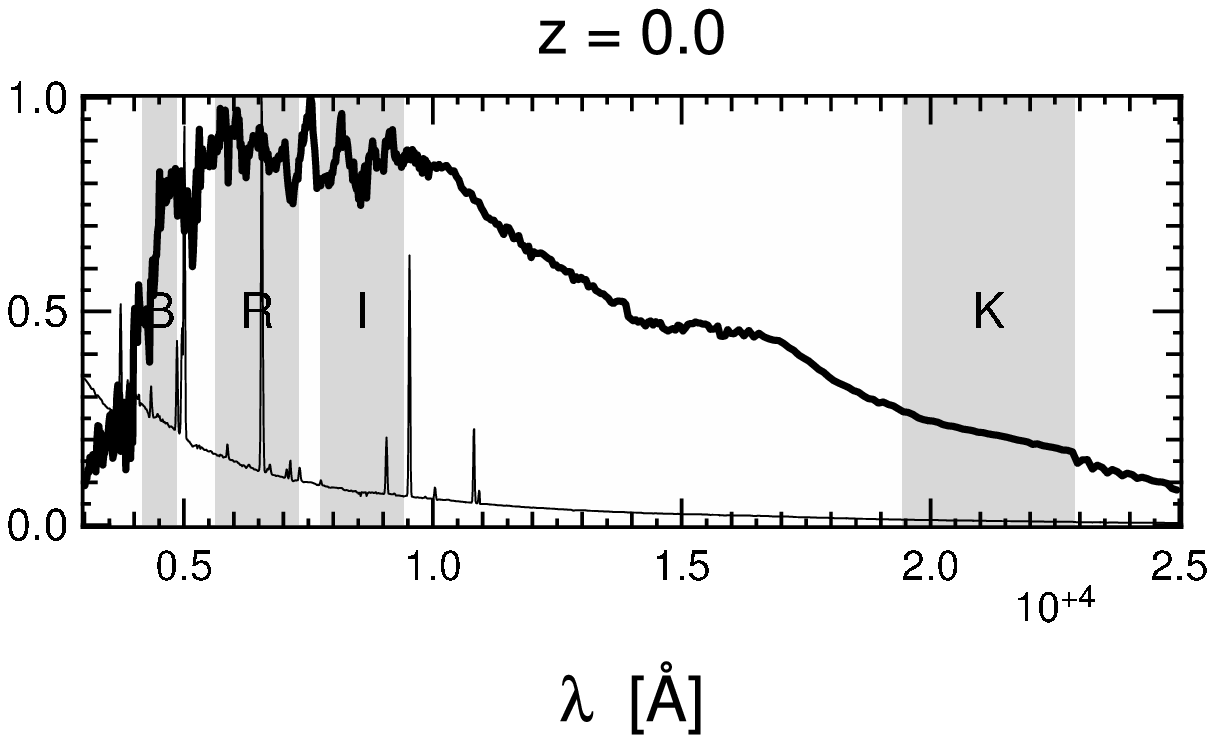,width=0.3\textwidth}
\hspace*{.5cm}
\epsfig{figure=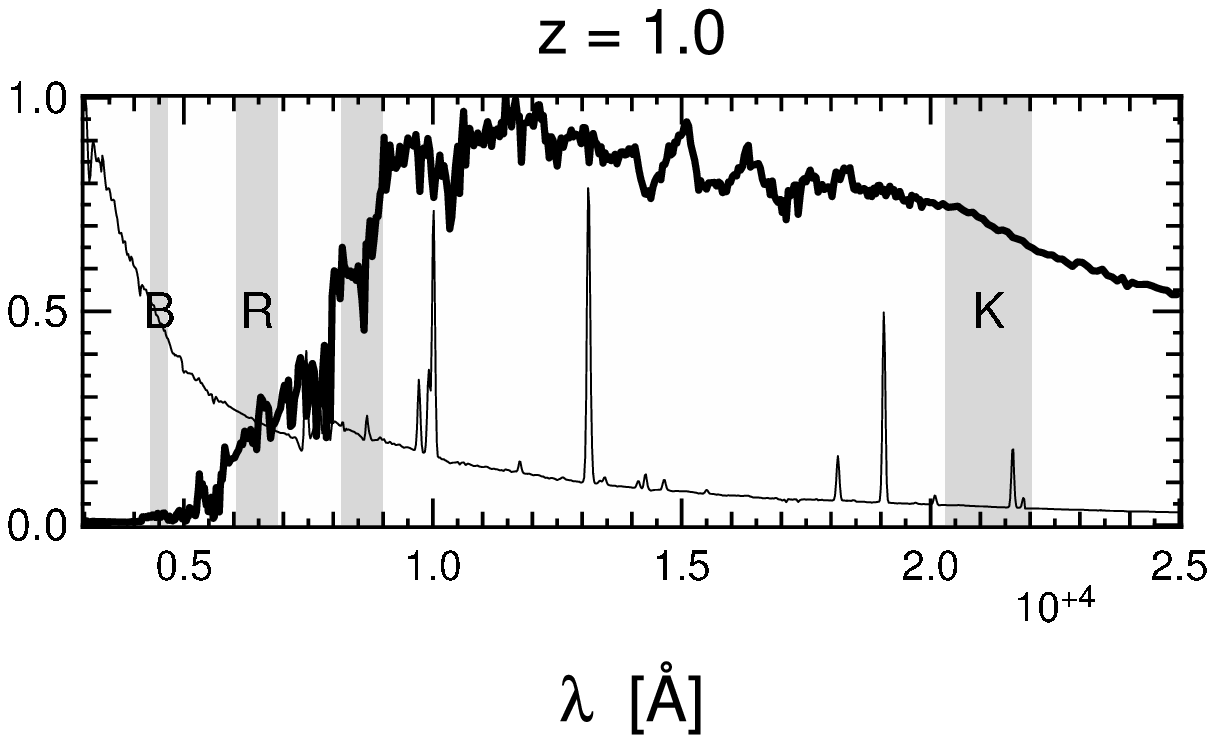,width=0.3\textwidth}
\hspace*{.5cm}
\epsfig{figure=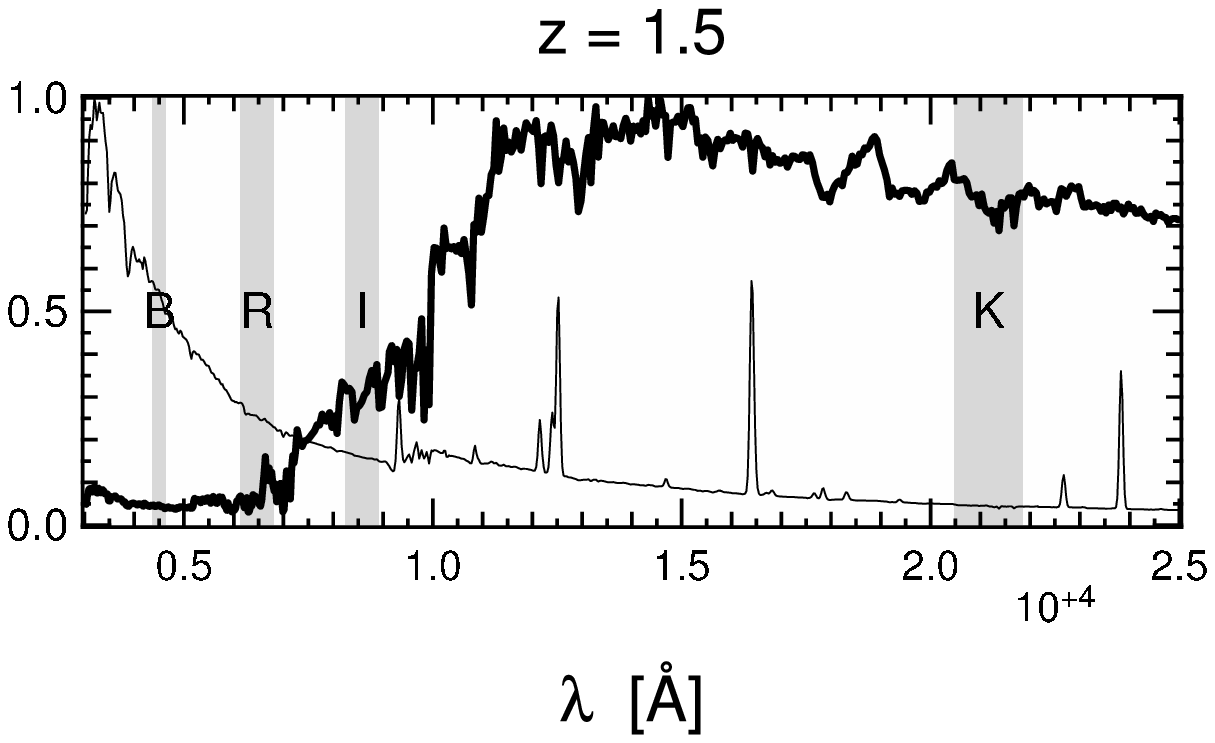,width=0.3\textwidth}

\vspace{.5cm}

\epsfig{figure=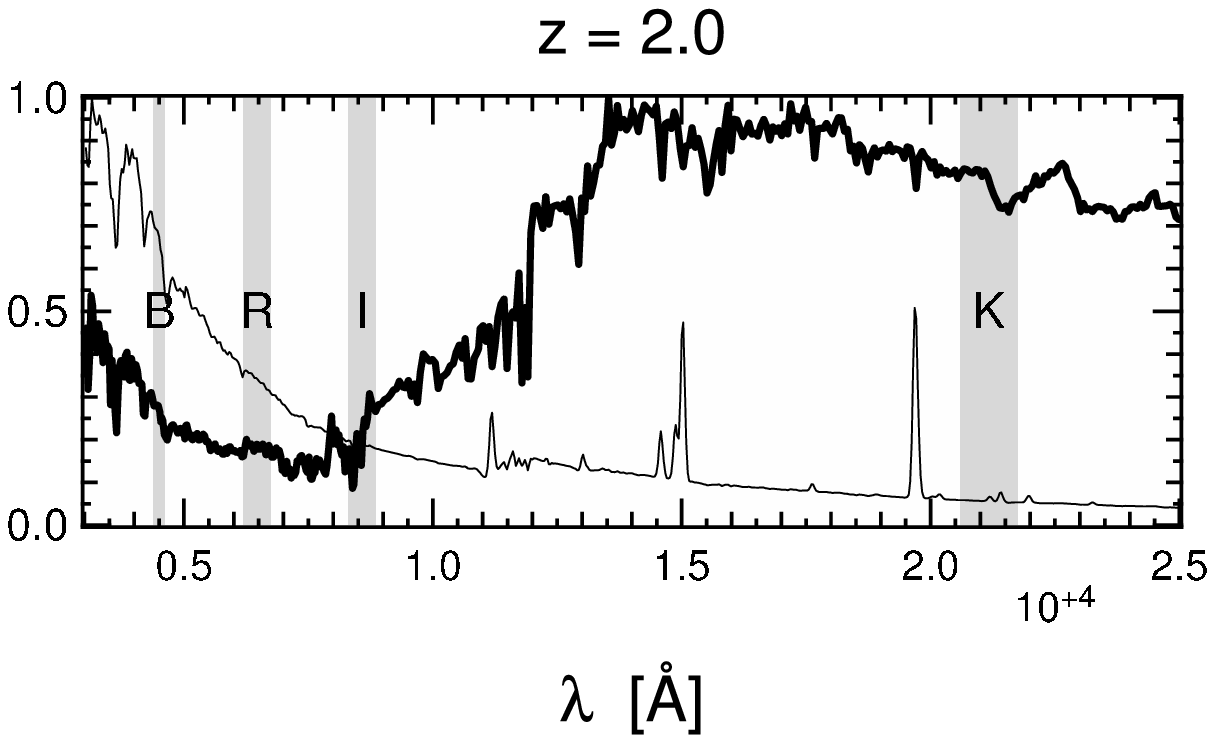,width=0.3\textwidth}
\hspace*{.5cm}
\epsfig{figure=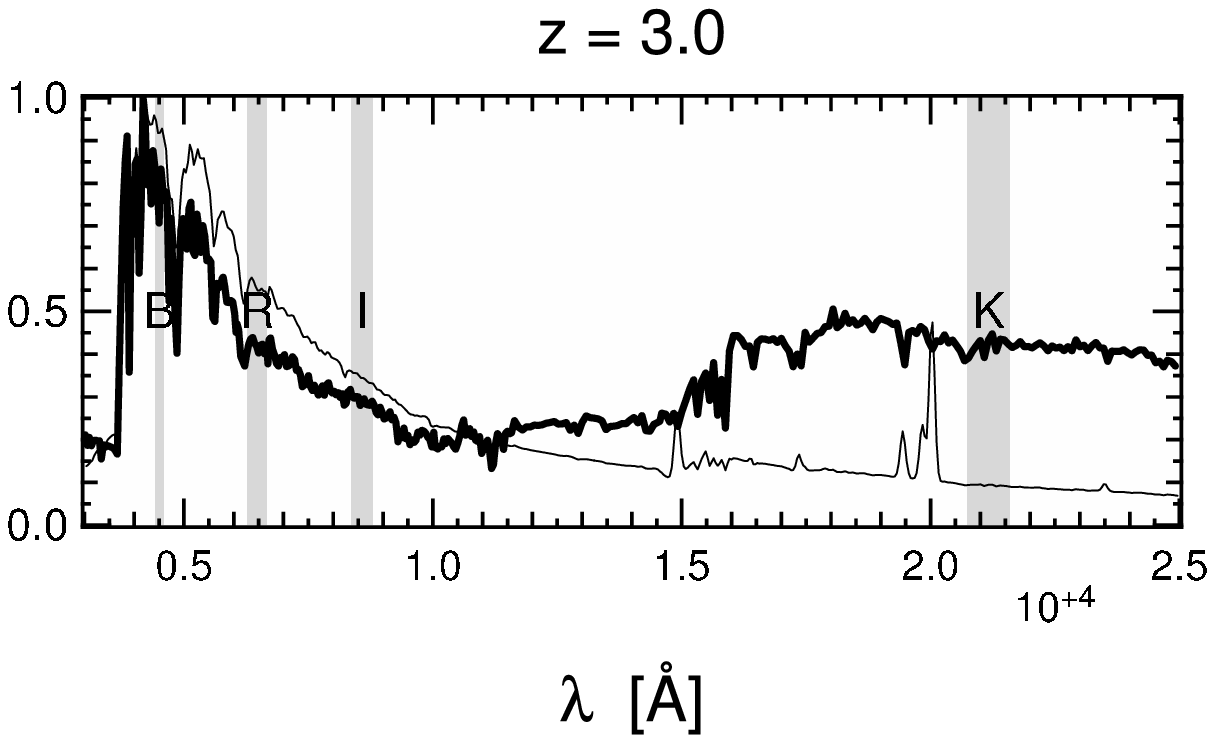,width=0.3\textwidth}
\hspace*{.5cm}
\epsfig{figure=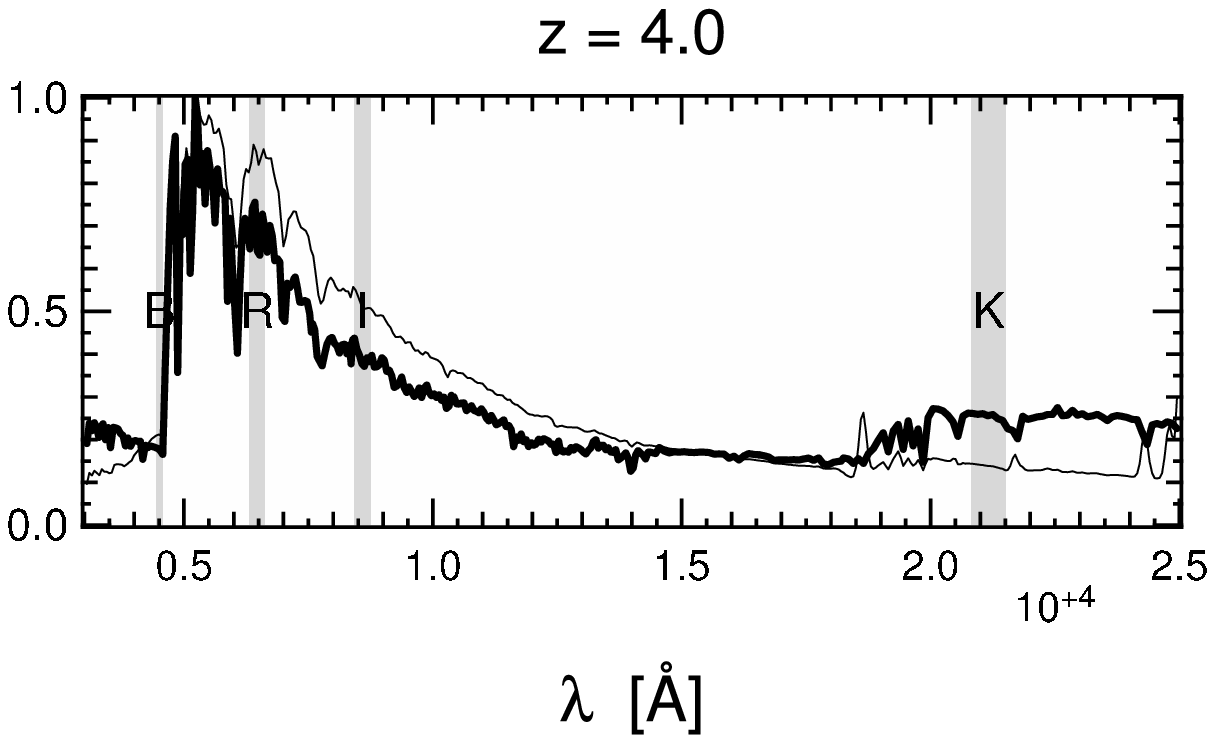,width=0.3\textwidth}

\caption{Redshifted spectral energy distributions (arbitrarily
  normalised) of an early-type galaxy (\textit{thick line}) and a
  star-burst galaxy (\textit{thin line}) and the selection filters
  $B$, $R$, $I$, and $K$ (shaded areas) for various redshifts. The
  early-type SED is constructed from \citet{BC2003} models
  and has an age comparable to the age of the universe at each
  redshift.}
\label{f:selection}

\end{figure*}

\subsection{Photometry}

Photometry was done in elliptical apertures the shape of which was
determined from the first and second moments of the light distribution
in the detection image, as described in \citet{yoda}, and
additionally in fixed size circular apertures of 5 and 7 arc seconds
diameter. To ensure measurement at equal physical scales in every
pass-band, the individual frames were convolved to the same seeing
FWHM, namely that of the image with the worst seeing in each
field.

\begin{figure*}

\epsfig{figure=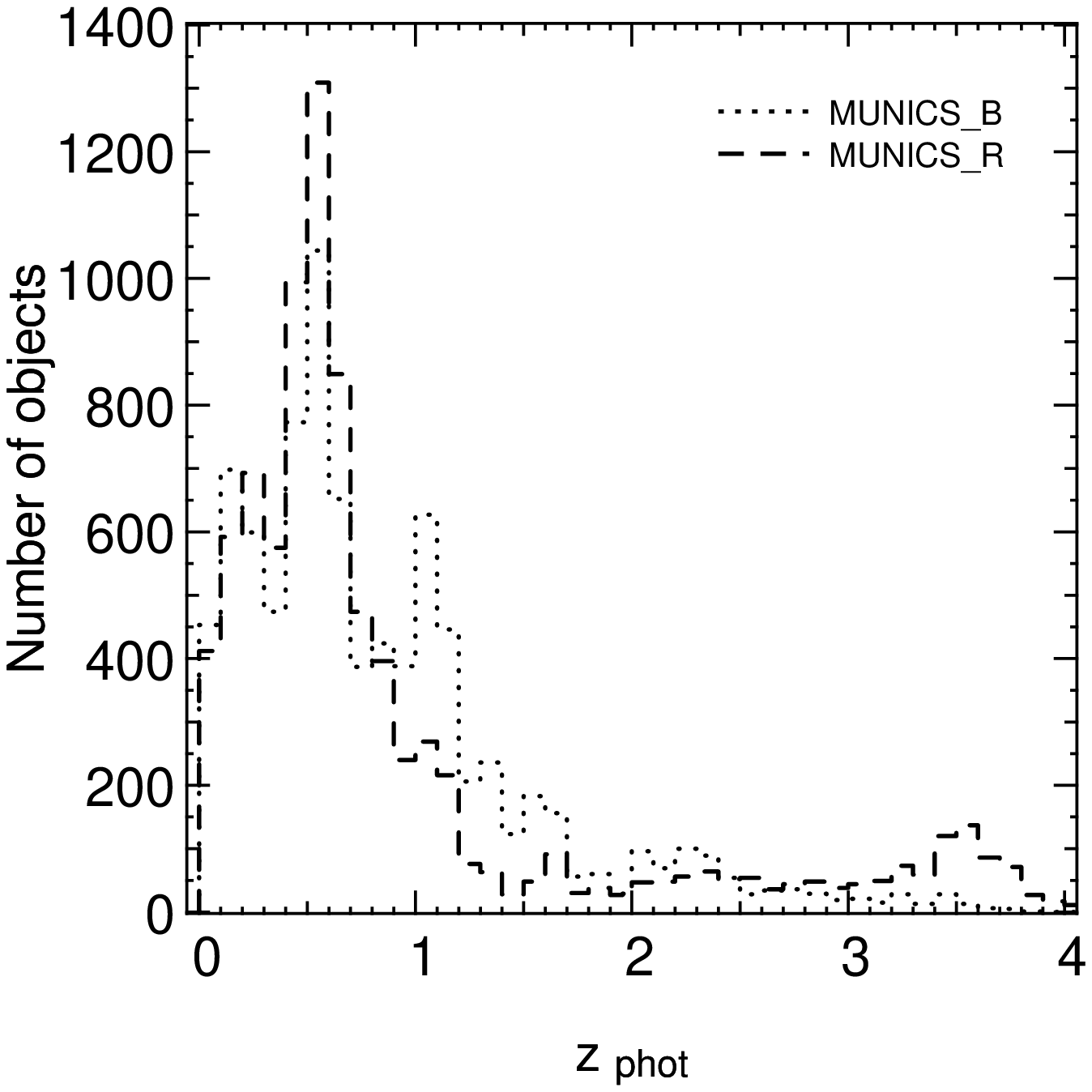,width=0.4\textwidth}
\hspace*{1cm}
\epsfig{figure=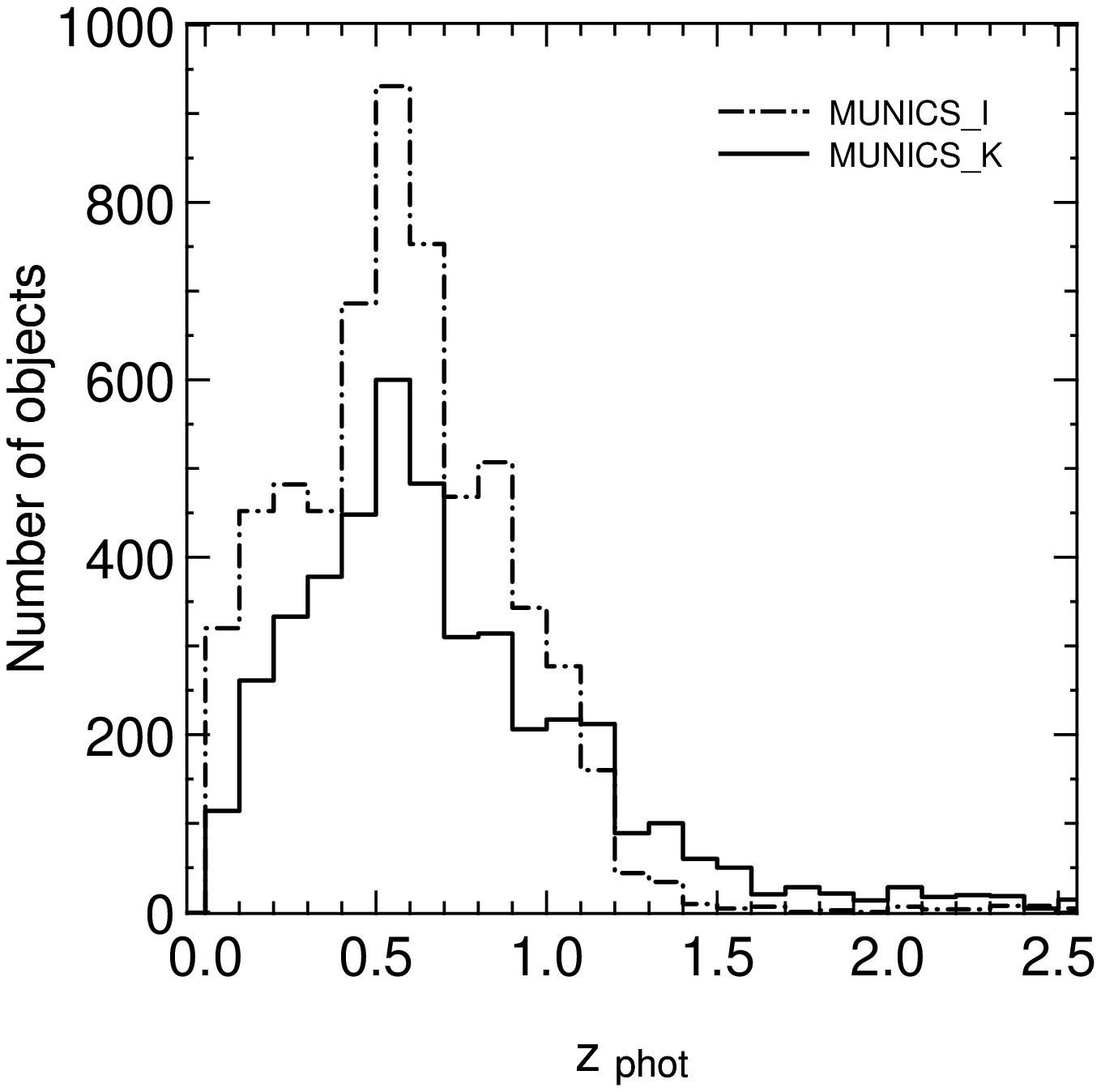,width=0.4\textwidth}

\caption{Histograms of photometric redshifts for MUNICS\_B
  (\textit{dotted line}), MUNICS\_R (\textit{dashed
      line}), MUNICS\_I (\textit{dot-dashed line}), and
  MUNICS\_K (\textit{solid line}).}
  \label{f:pzhist}

\end{figure*}

Aperture fluxes and magnitudes were computed for each object present
in the $B$, $R$, and $I$-band catalogues irrespective of a detection in
any other band. For this purpose the centroid coordinates of the
sources found in the detection images were transformed to the other
frames using geometric transformations between the image coordinate
systems. For this purpose, the images in the other five filters of
each field were registered against the image in the detection band by
matching the positions of \mbox{$\sim 200$} bright homogeneously
distributed objects in the frames and determining the coordinate
transform from the detection system to each image in the other five
pass-bands using the tasks \textsc{xyxymatch} and \textsc{geomap}
within \textsc{iraf}. The scatter in the determined solutions is less
than 0.1 pixels RMS in the transformation from $I$ or $B$ to the other
optical bands, and less than 0.2 pixels RMS from $I$ or $B$ to the
near-infrared frames. Note that the frames themselves are not
transformed to avoid artefacts from the re-sampling. We only determine
accurate transformations and apply these later to the apertures in the
photometry process, where the shapes of the apertures were transformed
using only the linear terms of the transformation.

As a result of this method, small differences in object centring and
differences in the object's shape in the different detection images
lead to slightly different photometry in the various
catalogues. Nevertheless, a direct comparison of magnitudes in the same
filter, but derived in the different selection catalogues is an
important consistency check. The comparison shows excellent agreement
between MUNICS\_B, MUNICS\_R, MUNICS\_I, and MUNICS\_K
\citep[comparison plots are published in][]{feulnerphd}.

\subsection{Photometric Redshifts}

Photometric redshifts are derived using the method presented in
\citet{photred}, a template matching algorithm rooted in Bayesian
statistics closely resembling the method presented by
\citet{Benitez00}. The application of this method to the $K$-selected
MUNICS sample is described in great detail in \citet{munics2}.

The left-hand panel of Figure~\ref{f:model-seds} shows the final
template spectral energy distribution (SED) library used to derive
photometric redshifts in what follows. It is the same set of SEDs also
used for the $K$-band selected catalogue, since these SEDs already
cover a vast range from very red old models to young star-bursting
models. As we show in the right-hand panel of
Figure~\ref{f:model-seds}, the difference between the differently
selected catalogue lies then in the distribution of selected SEDs:
While in the $K$-selected sample the algorithm picks preferentially
red galaxy types (and only very few heavily star-forming objects), the
distribution for the $I$- and $R$-selected samples is more balanced,
and indeed reversed for the $B$-selected sample.

Figure~\ref{f:zz} compares photometric and spectroscopic redshifts for
all $\sim 600$ objects with spectroscopic redshifts within the MUNICS
fields and shows the distribution of redshift errors.  The typical
scatter in the relative redshift error $\Delta z / (1+z)$ is
0.057. This is very similar to the results for the $K$-band selected
catalogue described in \citet{munics2}. The mean redshift bias is
negligible.  The distribution of the errors is roughly Gaussian. There
is no visible difference between the distributions among the survey
fields. Although this performance is encouraging, it is important to
say that the spectroscopic data become sparse at $z \ga 0.6$ and there
are only very few spectroscopic redshifts at $z > 1$. Note that many
quasars are among the few dramatic outliers which is to be expected
from the power-law like SEDs of active galactic nuclei.

Many objects in the MUNICS fields are, of course, detected in more
than one band. As a consistency check, we present comparisons of the
photometric redshifts of these objects derived in different catalogues
in Appendix~\ref{s:pzpz}, showing excellent agreement.

In Figure~\ref{f:BRIKz} we show the distribution of absolute $B$, $R$,
$I$, and $K$ magnitude versus photometric redshift for the four MUNICS
samples and for different model SEDs, ranging from early types (redder
colours) to late types (bluer colours). We also show the expected
curve for a early-type model SED (red dashed line) and a very late
type model SED (blue dashed line) at the limiting magnitudes of the
samples, i.e.\ $B = 24.5$, $R = 23.5$, $I = 22.5$, and $K' = 19.5$.

The object distribution is generally in very good agreement with the
expectations. Outlying objects with suspiciously high luminosities are
likely to be partly photometric-redshift outliers and active galactic
nuclei (AGN). Indeed, spectroscopically identified AGN are marked in
the Figure, and they are found in this region of the diagram.  Note
that the spectroscopic follow-up of MUNICS was pre-selected against
point sources \citep[see][]{munics5}, and that the existing AGN
spectra are mostly drawn from a sample of AGN selected for their red
colour (Wisotzki et al., in preparation). The lack of objects with $z
\ge 1.5$ in MUNICS\_I seems puzzling at first. From their apparent
magnitudes we should be able to detect blue galaxies at these
redshifts. To investigate this apparent failure, we have selected
objects in this redshift range from MUNICS\_R and plotted their
position in the $I$-band images. It turns out that these objects are
simply not detected due to the higher noise level in the $I$-band
images caused by fringing. The higher noise level reduces the overall
depth of the sample and affects galaxies at all redshift. Looking at
this effect in terms of the luminosity function (LF) at different
redshifts, the incompleteness cuts away the faint end of the LF in
each redshift bin, with the limit moving to higher luminosities with
increasing redshift. At redshift $z \sim 1.5$ the $I$-band magnitude
limit of MUNICS\_I reaches the bright end of the LF, i.e.\ only very
few galaxies in this redshift range are detected. We have verified
this by taking the $R$-selected catalogue and plotting that $I$-band
magnitude histogram of all galaxies beyond $z = 1.5$ which indeed
falls below the $I$-band detection limit. Also, the much deeper
$I$-selected FDF catalogue \citep{FDF1_short} cut at the MUNICS\_I
limit shows a redshift distribution similar to the one for
MUNICS\_I. Note that this lack of galaxies beyond $z \sim 1.5$ in the
$I$-selected sample does not cause any problems as long as one
excludes these higher redshifts from any analysis.

The distribution of SED types in Figure~\ref{f:BRIKz} clearly
illustrates the importance of selection effects at higher redshift:
While the $K$-selected sample traces the population of early-type
galaxies out to redshifts beyond $z \simeq 1.5$ (at the limiting
magnitudes of MUNICS; indeed, that is how the survey was designed), it
lacks late-type galaxies at all redshifts. The $B$-selected sample on
the other hand does not contain early-type galaxies beyond $z \simeq
0.8$ (again at the limiting magnitudes of MUNICS), but is much better
suited for tracing late-type galaxies at low and high redshift.

These selection effects can be easily understood by looking at the
position of redshifted SEDs with respect to the four detection filters
as shown in Figure~\ref{f:selection} where we plot redshifted SEDs for
an early-type and a star-burst galaxy. Early-type galaxies disappear
from a $B$-selected catalogue at $z \sim 1$ and from $R$ as well as
$I$-selected catalogues at $z \sim 1.5$, while they remain visible to
much higher redshifts in MUNICS\_K. Similarly, the detection of very
blue galaxies at redshifts $z \sim 3$ in MUNICS\_B and MUNICS\_R can
be explained by the strong rest-frame ultraviolet emission at
wavelengths longer than that of the Lyman break entering the detection
bands.

The photometric redshift histograms for MUNICS\_B, MUNICS\_R,
MUNICS\_I, and MUNICS\_K can be found in Figure~\ref{f:pzhist}. Note
the high redshift tails in the distributions for MUNICS\_B, MUNICS\_R,
and MUNICS\_K, caused by luminous red galaxies at $1 \; \lsim \; z \;
\lsim \; 2$ and the ultraviolet emission of blue galaxies shifted into
the optical bands, respectively.

\subsection{Star--Galaxy Separation}

For computing statistical properties of the field-galaxy population
like its luminosity function, it is necessary to remove all stars from
the analysis. In a photometric catalogue, this can be done in two
ways. Either by looking at an object's morphology in the image, i.e.\
whether it appears to be similar to the point-spread function (PSF) of
the image or extended, or by looking at the spectral energy
distribution (SED), in this case as traced by the six-filter
photometry, and comparing it to template spectra of stars and
galaxies. The disadvantage of the morphological approach is clearly
its tendency to fail at fainter magnitudes (because of the lower
signal-to-noise ratio and because distant galaxies look more and more
compact), while it works reasonably well at brighter magnitudes (see,
e.g., \citeauthor{munics1} 2001b).

For this investigation, we decided to take the second route. The
photometric redshift technique as described above compares the
photometry in the six MUNICS filters to template SEDs of stars and
galaxies. Each fit is assigned a $\chi^2$ value, and we can simply
discriminate between stars and galaxies by comparing the $\chi^2$
values of the best-fitting stellar and the best-fitting galactic
SED. More specifically, we classify all objects as stars for which
$\chi^2_\mathrm{star} < \chi^2_\mathrm{galaxy}$.

\begin{figure}

\centerline{\epsfig{figure=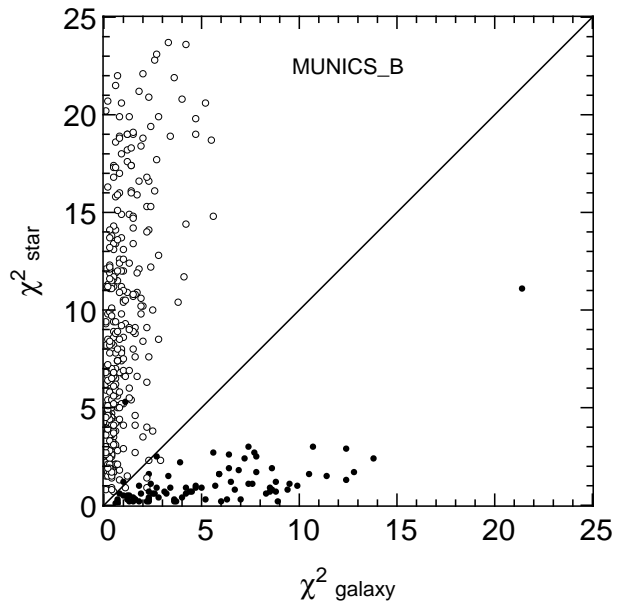,width=0.45\textwidth}}

\caption{This diagram shows the loci of spectroscopically classified
  galaxies (open circles) and stars (crosses) in a plane defined by
  the $\chi^2$ values of the best-fitting galactic and stellar
  spectral-energy distributions (SEDs) for MUNICS\_B. The line used to
  discriminate the two classes is also indicated. The
    corresponding diagrams for MUNICS\_R, MUNICS\_I, and MUNICS\_K
    look very similar.}
\label{f:stargalsep}

\end{figure}

This procedure can be tested by checking against the spectroscopic
sample described in \citet{munics5}. This is illustrated in
Figure~\ref{f:stargalsep}, where we show the $\chi^2$ values of
objects spectroscopically classified as stars or galaxies,
respectively, for the $B$-selected catalogue. The diagrams for the
other samples are very similar. Clearly, the SED classifier does work
very well. Note that in all comparisons between MUNICS\_B, MUNICS\_R,
MUNICS\_I, and MUNICS\_K presented in this paper we have used the same
$\chi^2$-based classification method.

%
% GALAXY NUMBER COUNTS
%
\section{Galaxy Number Counts}
\label{s:nc}

\begin{figure*}

\epsfig{figure=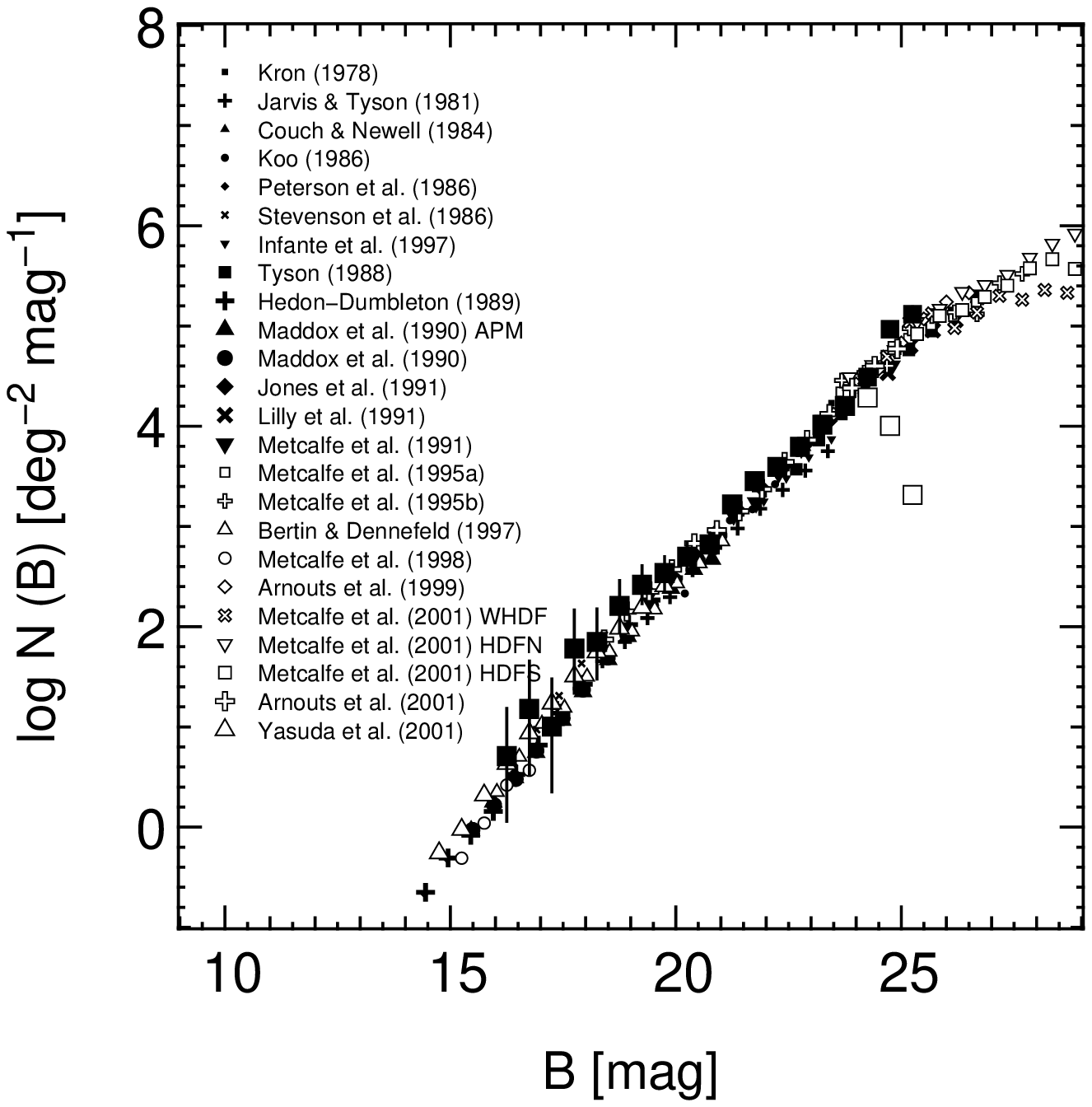,width=0.4\textwidth}
\hspace*{1cm}
\epsfig{figure=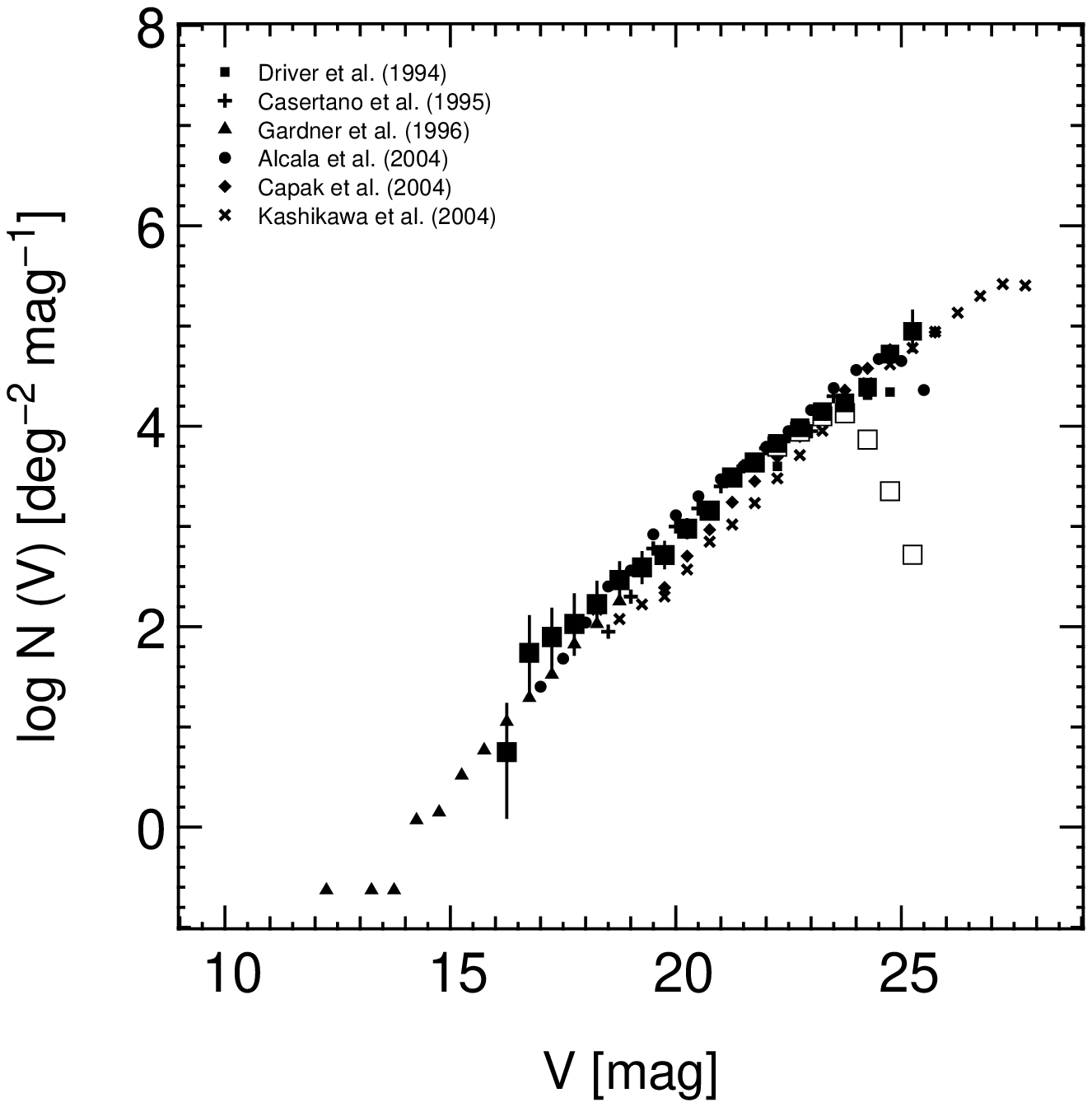,width=0.4\textwidth}

\vspace*{.5cm}

\epsfig{figure=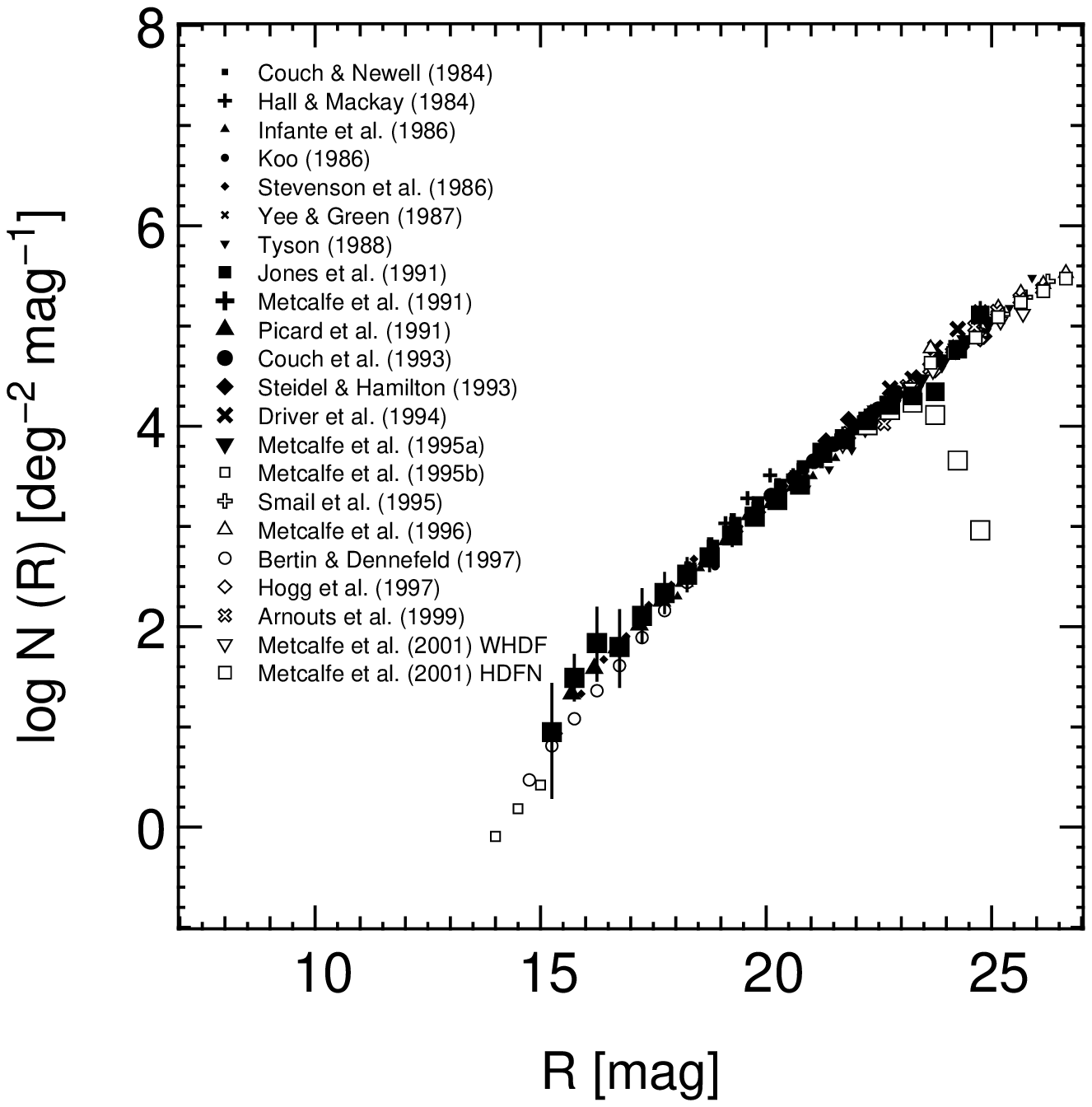,width=0.4\textwidth}
\hspace*{1cm}
\epsfig{figure=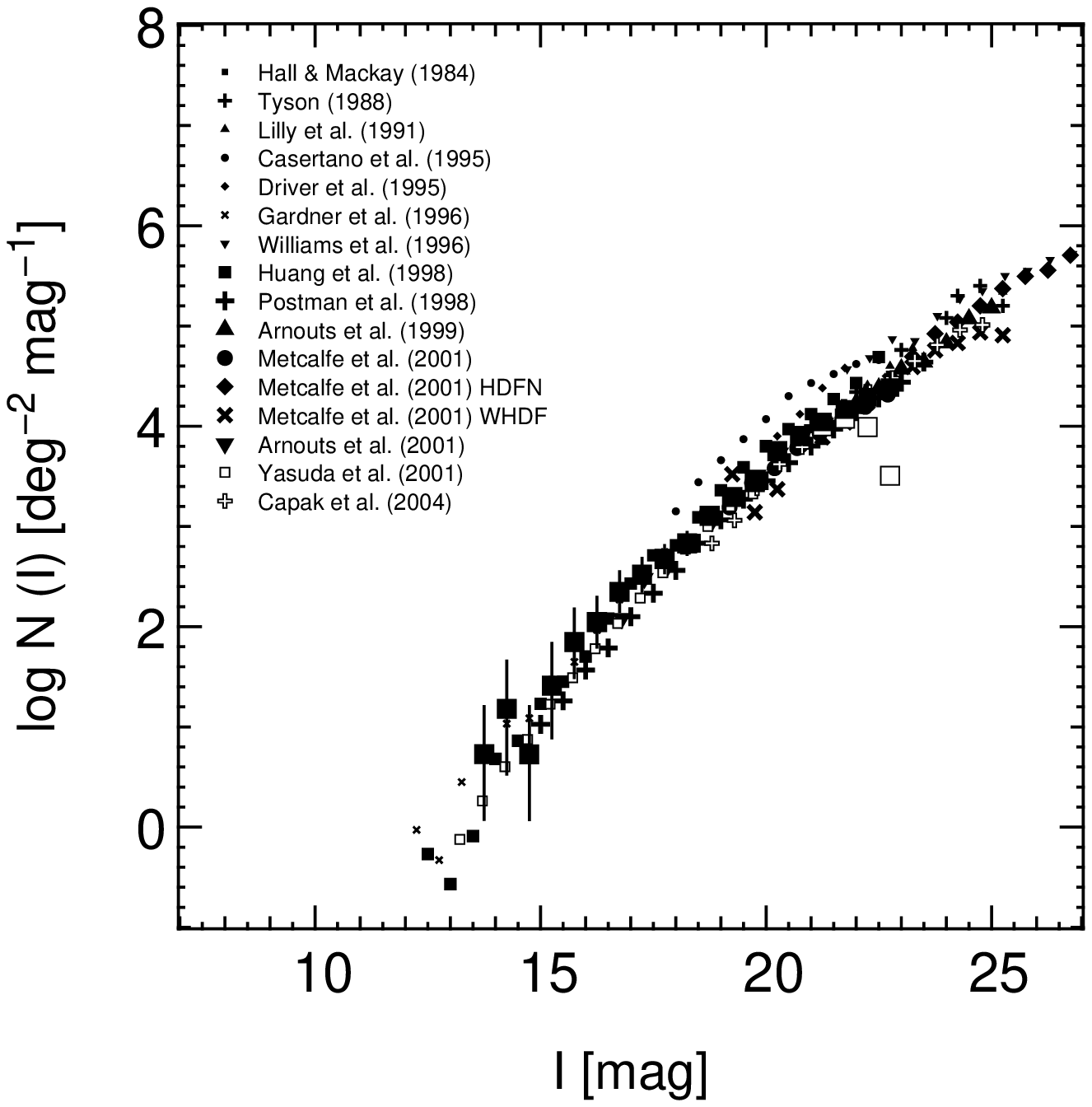,width=0.4\textwidth}

\vspace*{.5cm}

\epsfig{figure=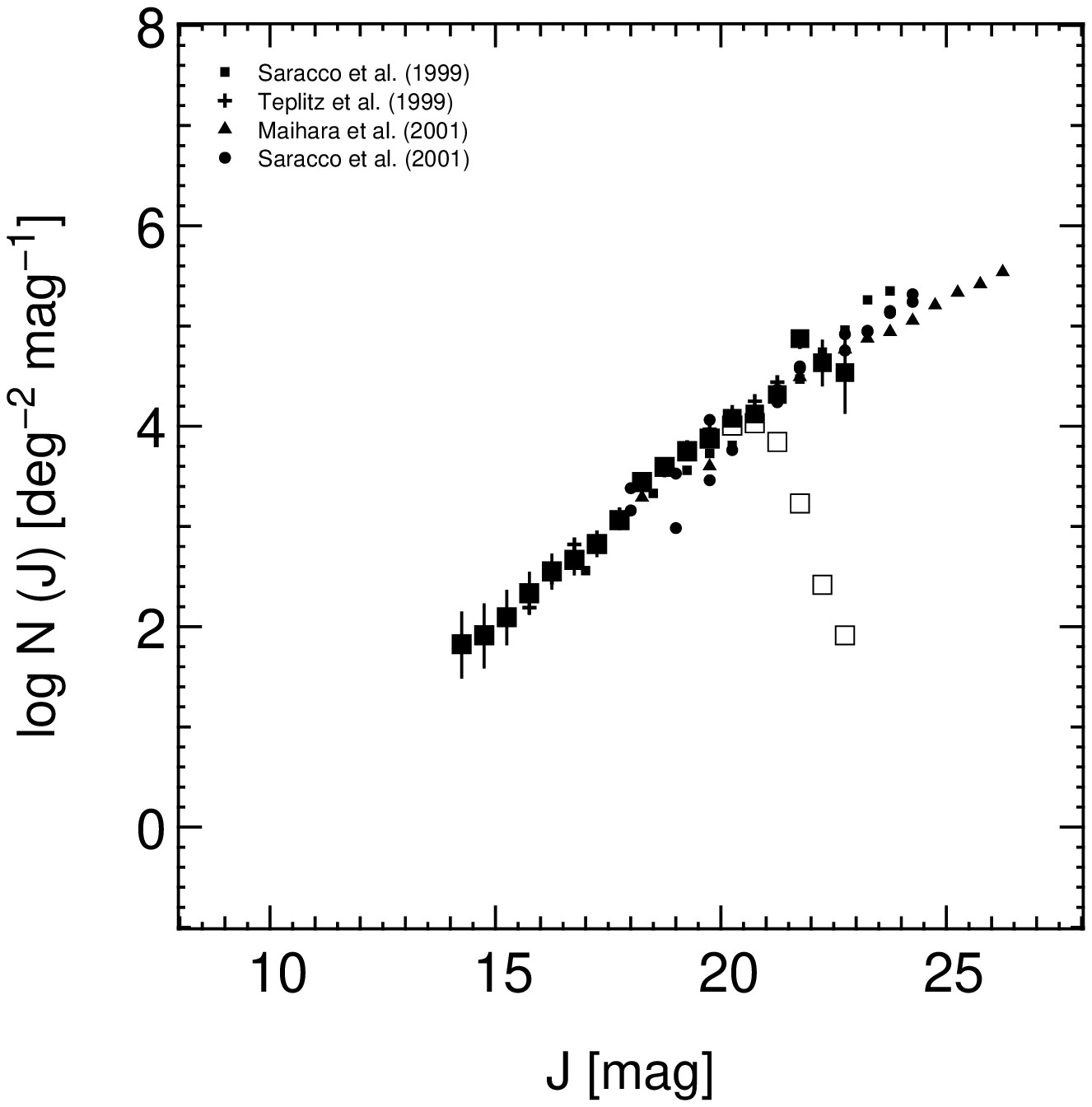,width=0.4\textwidth}
\hspace*{1cm}
\epsfig{figure=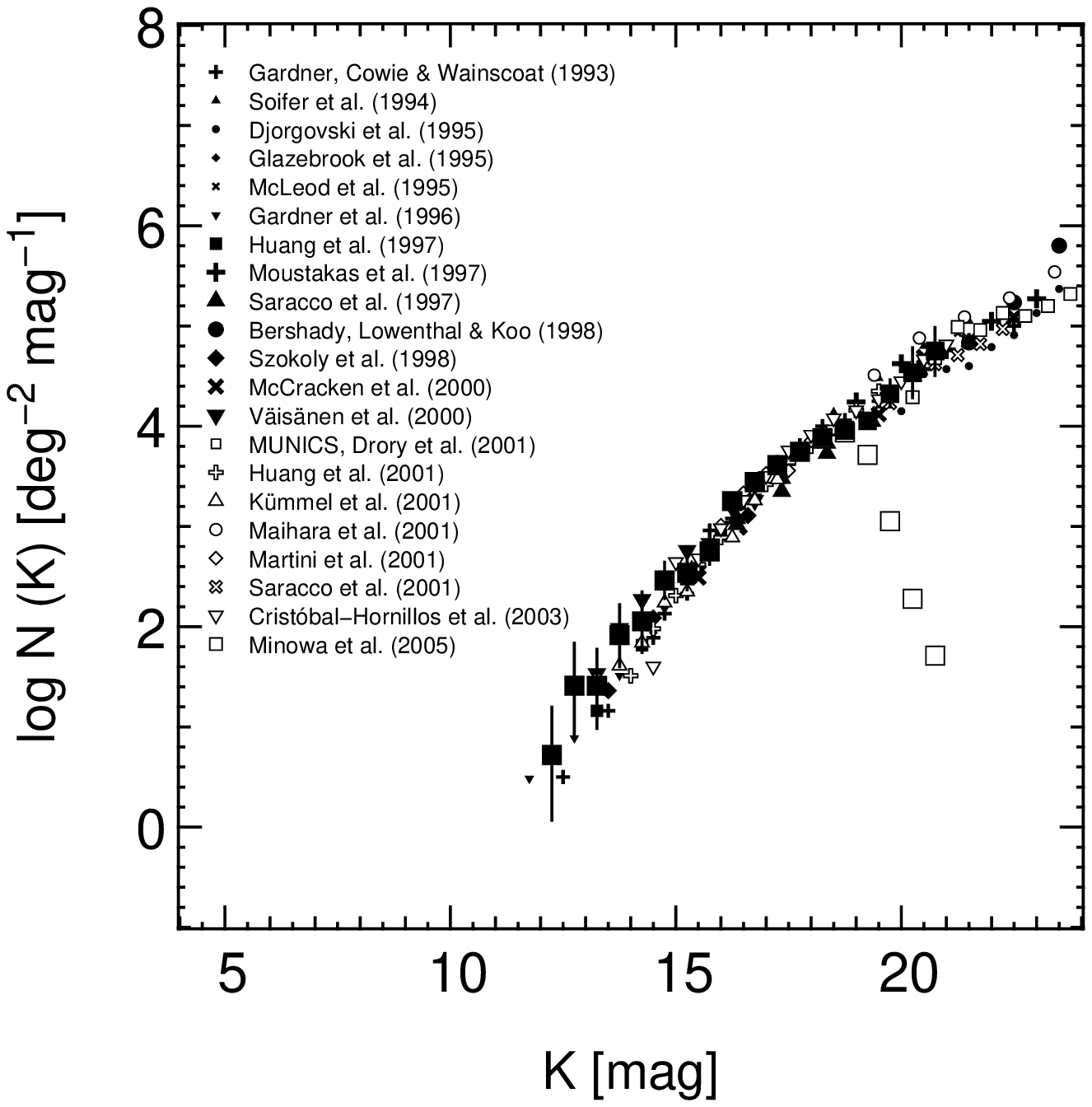,width=0.4\textwidth}

\caption{Galaxy number counts from MUNICS (\textit{large squares}) in $B$
  \textit{(upper left-hand panel)}, $V$ \textit{(upper right-hand
  panel)}, $R$ \textit{(middle left-hand panel)}, $I$ \textit{(middle
  right-hand panel)}, $J$ \textit{(lower left-hand panel)} and $K$
  \textit{(lower right-hand panel)}. The filled squares indicated
  completeness-corrected counts, whereas the open squares
  represent the uncorrected values. Literature counts are shown for
  comparison.}
\label{f:nc}

\end{figure*}

Galaxy number counts, although not as widely accepted as important
tool for the study of galaxy evolution and cosmology as in the past,
still serve as an interesting probe of a survey's data quality. Since
the publication of galaxy number counts from MUNICS in
\citeauthor{munics1} (2001b), the survey has considerably increased in
size and quality. Moreover, $B$-band number counts have never been
published for the MUNICS project, so we show galaxy number counts in
all six filters $B$, $V$, $R$, $I$, $J$ and $K$ in
Figure~\ref{f:nc}. The counts were derived from detection catalogues
constructed on all filter images. In contrast to the analysis
presented later in this paper, star--galaxy separation is based on the
morphological classifier described in detail in \citeauthor{munics1}
(2001b). The reason for this is that no photometric redshift data are
available for the $V$, $R$ and $J$-band detections, so we cannot rely
on the $\chi^2$-based classification. However, the two classification
methods agree very well. Completeness-corrected galaxy number counts
from MUNICS are presented here for the first time. The completeness
correction is based on the simulations described in
Section~\ref{s:det}. All errors given are Poissonian errors.

The galaxy number counts presented in Figure~\ref{f:nc} for all six
MUNICS filters show excellent agreement with existing data and clearly
demonstrate the quality of our dataset. The completeness-corrected
values for the number counts are also summarised in Table~\ref{t:nc}.

% B-band literature counts:
\nocite{MSFJ91, AARS, MFS95, MSCMF2001, BD97, ADCZFG99, MSFR95, Koo86,
CN84, IPQ86, SSF86, Tyson88, LCG91, HCM89, Kron78, JFSEP91, FOCAS81,
MSELP90, Yasuda2001s, MRSF98, APM90, EISCDFS2001}

% V-band literature counts:
\nocite{DPDMD94, CRGINOW95, gardner1996, Alcala2004s, Capak2004,
Kashikawa2004s}

% R-band literature counts:
\nocite{CN84, HM84, IPQ86, Koo86, SSF86, YG87, Tyson88, JFSEP91,
MSFJ91, P91, CJB93, Steidel1993, DPDMD94, MFS95, MSFR95, SHYC95, MSCFG96,
BD97, HPMCBSS97, ADCZFG99, MSCMF2001}

% I-band literature counts:
\nocite{HM84, Tyson88, LCG91, CRGINOW95, DWOKGR95, HDFN, HCL98, PL98,
ADCZFG99, MSCMF2001, EISCDFS2001, Yasuda2001s, Capak2004}

% J-band literature counts:
\nocite{Saracco99, saracco2001, maihara2001s, Teplitz99}

% K-band literature counts:
\nocite{gardner1993, soifer1994, djorgovski1995, glazebrook1995,
mcleod1995, gardner1996, huang1997, moustakas1997, saracco1997,
bershady1998, szokoly1998, mccracken2000, vaisanen2000, munics1,
huang2001, kuemmel2001, maihara2001s, martini2001, saracco2001,
cristobal2003, minowa2005}

\begin{table*}

\caption{Completeness corrected galaxy number counts from MUNICS in the
$B$, $V$, $R$, $I$, $J$ and $K$ bands. $\log N$ and $\sigma_{\log N}$ are 
given, where $N$ is in units of $\mathrm{mag}^{-1} \mathrm{deg}^{-2}$.}
\label{t:nc}
\begin{center}
\begin{tabular}{ccccccccccccc}
\hline
& $B$ & & $V$ & & $R$ & & $I$ & & $J$ & & $K$ \\
$m$ & $\log N$ & $\sigma_{\log N}$ & $\log N$ & $\sigma_{\log N}$ & $\log N$ &
$\sigma_{\log N}$ &$\log N$ & $\sigma_{\log N}$ &$\log N$ & $\sigma_{\log N}$ &$\log N$ &
$\sigma_{\log N}$ \\\hline

 12.25 &       &       &       &       &       &       &       &       &       &       &  0.72 &  0.60 \\
 12.75 &       &       &       &       &       &       &       &       &       &       &  1.41 &  0.51 \\
 13.25 &       &       &       &       &       &       &       &       &       &       &  1.41 &  0.42 \\
 13.75 &       &       &       &       &       &       &       &       &       &       &  1.92 &  0.33 \\
 14.25 &       &       &       &       &       &       &  1.18 &  0.60 &  1.82 &  0.34 &  2.05 &  0.32 \\
 14.75 &       &       &       &       &       &       &  0.73 &  0.60 &  1.91 &  0.33 &  2.46 &  0.19 \\
 15.25 &       &       &       &       &  0.95 &  0.60 &  1.41 &  0.51 &  2.09 &  0.28 &  2.54 &  0.17 \\
 15.75 &       &       &       &       &  1.49 &  0.23 &  1.85 &  0.37 &  2.33 &  0.21 &  2.75 &  0.13 \\
 16.25 &  0.71 &  0.60 &  0.75 &  0.60 &  1.84 &  0.39 &  2.05 &  0.26 &  2.55 &  0.17 &  3.25 &  0.07 \\
 16.75 &  1.18 &  0.60 &  1.74 &  0.41 &  1.80 &  0.41 &  2.35 &  0.21 &  2.67 &  0.14 &  3.44 &  0.05 \\
 17.25 &  1.00 &  0.60 &  1.90 &  0.30 &  2.11 &  0.28 &  2.52 &  0.17 &  2.82 &  0.12 &  3.61 &  0.04 \\
 17.75 &  1.78 &  0.45 &  2.03 &  0.32 &  2.34 &  0.20 &  2.68 &  0.14 &  3.06 &  0.09 &  3.74 &  0.04 \\
 18.25 &  1.85 &  0.37 &  2.22 &  0.23 &  2.52 &  0.17 &  2.83 &  0.11 &  3.44 &  0.05 &  3.89 &  0.03 \\
 18.75 &  2.21 &  0.27 &  2.47 &  0.18 &  2.69 &  0.13 &  3.11 &  0.08 &  3.59 &  0.04 &  3.96 &  0.03 \\
 19.25 &  2.42 &  0.19 &  2.59 &  0.15 &  2.91 &  0.10 &  3.29 &  0.06 &  3.75 &  0.04 &  4.05 &  0.05 \\
 19.75 &  2.53 &  0.17 &  2.72 &  0.13 &  3.10 &  0.08 &  3.44 &  0.05 &  3.88 &  0.03 &  4.32 &  0.14 \\
 20.25 &  2.70 &  0.14 &  2.98 &  0.09 &  3.26 &  0.06 &  3.75 &  0.04 &  4.08 &  0.03 &  4.53 &  0.26 \\
 20.75 &  2.82 &  0.12 &  3.16 &  0.07 &  3.42 &  0.05 &  3.90 &  0.03 &  4.12 &  0.03 &  4.75 &  0.25 \\
 21.25 &  3.22 &  0.07 &  3.49 &  0.05 &  3.73 &  0.04 &  4.04 &  0.03 &  4.32 &  0.03 &       &       \\
 21.75 &  3.45 &  0.05 &  3.64 &  0.04 &  3.87 &  0.03 &  4.17 &  0.03 &  4.87 &  0.06 &       &       \\
 22.25 &  3.59 &  0.04 &  3.83 &  0.03 &  4.05 &  0.03 &  4.29 &  0.06 &  4.63 &  0.23 &       &       \\
 22.75 &  3.79 &  0.03 &  3.99 &  0.03 &  4.21 &  0.02 &  4.39 &  0.14 &       &       &       &       \\
 23.25 &  4.02 &  0.03 &  4.15 &  0.02 &  4.30 &  0.02 &       &       &       &       &       &       \\
 23.75 &  4.20 &  0.02 &  4.23 &  0.03 &  4.34 &  0.03 &       &       &       &       &       &       \\
 24.25 &  4.49 &  0.02 &  4.39 &  0.06 &  4.77 &  0.06 &       &       &       &       &       &       \\
 24.75 &  4.97 &  0.03 &  4.72 &  0.08 &  5.11 &  0.12 &       &       &       &       &       &       \\
 25.25 &  5.12 &  0.07 &  4.95 &  0.21 &       &       &       &       &       &       &       &       \\\hline

\end{tabular}
\end{center}

\end{table*}

\begin{figure*}

\epsfig{figure=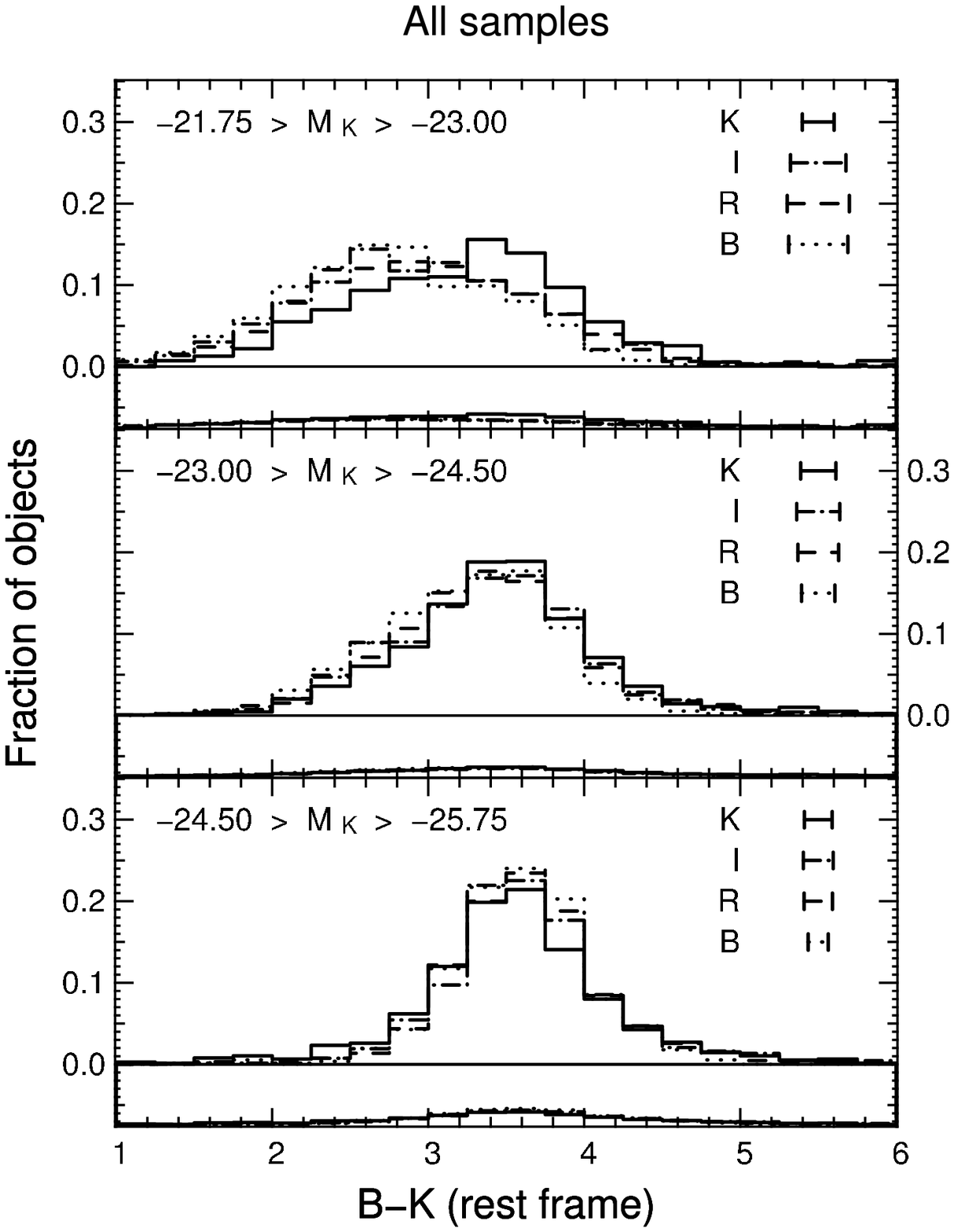,width=0.4\textwidth}
\hspace*{1cm}
\epsfig{figure=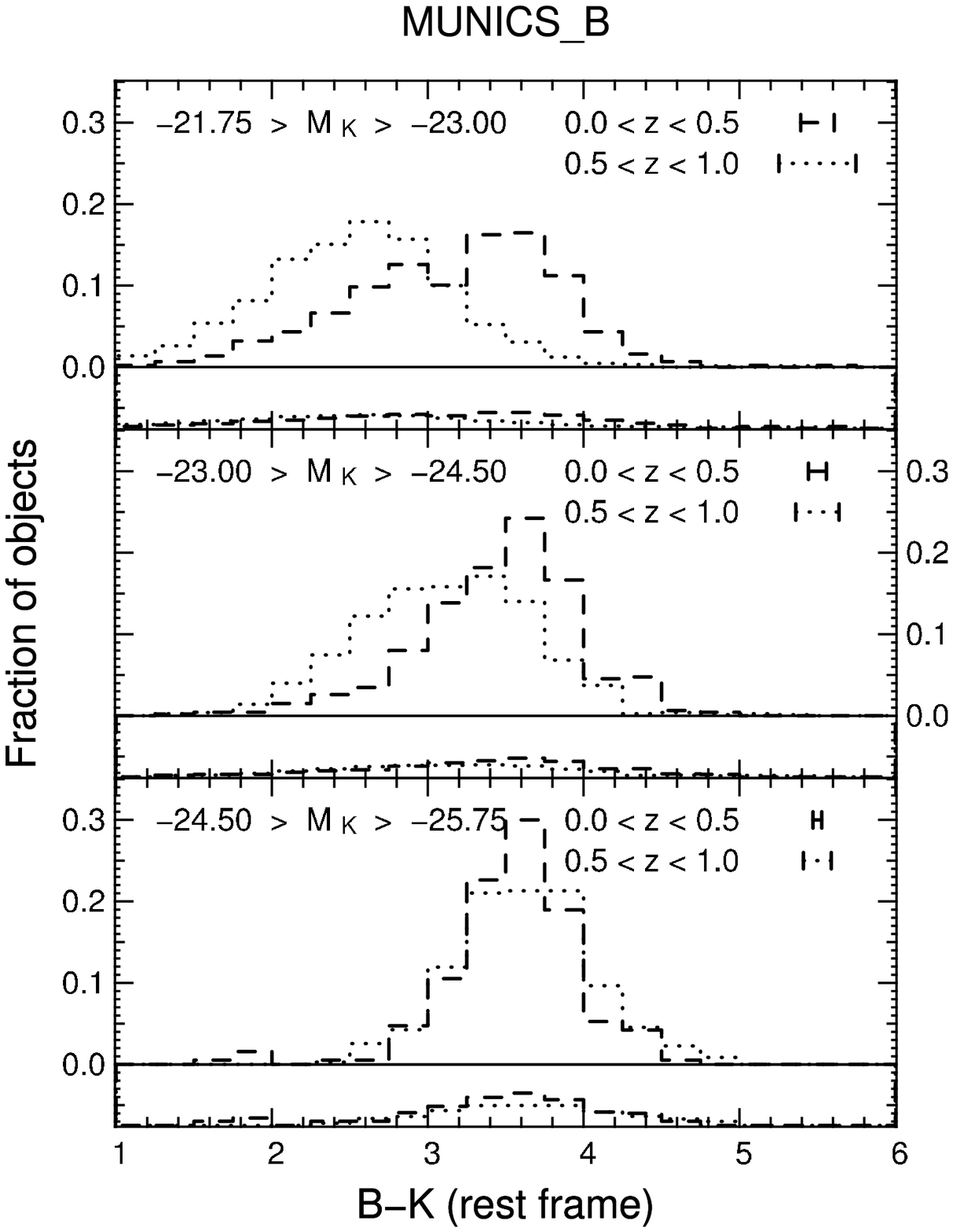,width=0.4\textwidth}

\vspace*{.5cm}

\epsfig{figure=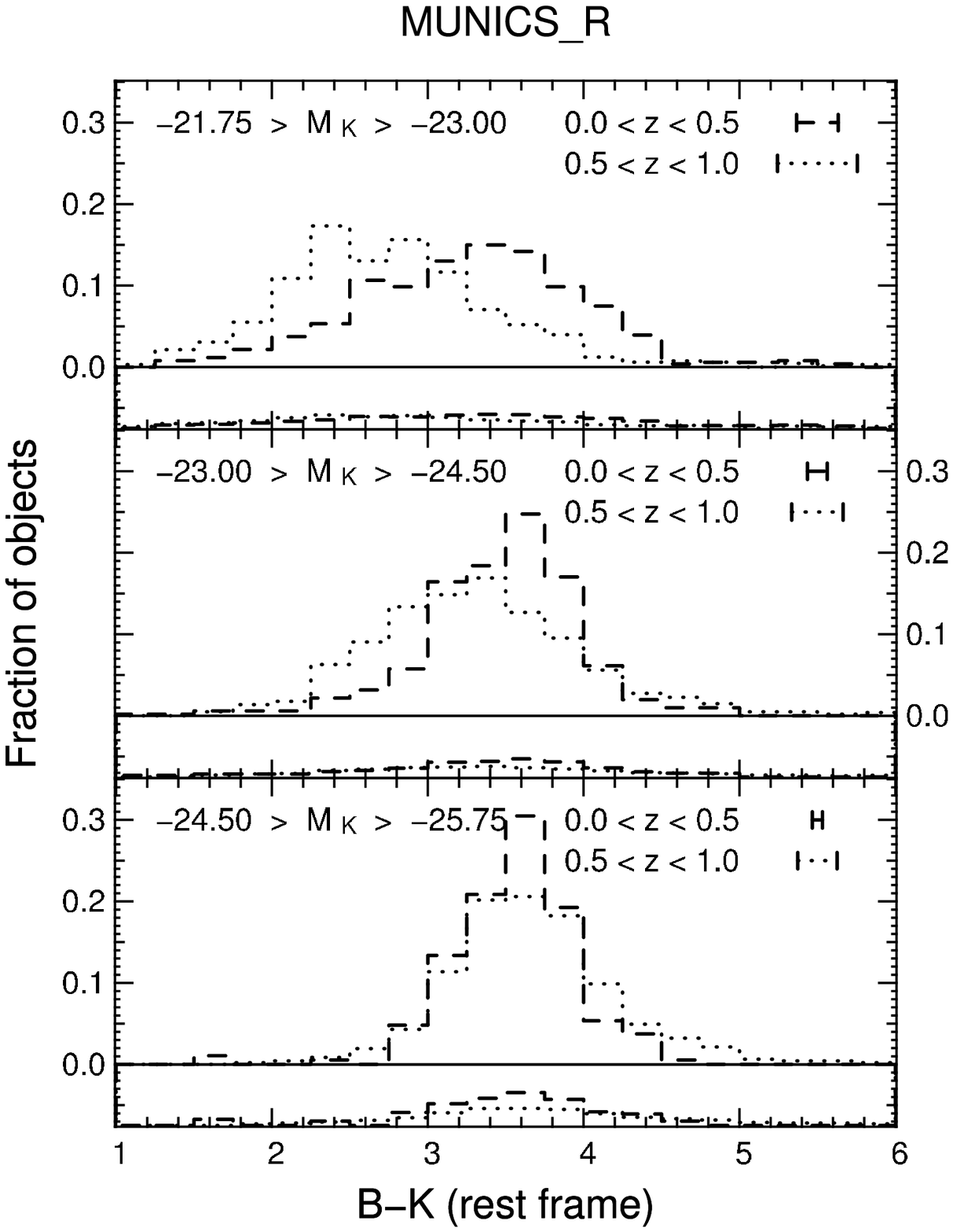,width=0.4\textwidth}
\hspace*{1cm}
\epsfig{figure=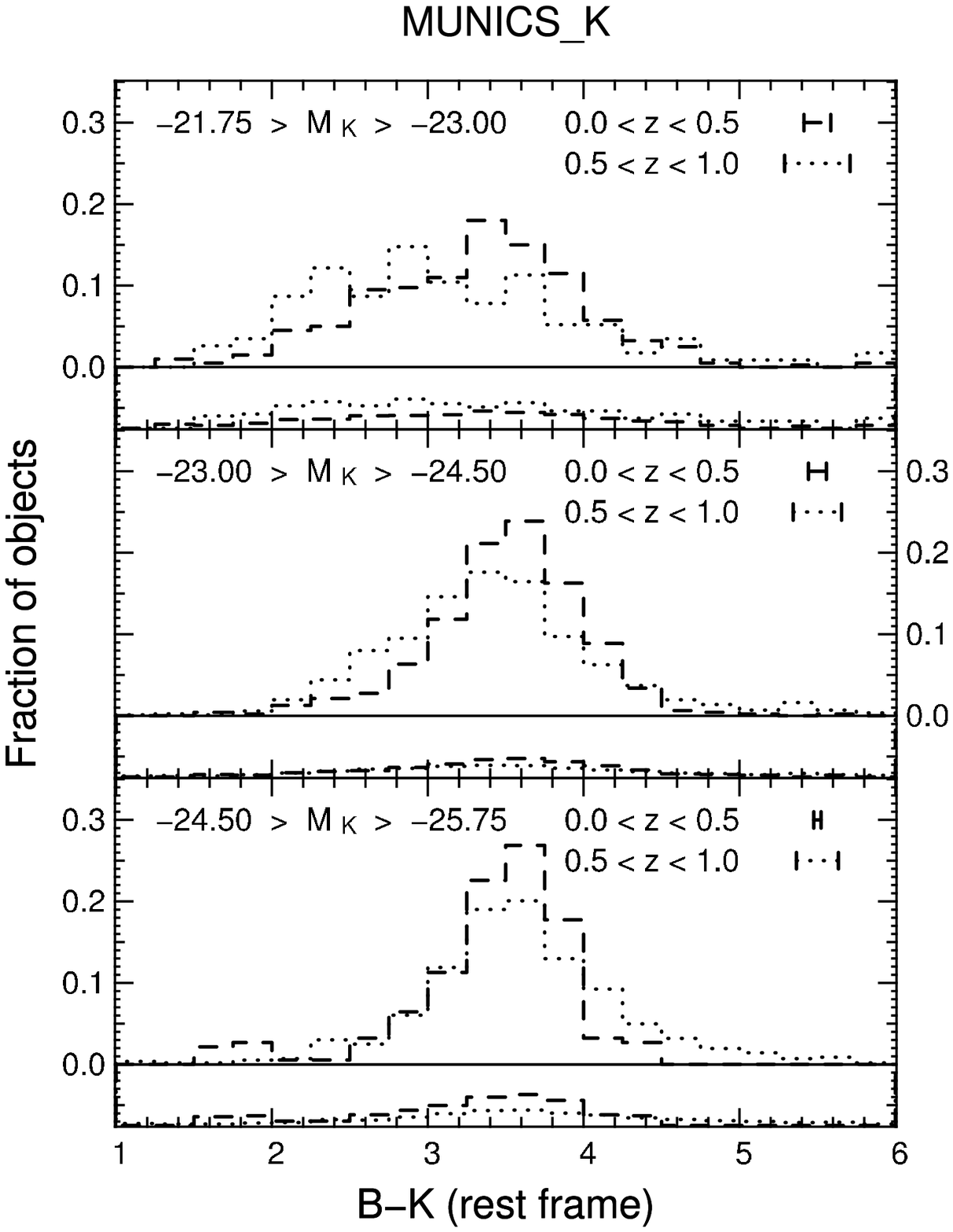,width=0.4\textwidth}

\caption{Distribution of rest-frame $B-K$ colours as a
    function of selection band, luminosity, and
    redshift. \textit{Upper left-hand panel:} All galaxies with $0 \le
    z \le 1$ in the $K$-band selected (solid line), $I$-band selected
    (dot-dashed line), $R$-band selected (dashed line) and $B$-band
    selected (dotted line) MUNICS samples, divided into three
    different $K$-band luminosity classes.  \textit{All other panels:}
    Sub-divided into galaxies with $0 \le z \le 0.5$ (dashed line) and
    $0.5 \le z \le 1.0$ (dotted line) for MUNICS\_B (\textit{upper
      right-hand panel}), MUNICS\_R (\textit{lower left-hand panel}),
    and MUNICS\_K (\textit{lower right-hand panel}). In each plot,
  the distribution of Poisson errors is also shown below the object
  histograms. Furthermore, typical $B-K$ error bars are also indicated
  in every plot.}
\label{f:colordist}

\end{figure*}

%
% COLOUR DISTRIBUTIONS
%
\section{Distribution of Rest-Frame Colours}
\label{s:colours}

In this paper, we want to investigate the differences of the galaxy
populations in $K$, $I$, $R$, and $B$-selected samples and follow
their evolution with redshift. One simple, but rewarding way of doing
this is to study rest-frame colour distributions as a function of
luminosity and redshift. \citet{Coleetal2001s} divided their
$K$-selected sample of local galaxies into three luminosity classes
and looked at the distribution of rest-frame $b_J-K_\mathrm{s}$ and
$J-K_\mathrm{s}$ colours. Since the $K$-band luminosity is a measure
of a galaxy's stellar mass, the luminosity classes correspond to a
binning in stellar mass. Their main finding is that at smaller
$K$-luminosities (i.e.\ smaller masses) there is a more and more
prominent population of blue, star forming galaxies. It is interesting
to test this also for higher redshifts and different selection bands.

The solid line in the upper left-hand panel of
Figure~\ref{f:colordist} shows the $B-K$ rest-frame colour
distributions of $K$-selected MUNICS galaxies at $0 \le z \le 1$,
confirming the result of \citet{Coleetal2001s}. Moreover, we present
in this Figure the same distributions for $I$-band (dot-dashed line)
as well as $R$-band selected MUNICS galaxies (dashed line), showing
clearly that the main difference between near-infrared and optically
selected samples are the bluer colours of low-luminosity (low-mass)
galaxies. This is even more pronounced for the $B$-selected sample
shown as a dotted line.

Four conclusions can be immediately drawn from this
distributions. Firstly, as already noted by \citet{Coleetal2001s}, a
population of blue, star forming galaxies contributes more strongly at
fainter luminosities. Secondly, this populations becomes more numerous
at bluer selection wavelengths. Thirdly, the colour distribution seems
to be wider at fainter absolute $K$-band magnitudes (although part of
this effect may be attributed to the larger errors in the rest-frame
$B-K$ colour for these objects). Fourthly, and maybe most importantly,
the colour distributions of high-mass galaxies change very little as a
function of redshift. Hence one would not expect much variation of the
bright end of the galaxy luminosity function in different
\textit{selection} bands like $K$ and $I$. The faint end, however,
will be certainly affected in the sense that in bluer selection bands
the faint-end slope will likely be steeper.

In the other three panels of Figure~\ref{f:colordist} we present the
rest-frame $B-K$ colour distributions in the two redshift intervals
$[0.0,0.5]$ and $[0.5,1.0]$ for MUNICS\_B (upper right-hand panel),
MUNICS\_R (lower left-hand panel) and MUNICS\_K (lower right-hand
panel). First, it is very interesting that the colour distributions
agree so well for the highest luminosity objects (irrespective of the
selection filter). Furthermore, there is a clear trend with redshift
in the sense that lower-luminosity objects get bluer. That means that
even at redshift $z \sim 1$ the increased star-formation rate is
dominated by low-luminosity (low-mass) systems. Finally, it is worth
noting that this evolutionary trend with redshift is barely visible in
the $K$-selected sample, clearly evident in the $R$-selected sample,
but gets very large in the $B$-selected sample: Indeed, for the
low-mass objects with $-21.75 \; > \; M_K \; > \; -23.00$, the shift
between redshift $z \; \sim \; 0.25$ and $z \; \sim \; 0.75$ is
$\Delta (B-K) \sim 0.3$ for MUNICS\_K, $\Delta (B-K) \sim 0.7$ for
MUNICS\_R, but as large as $\Delta (B-K) \sim 1.0$ for MUNICS\_B.

How can this be interpreted? The rest-frame colour of a galaxy is
governed by the age of its stellar population, its star-formation
activity, the metalicity and the dust content. Let us neglect the
influence of metalicity and dust for the moment. The lack of redshift
evolution of the rest-frame colours of high $K$-band luminosity
galaxies over $0 \; \lsim \; z \; \lsim \; 1$ means that these objects
must have built up their stellar mass at earlier times. The fact that
also at lower $K$ luminosities galaxies of similar rest-frame colour
can be found even in the higher redshift bin indicates that also part
of the lower mass objects formed early, although the majority of them
had (or has) major star-formation activity at $z \; \lsim \; 1$. It is
also clear from these colour distributions that the rise of the
star-formation rate to redshift unity is mostly driven by lower-mass
galaxies. Note that `lower mass' in this context includes
intermediate-mass spirals for which there is evidence for increased
star-formation activity in the past \citep{Bell2005}.

Overall, this picture of different paths of evolution for the most
massive and the lower-mass galaxies is in good agreement with recent
findings on the evolution of the specific star-formation rate (SSFR),
i.e.\ the star formation rate per unit stellar mass (see, e.g.,
\citealt{fdfssfr}a and references therein). We will discuss the SSFR
in more detail in Sections~\ref{s:ssfrsel} and \ref{s:ssfrenv} of this
paper.

%
% SSFR SELECTION EFFECTS
%
\section{The Specific Star Formation Rate and Selection Effects}
\label{s:ssfrsel}

\begin{figure*}

\epsfig{figure=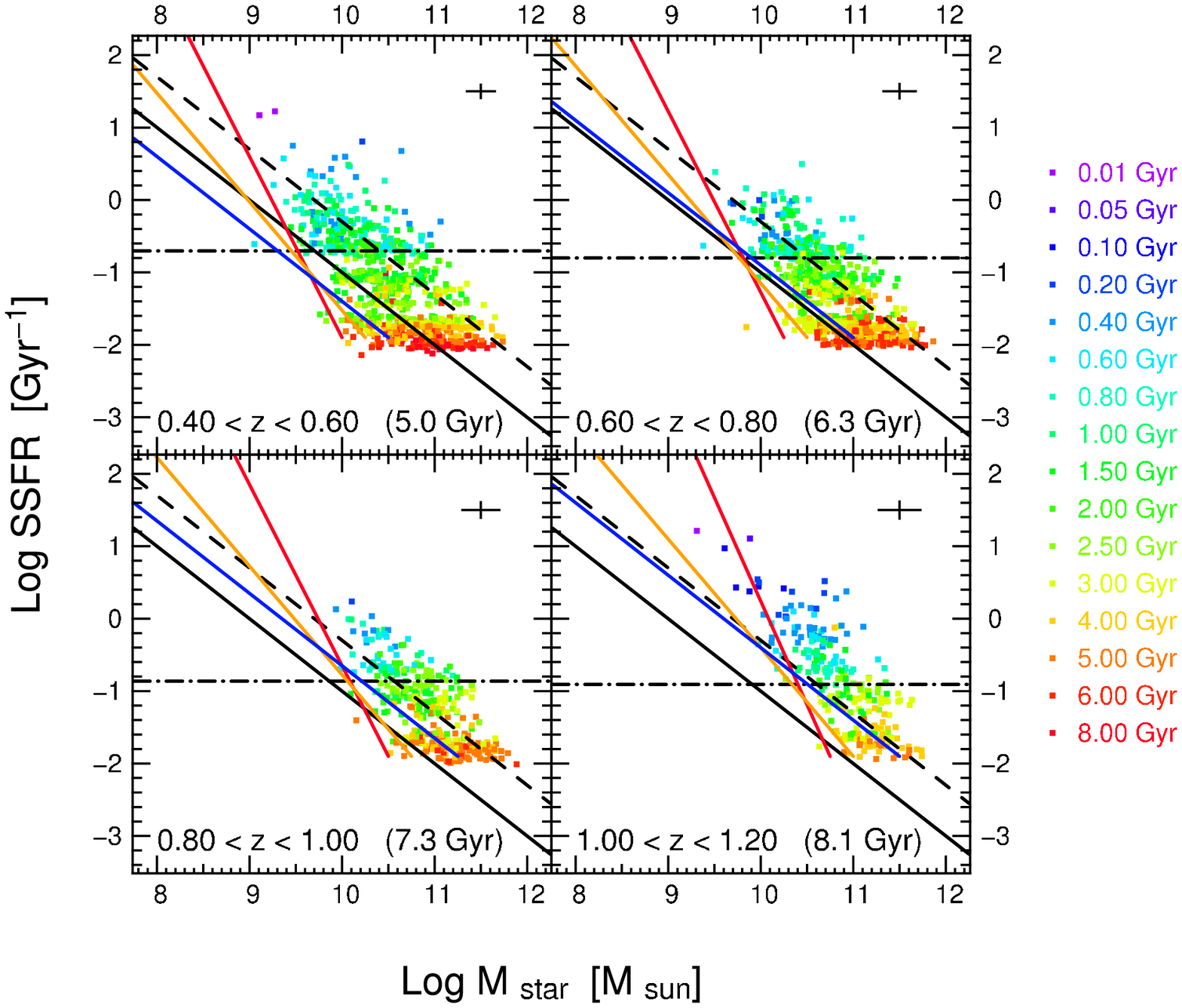,width=0.8\textwidth}

\caption{SSFR versus stellar mass for galaxies in MUNICS\_K
  (symbols). Objects are coloured according to the age of the
  composite stellar-population synthesis model fit to their
  photometry, ranging from 0.01~Gyr (purple) to 8~Gyr (red). The solid
  and dashed black lines correspond to SFRs of $1 \; M_\odot \;
  \mathrm{yr}^{-1}$ and $5 \; M_\odot \; \mathrm{yr}^{-1}$,
  respectively. The dot-dashed line is the SSFR required to double a
  galaxy's mass between each redshift epoch and today, assuming
  constant SFR. The corresponding look-back time as well as an
  estimate of the typical error bars are indicated in each
  panel. Selection effects in this diagram are visualised for
  MUNICS\_K \textit{(red line)}, MUNICS\_I \textit{(orange line)}, and
  MUNICS\_B \textit{(blue line)}.}
\label{f:ssfrsel}

\end{figure*}

In this Section we will discuss the influence of the selection band on
the specific star formation rate (SSFR), i.e. the star formation rate
per unit stellar mass.

As in \citeauthor{munics7} (2005b), we estimate the star formation
rates (SFRs) of our galaxies from the SEDs by deriving the luminosity
at $\lambda = 2800 \pm 100$\AA\ and converting it to an SFR as
described in \citet{Madau1998} assuming a Salpeter initial mass
function (IMF; \citealt{Salpe55}).  We have convinced ourselves that
these photometrically derived SFRs are in reasonable agreement with
spectroscopic indicators for objects with available spectroscopy. Note
that since our bluest band is $B$, this is an extrapolation for $z <
0.4$. Hence we restrict any further analysis to redshifts $z > 0.4$,
where the ultraviolet continuum at $\lambda \simeq 2800$\AA\ is
shifted into or beyond the $B$ band.

Stellar masses are computed from the multi-colour photometry using a
method similar to the one used in \citet{munics6}. It is described in
detail and tested against spectroscopic and dynamical mass estimates
in Drory, Bender \& Hopp (\citeyear{masscal}). In brief, we derive
stellar masses by fitting a grid of stellar population synthesis
models by \citet{BC2003} with a range of star formation histories
(SFHs), ages, metalicities and dust attenuations to the broad-band
photometry. We describe star formation histories (SFHs) by a
two-component model consisting of a main component with a smooth SFH
$\propto \exp (-t/\tau)$ and a burst. We allow SFH time-scales $\tau
\in [0.1,\infty]$~Gyr, metalicities $[\mathrm{Fe/H}] \in [-0.6,0.3]$,
ages between 0.5~Gyr and the age of the universe at the objects
redshift, and extinctions $A_V \in [0,1.5]$. The SFRs derived from
this model fitting is in good agreement with the ones from the
ultraviolet continuum. Note that we apply the extinction correction
derived from this fitting also to the SFRs.

Figure~\ref{f:ssfrsel} presents the SSFR versus stellar mass diagram
for four redshift bins as derived from MUNICS\_K. The influence of the
limiting apparent magnitude of the $K$-selected catalogue is clearly
visible in the lower left-hand part of the panels. The boundary for
MUNICS\_K is marked by the red solid line, while the orange and blue
lines give the limits of the point distributions for MUNICS\_I and
MUNICS\_B, respectively.

The selection band clearly affects the slope of this boundary. While
for the $K$-selected catalogue the line is steeper (i.e.\ closer to a
selection in stellar mass), it gets gradually shallower for selection
bands at shorter wavelengths. For MUNICS\_B, the ridge runs
essentially parallel to lines of constant SFR, demonstrating that
$B$-band selection indeed selects galaxies according to their
star-formation activity.

\begin{figure}

\centerline{\epsfig{figure=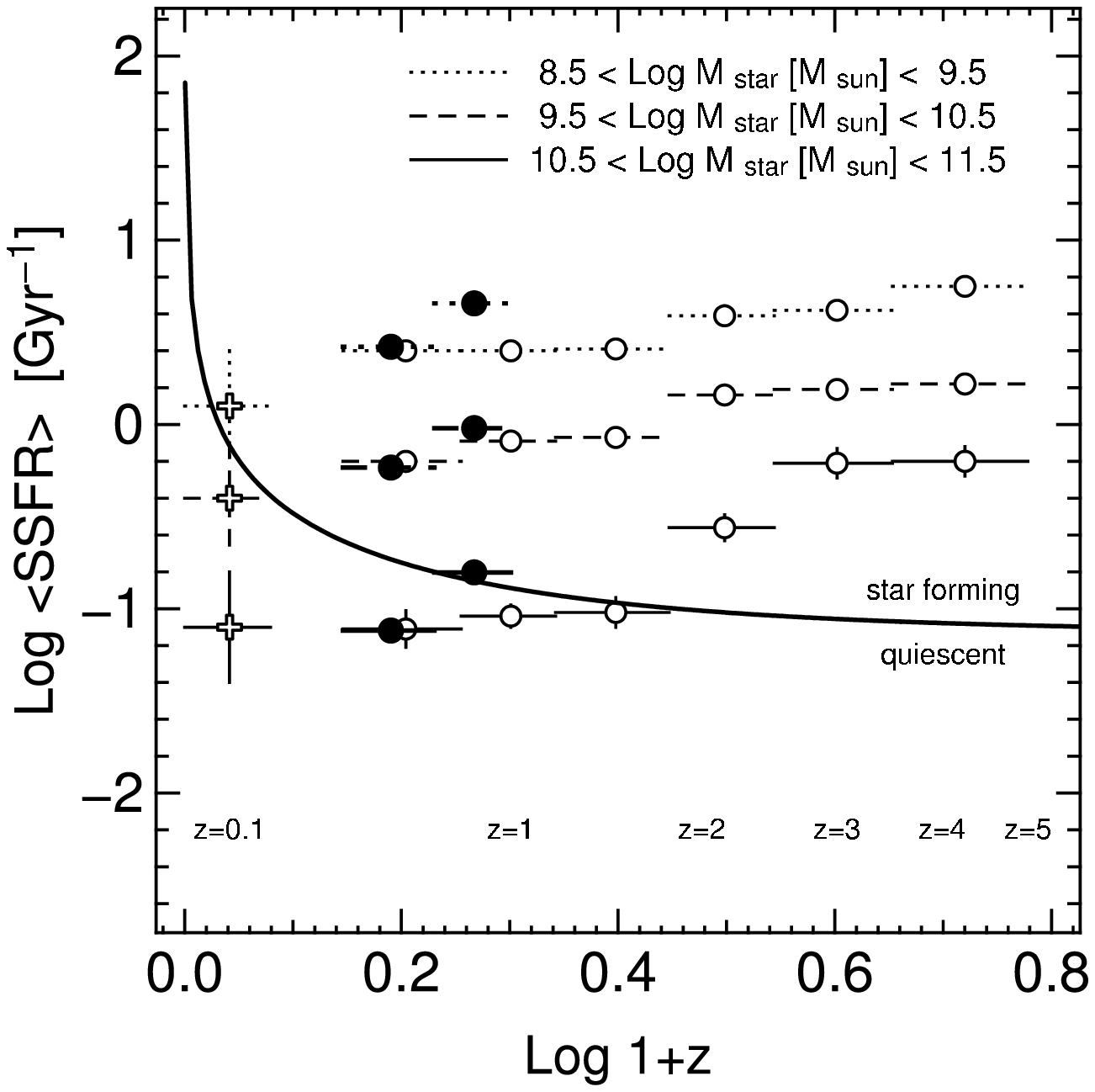, width=0.45\textwidth}}

\caption{The average SSFR of galaxies in three mass intervals (see
  legend) as a function of redshift for galaxies in MUNICS\_B (filled
  circles), FDF/GOODS-S (open circles; \citealt{fdfssfr}), and SDSS
  (open crosses; \citealt{Brinchmann2004}).}
\label{f:ssfrmeanz}

\end{figure}

For massive galaxies ($\log M_\ast > 9.5$) with non-negligible
star-formation activity ($\mathrm{SFR} \; > \; 1 \; M_\odot \;
\mathrm{yr}^{-1}$) at redshifts $z < 1$ there is little difference
between the three selection bands. This can also be seen in
Figure~\ref{f:ssfrmeanz} where we compare the average SSFR as a
function of redshift for galaxies in three different mass bins as
derived from MUNICS\_B and a combined sample from the $I$-selected
FDF, and the $K$-selected GOODS catalogue \citep{fdfssfr}. We chose
MUNICS\_B for this exercise because it is the only sample not affected
by incompleteness for the lowest mass bin. Values from MUNICS\_B and
FDF/GOODS-S are in remarkable agreement. The good agreement between
the $I$-selected FDF and the $K$-selected GOODS-S catalogue has
already been demonstrated in \citeauthor{fdfssfr} (2005a). We also
show local values as derived by \citet{Brinchmann2004} for the Sloan
Digital Sky Survey (SDSS). We have shifted these by $\Delta
\mathrm{SSFR} \: = \: +0.6$ to account for differences in the
normalisation of the SFR and in the dust correction.

\begin{figure*}

\epsfig{figure=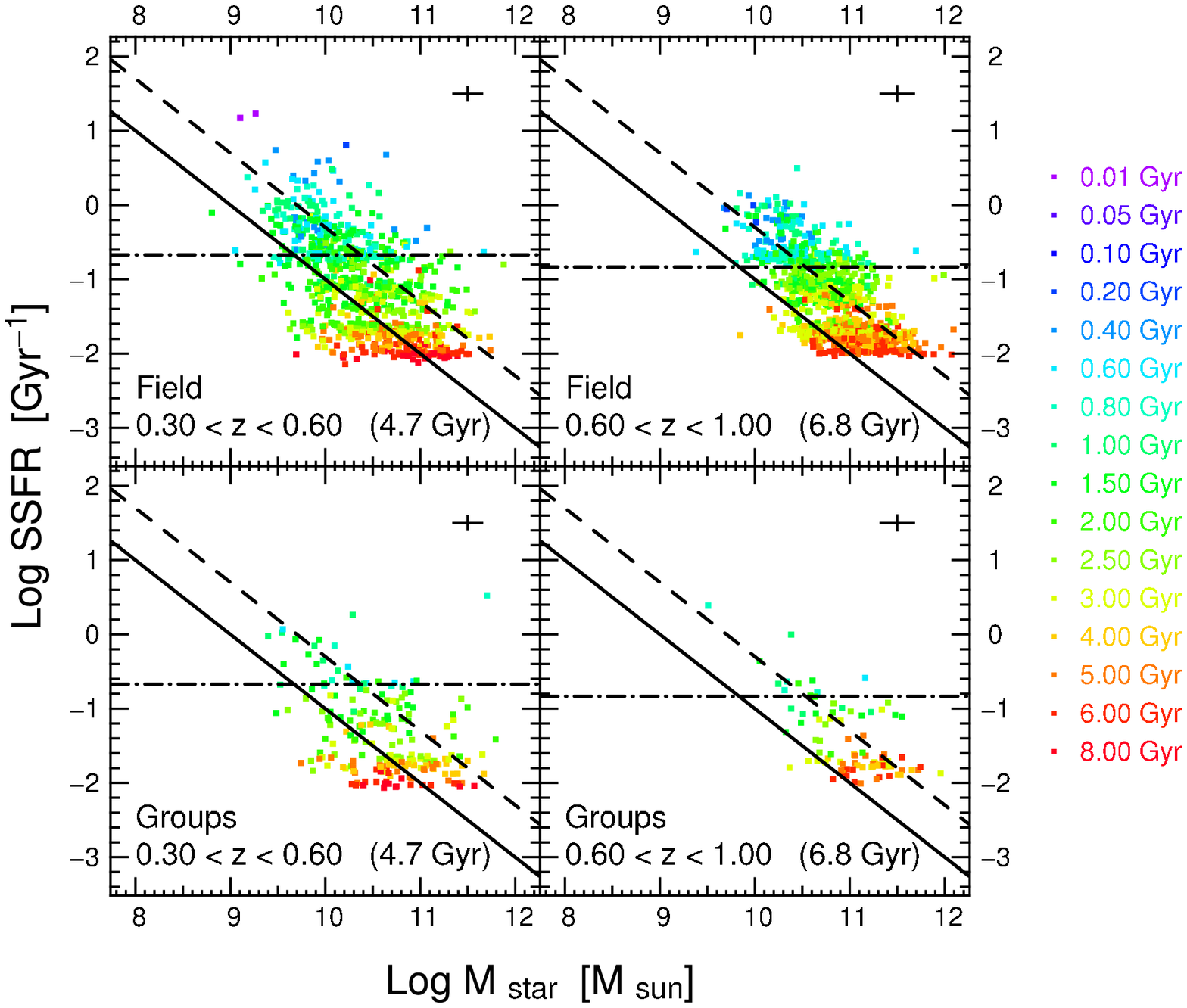,width=0.8\textwidth}

\caption{SSFR versus stellar mass for galaxies in MUNICS\_K (symbols)
  for field galaxies \textit{(upper panels)} and galaxies in MUNICS
  groups \textit{(lower panels)}. See caption of Figure~\ref{f:ssfrsel}
  for details.}
\label{f:ssfrenv}

\end{figure*}

%
% SSFR AS A FUNCTION OF ENVIRONMENT
%
\section{The Specific Star Formation Rate in Different Environments}
\label{s:ssfrenv}

To study the influence of the environment on the SSFRs of galaxies we
investigate its behaviour for field galaxies and for galaxies in
groups. Group membership in the $K$-selected MUNICS catalogue is
assigned according to a modified version of the friends-of-friends
algorithm, specifically designed to cope with photometric redshift
datasets \citep{Botzler2004}. In brief, the algorithm works in
redshift slices, and corresponding structures in adjacent slices are
combined a posteriori. Additionally to the traditional linking
parameters in projected distance ($D_L$) and velocity space ($V_L$), a
redshift pre-selection is performed to exclude objects with large
photometric redshift errors. The linking criteria used for the MUNICS
sample are $D_L = 0.125 \,h^{-1} \,\mathrm{Mpc}$ and $V_L = 1000\,
\mathrm{km\,s}^{-1}$; a minimum number of 3 objects is required to
form a group or cluster \citep{munics8}. Expected completeness
fractions and contamination rates are roughly 90 per cent and 40 per
cent, respectively.

The resulting structure catalogue is presented in \citet{chrisphd} and
\citet{munics8} and comprises 162 structures (mostly groups)
containing 890 galaxies in total. This group sample has already been
used to study the integrated SSFR of groups in comparison with field
galaxies and clusters \citep{intssfr}.

A comparison of field galaxies and group members in the SSFR--stellar
mass plane in two redshift intervals is presented in
Figure~\ref{f:ssfrenv}. In contrast to the field population, group
galaxies tend to populate the region of passive galaxies below the
doubling line only. This, of course, is in agreement with previous
findings that galaxies in higher density environments tend to have
lower star-formation activity. Interestingly, however, passive
galaxies are by no means limited to the group environment. This could
be partially explained by group selection effects, but is in agreement
with recent findings by \citet{Gerke2007} who identified a significant
population of red galaxies in the field, with morphologies
characteristic for passive galaxies.

\begin{figure}

\centerline{\epsfig{figure=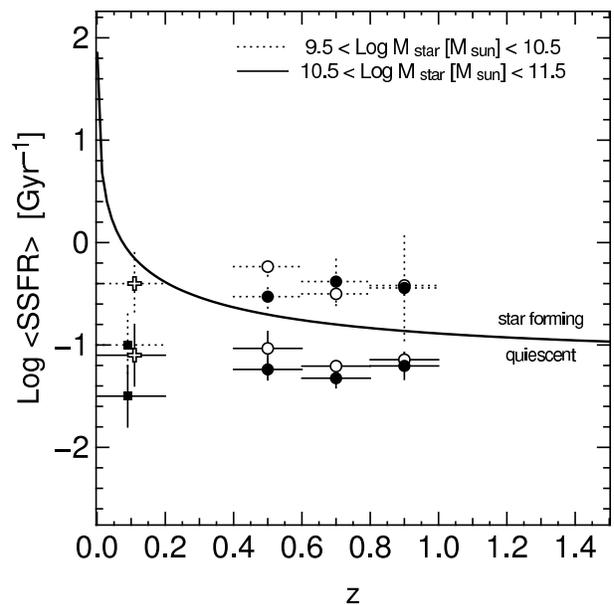, width=0.45\textwidth}}

\caption{Same as Figure~\ref{f:ssfrmeanz} but for MUNICS field galaxies
  (open circles) and group members (filled circles). The local
  comparison values are taken from \citet{Weinmann2006} for SDSS
  groups (filled squares) and from \citet{Brinchmann2004} for SDSS
  field galaxies (open crosses).}
\label{f:ssfrmeanzenv}

\end{figure}

This behaviour is also obvious from Figure~\ref{f:ssfrmeanzenv} where we
show the redshift evolution of the average SSFR of galaxies in two
different mass intervals, comparing field and group galaxies. At $z
\sim 1$ there seems to be no marked difference between field galaxies
and galaxies in groups, with a tendency of the difference increasing
with decreasing redshift in the sense that group galaxies are rather
more inactive than field galaxies. Note that this difference is larger
for less massive galaxies than for the most massive objects. This is
in principal agreement with earlier findings on the dependence of SSFR
on density for galaxies in the local universe \citep{Kauffmann2004}.

The fact that we do not see a more pronounced difference between field
and groups could be at least partially attributed to two
effects. First, incompleteness of the group catalogue as well as
contamination with false structures is likely to reduce the
signal. Secondly, trends of the photometric redshift errors with SFR
and redshift might affect the analysis, especially in the higher
redshift bins. Photometric redshift errors in the MUNICS catalogue
increase with redshift (from $\sim 0.08$ at $z = 0.5$ to $\sim 0.12$
at $z = 1$) and with SFR (they are typically 20 per cent larger for
star-forming galaxies as compared to passive galaxies). In reality,
the difference in SSFR between groups and the field may be
more significant.

%
% LF
%
\section{Evolution of the Luminosity Function}
\label{s:lf}

\subsection{Computing the Galaxy Luminosity Function}

As for the near-infrared luminosity function (LF) presented in
\citet{munics5} and \citet{munics2}, the LF is computed using the
non-parametric $V_\mathrm{max}$ formalism \citep{Schmi68}. In brief,
the $V_\mathrm{max}$ formalism accounts for the fact that some fainter
galaxies are not visible in the whole survey volume. In a
volume-limited redshift survey, each galaxy $i$ in a given redshift
bin $[z_\mathrm{lower},z_\mathrm{upper}]$ contributes to the number
density an amount inversely proportional to the volume $V_i$ of the
survey:

\begin{equation}
  V_i \; = \; \int\limits_{z_\mathrm{lower}}^{z_\mathrm{upper}} \:
  \frac{dV}{dz} \: dz ,
\end{equation}

where $dV = d\Omega \, r^2 \, dr$ is the co-moving volume element and
$d\Omega$ is the solid angle covered by the survey. However, due to
the fact that we have to deal with magnitude-limited surveys, a faint
galaxy may not be visible in the whole survey volume. Assuming that in
a survey with given limiting magnitude galaxy $i$ can be seen out to
redshift $z_\mathrm{max}$, we have to correct the volume factor by
$V_i^\mathrm{max} / V_i$, where

\begin{equation}
  V_i^\mathrm{max} \; = \;
  \int\limits_{z_\mathrm{lower}}^{\mathrm{min}(z_\mathrm{upper},
  z_\mathrm{max})} \: \frac{dV}{dz} \: dz .
\end{equation}

Obviously, we have $z_\mathrm{max} \ge z_\mathrm{upper}$ for a galaxy
which is bright enough to be seen in the whole volume in
investigation, and the correction factor is one. Otherwise,
$z_\mathrm{max} < z_\mathrm{upper}$, and the volume is smaller than
the volume corresponding to the redshift range in which we compute the
LF.  We have made sure that the effect of the volume correction is of
importance only in the faintest bin in absolute magnitude, and that
even in this case the correction is at most a factor of 5.

Additionally, the contribution of each galaxy $i$ is weighted by the
inverse of the detection probability $P(m_{\mathrm{det},i})$ of the
parent catalogue. The LF $\Phi (M)$ is then computed according to the
formula

\begin{equation}
  \Phi (M) \: dM \; = \;\sum\limits_i \: \frac{V_i}{V_i^\mathrm{max}} \:
  \: \frac{1}{V_i} \: \frac{1}{P (m_{\mathrm{det},i})} \: dM ,
\end{equation}

where the sum runs over all objects $i$ in the redshift range for
which we want to calculate the LF. Naturally the volume terms can be
simplified to $1 / V_i^\mathrm{max}$.

We include two effects in the total error budget of the luminosity
function. Firstly, the limited number of objects in each magnitude bin
produces statistical uncertainties. Secondly, the errors in the
photometric redshift estimates have some influence on the luminosity
function.

The statistical errors are derived using Poissonian statistics,
following the methods described in \citet{Gehrels1986},
\citet{Ebeling2003}, and \citet{Ebeling2004} for approximation of
Poissonian errors for small numbers.

To investigate the influence of photometric redshift errors on the
luminosity function, we perform Monte-Carlo simulations. In these
simulations, we take the original input catalogue used for deriving
the luminosity function, and assign to each object a redshift within
the redshift error distribution given by the photometric-redshift
algorithm. In principle, there are two ways of doing this. Either one
uses the redshift probability distribution of the best-fitting SED, or
the sum of the probability distributions of all SEDs (the total
probability distribution). One might expect the total probability
distribution to have the advantage that it accounts better for
systematic uncertainties in the SED fitting. However, we have carried
out careful tests which show that the errors derived from the total
distribution are not significantly different from the ones for the
best distribution. This is due to the fact that the probability
distribution for the best-fitting SED usually dominates the total
distribution. Since the total distribution also takes more Monte-Carlo
realisations to converge, we decided to use the probability
distribution of the best-fitting SED for our simulations. The median
of the redshift error assigned by the photometric redshift algorithm
to galaxies in MUNICS\_R is $\langle\sigma\rangle_\mathrm{median}
\simeq 0.08$, in good agreement with the measured deviation of
photometric and spectroscopic redshifts of $\delta z / (1+z) \simeq
0.057$. Hence these error distributions are a good representation of
the true error distribution.

The errors are then computed as follows. The Poisson error and the
standard deviation around the mean derived from the Monte-Carlo
simulations are summed quadratically. In addition, any difference
between the measured value for the LF in a magnitude bin and the mean
from the Monte-Carlo simulations is considered a measure of systematic
errors. This is also added quadratically, but only in one direction,
i.e.\ if the Monte-Carlo mean is higher than the measured value the
upper error bar is enlarged, but not the lower one. All our $\chi^2$
fitting routines, both for the Schechter parameters and for the LF
evolution, can handle asymmetric errors.

\begin{figure*}

\epsfig{figure=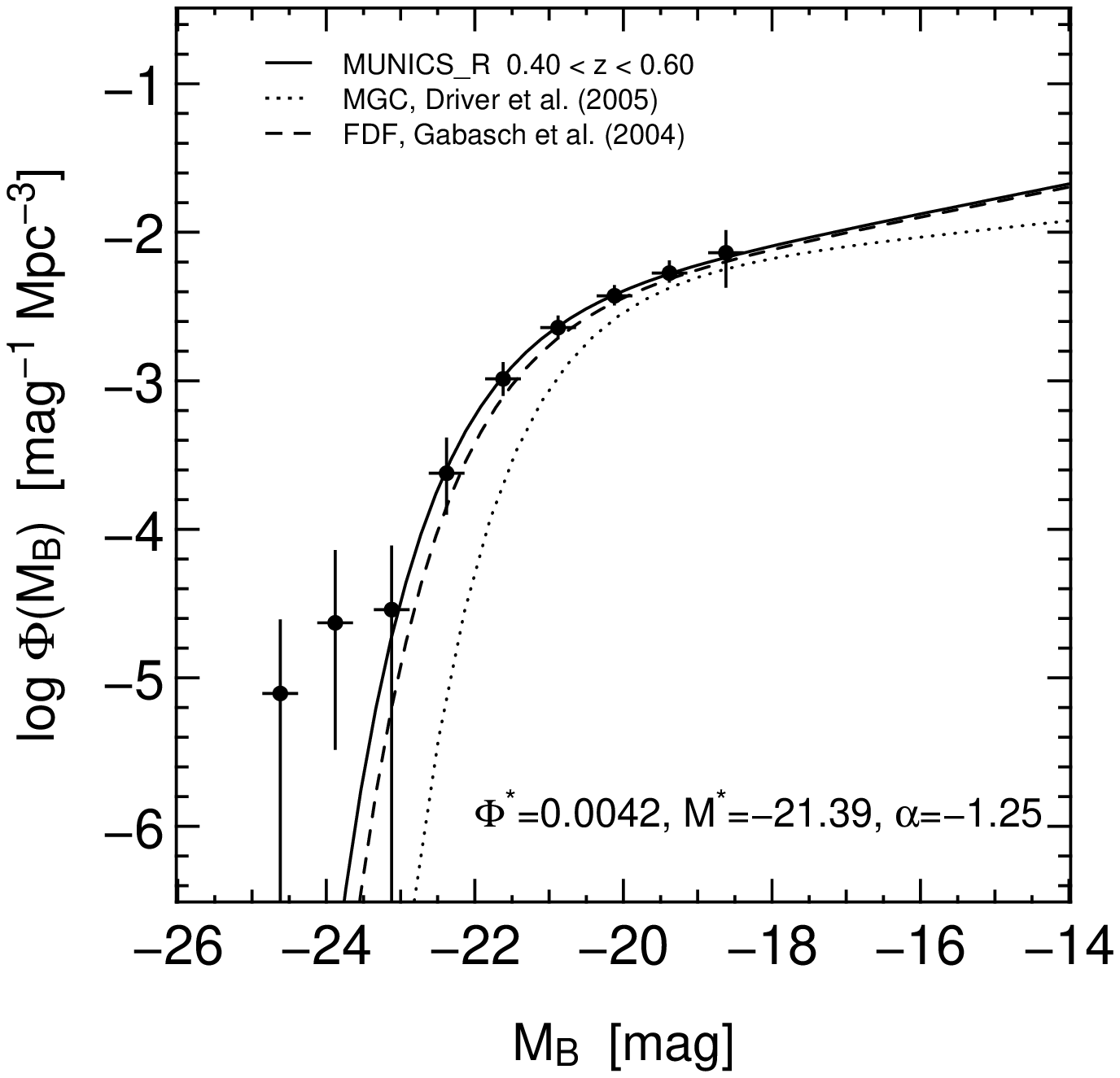,width=0.4\textwidth}
\hspace*{1cm}
\epsfig{figure=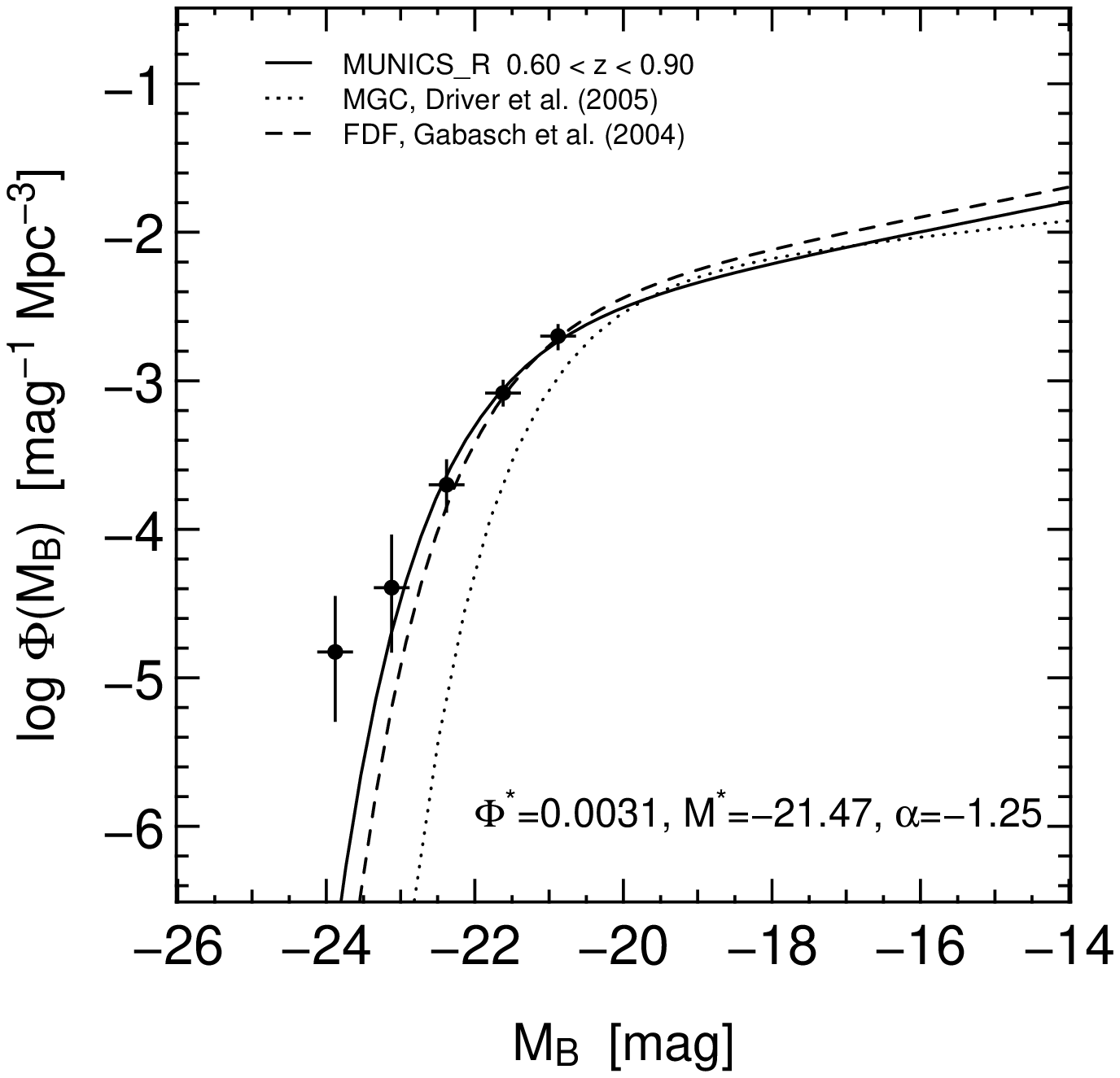,width=0.4\textwidth}

\vspace*{.5cm}

\epsfig{figure=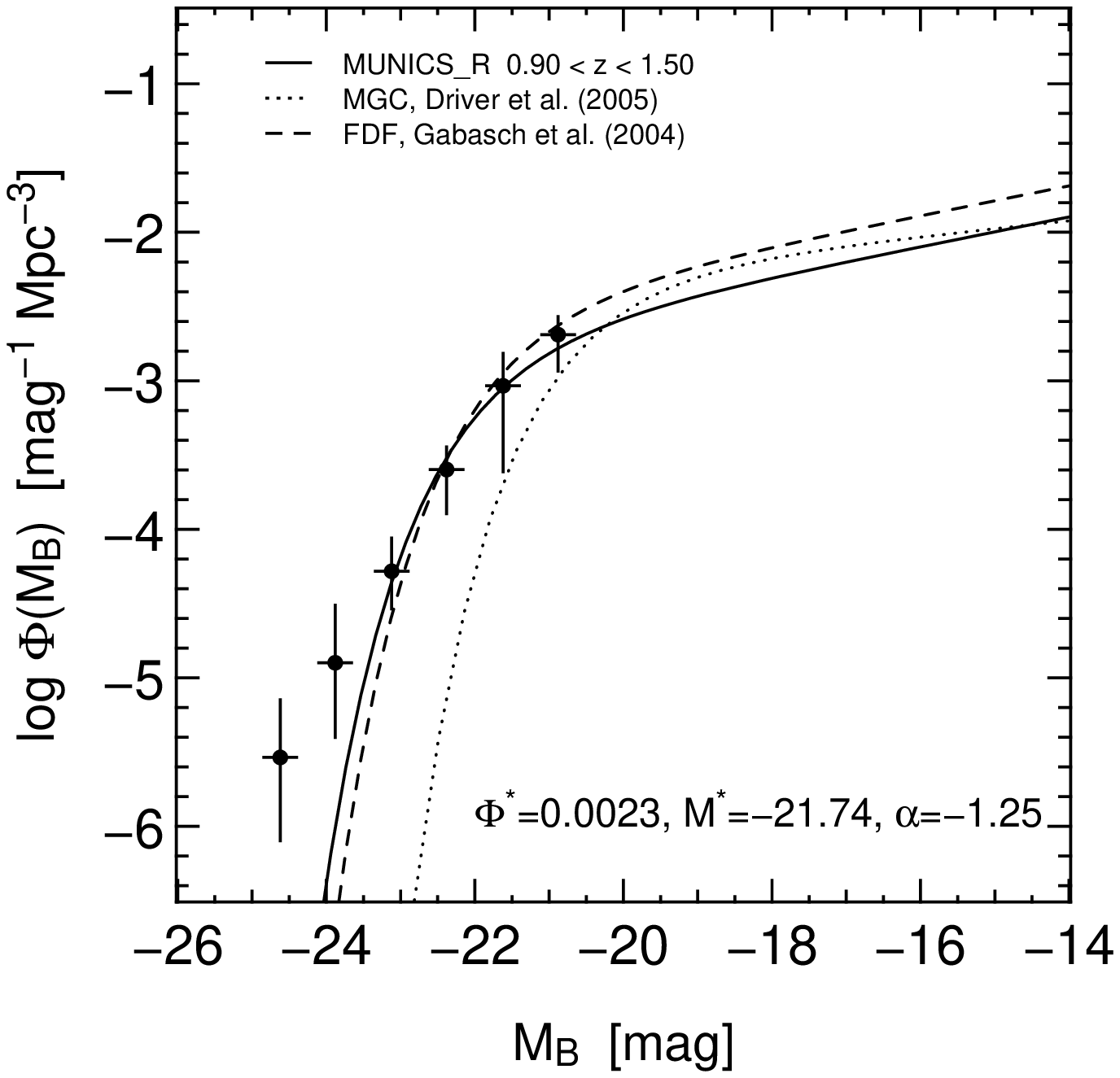,width=0.4\textwidth}
\hspace*{1cm}
\epsfig{figure=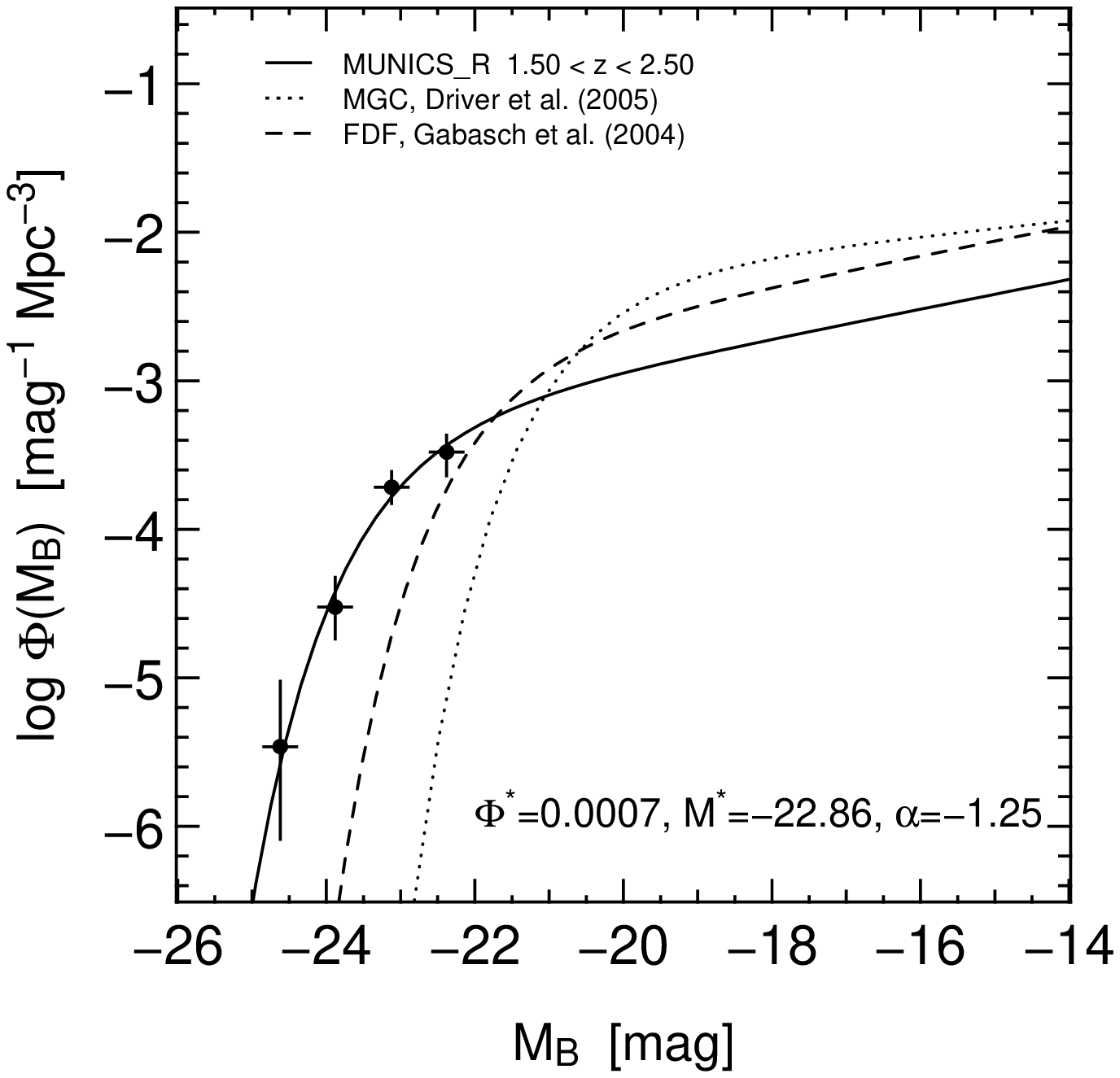,width=0.4\textwidth}

\caption[MUNICS\_R: $B$-band LF]{The $B$-band LF from the
  $R$-selected catalogue at redshifts $z=0.50$ (\textit{upper left-hand
  panel}), $z=0.75$ (\textit{upper right-hand panel}), $z=1.20$
  (\textit{lower left-hand panel}) and $z=2.00$ (\textit{lower
  right-hand panel}). One can clearly see the effect of brightening
  and decreasing number density with increasing redshift $z$. The local
  $B$-band LF from \citet{Driver2005} is shown as dotted line, the FDF
  LF from \citet{fdflf1} in the redshift bins [0.45,0.81],
  [0.45,0.81], [0.81,1.11] and [2.15,2.91] as dashed line.}
\label{f:plf_r_b}

\end{figure*}

\begin{figure*}

\epsfig{figure=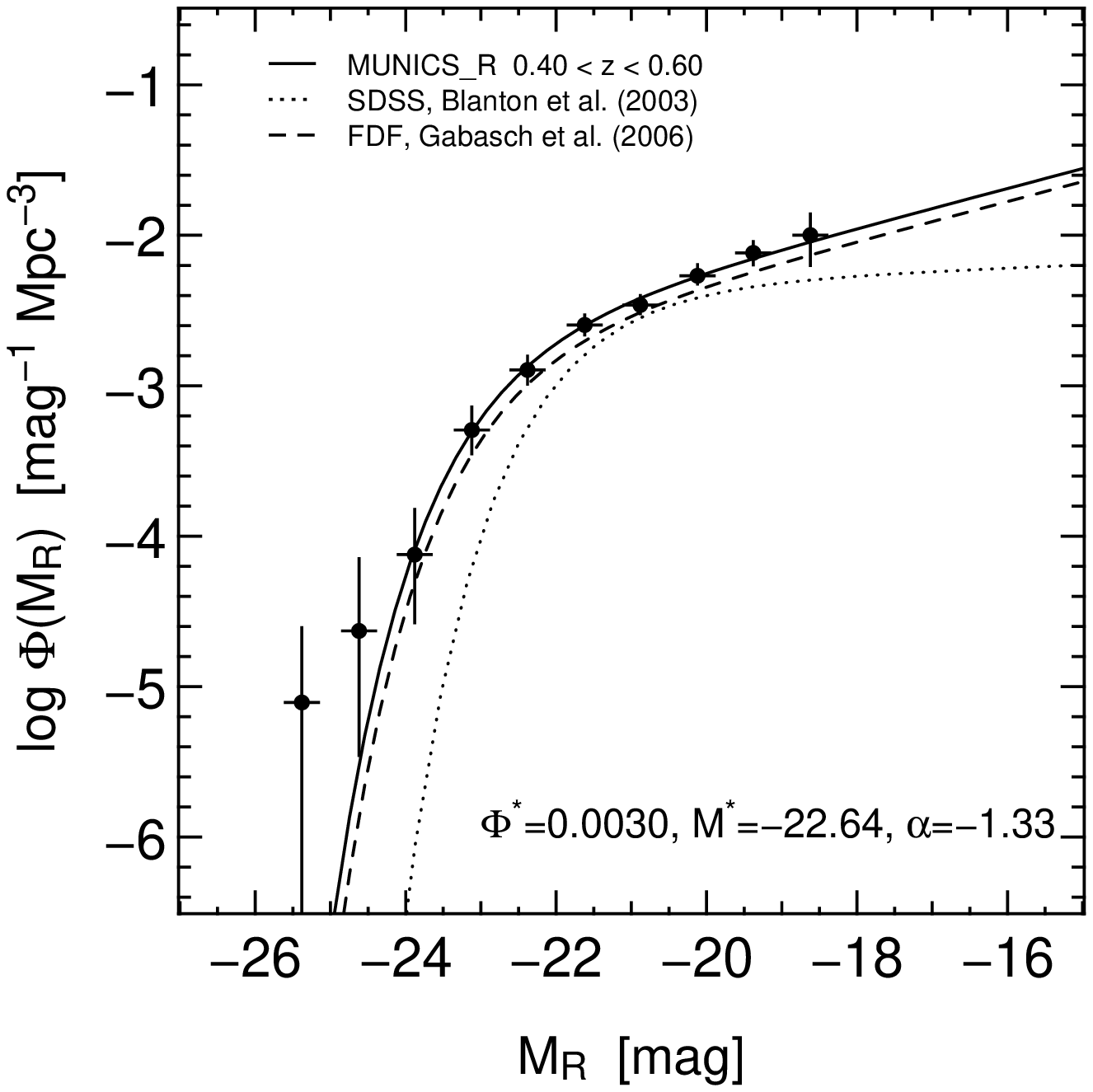,width=0.4\textwidth}
\hspace*{1cm}
\epsfig{figure=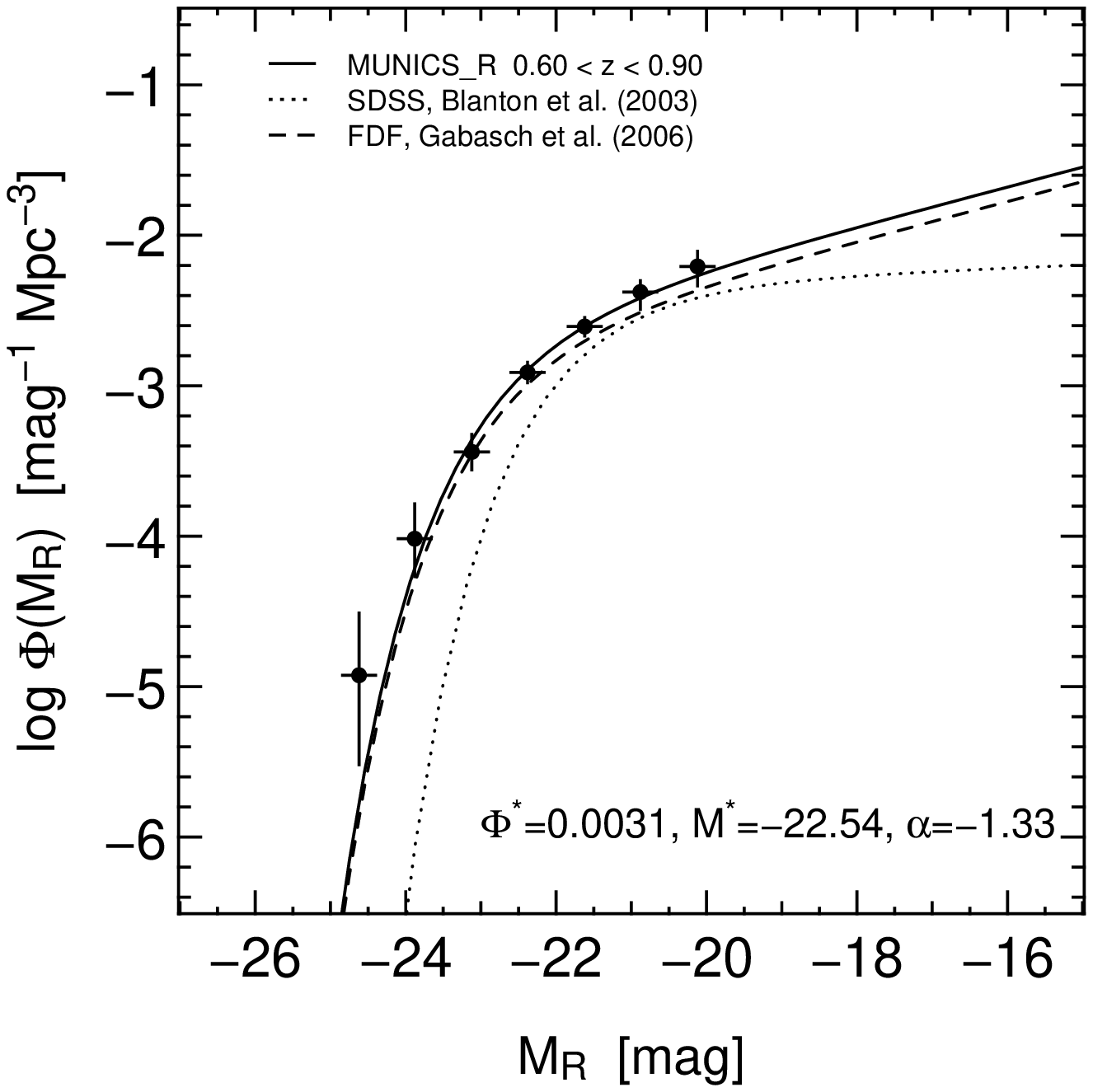,width=0.4\textwidth}

\vspace*{.5cm}

\epsfig{figure=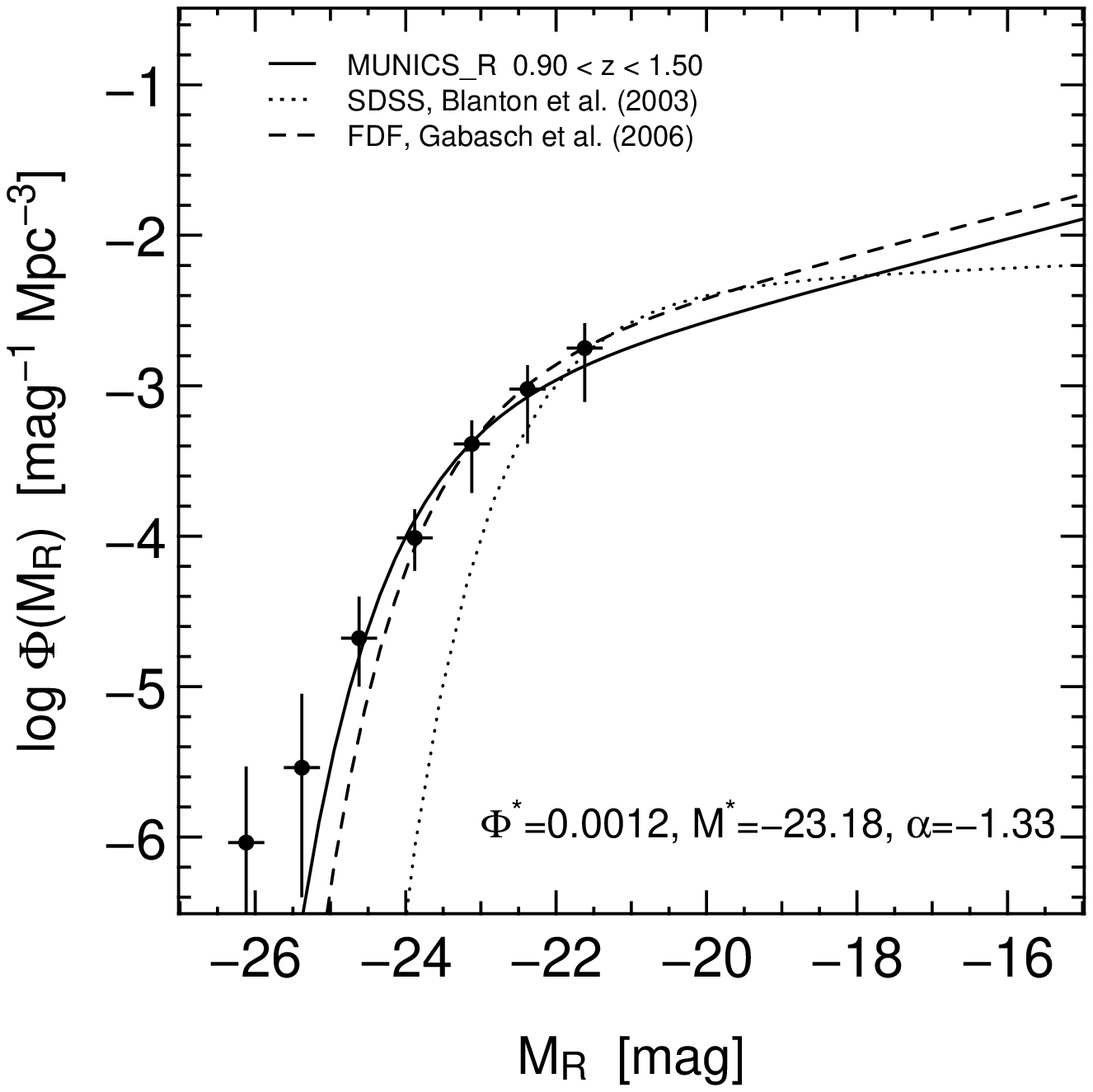,width=0.4\textwidth}
\hspace*{1cm}
\epsfig{figure=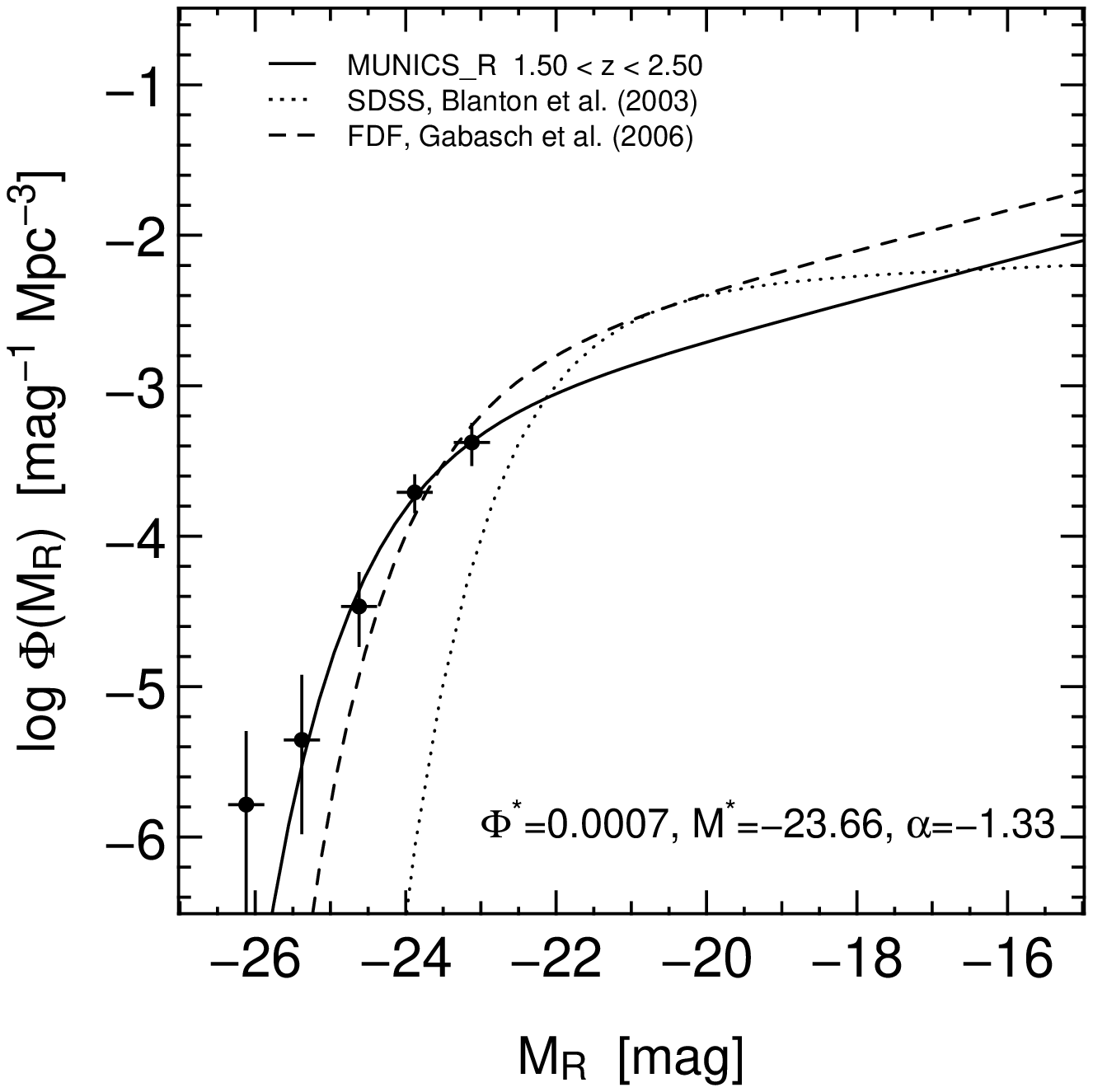,width=0.4\textwidth}

\caption[MUNICS\_R: $R$-band LF]{The $R$-band LF from the $R$-selected
  catalogue at redshifts $z=0.50$ (\textit{upper left-hand panel}),
  $z=0.75$ (\textit{upper right-hand panel}), $z=1.20$ (\textit{lower
    left-hand panel}) and $z=2.00$ (\textit{lower right-hand
    panel}). One can clearly see the effect of brightening and
  decreasing number density with increasing redshift $z$. The local
  $r'$-band LF from \citet[][ transformed to the $R$
    band]{Blanton2003s} is shown as dotted line, the FDF $r$-band LF
  from \citet[][ transformed to the $R$ band]{fdflf2} in the redshift
  bins [0.45,0.85], [0.45,0.85], [0.85,1.31] and [1.91,2.61] as dashed
  line.}
\label{f:plf_r_r}

\end{figure*}

\subsection{$R$-Band Selected Luminosity Functions}

In this Section we present luminosity functions from the $R$-selected
MUNICS sample in the redshift intervals $[0.4,0.6]$, $[0.6,0.9]$,
$[0.9,1.5]$, and $[1.5,2.5]$. The luminosity function results together
with a Schechter approximation \citep{Schechter76} and a local
comparison function are presented in Figure~\ref{f:plf_r_b} for the
$B$ band and in Figure~\ref{f:plf_r_r} for the $R$ band,
respectively. We also compare our results to the LF derived in the FDF
in \citet{fdflf1, fdflf2}. The agreement is very good in general,
although we see a slight excess of very bright galaxies in MUNICS,
resulting in somewhat larger characteristic luminosities and, because
of the degeneracy of the Schechter paramters $\Phi^*$ and $M^*$, lower
characteristic densities. The excess of bright galaxies might
be caused by three effects. First, quasars which cannot be easily
discarded from our photometric catalogue are expected to populate the
bright end of the LF \citep[e.g.][]{Jahnke2003}. Secondly, a slight
excess of bright galaxies as compared to the Schechter function has
been noted in spectroscopic surveys in the local universe
\citep[e.g.][]{Blanton2003s, Jones2006}. Finally, it may be at least
partially attributed to photometric redshift errors s \citep[see,
e.g., the simulations in][]{munics2}, although in this case the error
bars at the bright end of the LF seem to be smaller than the true
errors which might point to an underestimation of the errors for
individual galaxies, although they are in good agreement with the
measured errors on average. In the highest redshift bin, the MUNICS
data do not sample the knee of the LF very well, resulting in rather
low values for $\Phi^*$.

\begin{table*}
\caption[MUNICS\_R: Schechter parameters]{Schechter parameters
$\Phi^*$ and $M^*$ and $\alpha$ for the LFs in the $B$ and $R$
band derived from MUNICS\_R.}
\label{t:plf_r_schechter}
\begin{center}
\begin{tabular}{ccccccc}
\hline
Band & $\langle z \rangle$ & $\Phi^*$ & $M^*$ & $\alpha$ & $M_\mathrm{lim}$ & 
$\min ( \chi^2_\mathrm{red} )$ \\
    &  & [$10^{-3}$ Mpc$^{-3}$] & [mag] & & [mag] \\\hline
$B$ & 0.50 & $4.20 \pm 0.60$ & $-21.39 \pm 0.18$ & $-1.25$ & $-18.0$ & $0.37$\\
    & 0.75 & $3.12 \pm 1.08$ & $-21.47 \pm 0.25$ & $-1.25$ & $-20.5$ & $1.11$\\
    & 1.20 & $2.32 \pm 1.28$ & $-21.74 \pm 0.29$ & $-1.25$ & $-20.5$ & $0.28$\\
    & 2.00 & $0.68 \pm 0.22$ & $-22.86 \pm 0.22$ & $-1.25$ & $-22.0$ & $0.41$\\
\hline
$R$ & 0.50 & $2.96 \pm 0.36$ & $-22.64 \pm 0.18$ & $-1.33$ & $-18.0$ & $0.37$\\
    & 0.75 & $3.12 \pm 0.62$ & $-22.54 \pm 0.18$ & $-1.33$ & $-20.0$ & $0.58$\\
    & 1.20 & $1.16 \pm 0.60$ & $-23.18 \pm 0.30$ & $-1.33$ & $-21.5$ & $0.28$\\
    & 2.00 & $0.72 \pm 0.30$ & $-23.66 \pm 0.26$ & $-1.33$ & $-23.0$ & $0.22$\\
\hline
\end{tabular}
\end{center}
\end{table*}

\begin{figure*}

\epsfig{figure=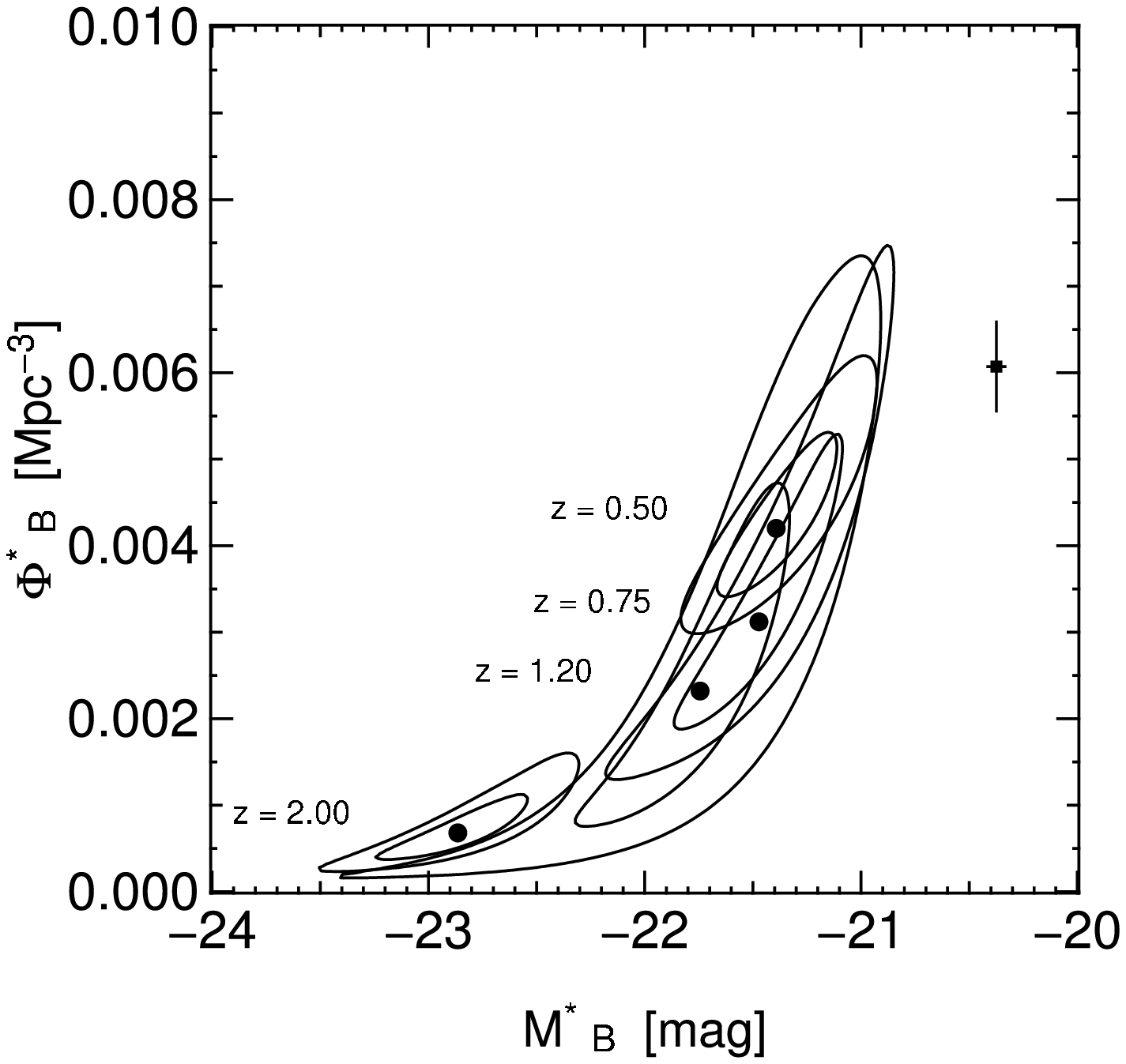,width=0.31\textwidth}
\hspace*{.5cm}
\epsfig{figure=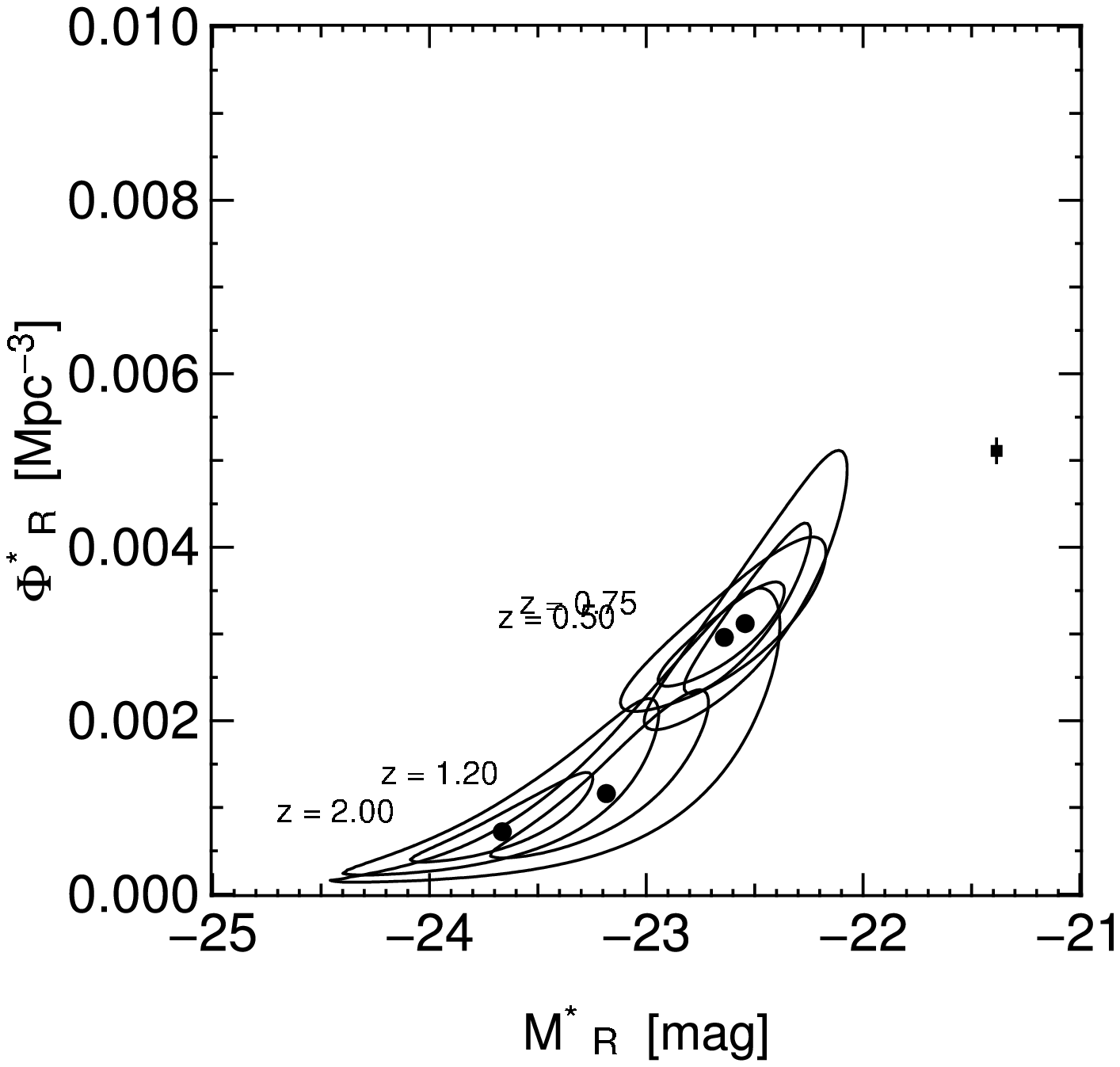,width=0.31\textwidth}

\caption{Schechter contours in the $\Phi^*$--$M^*$ plane for the
    $B$-band LF (\textit{left-hand panel}) and the $R$-band LF
    (\textit{right-hand panel}).}
\label{f:plf_r_schechter}

\end{figure*}

We summarise the parameters $\Phi^*$, $M^*$ and $\alpha$ of the
Schechter fits in Table~\ref{t:plf_r_schechter} and show the
corresponding Schechter contours in
Figure~\ref{f:plf_r_schechter}. Note that we have kept the faint-end
slope $\alpha$ fixed to the value $-1.25$ ($B$ band) and $-1.33$ ($R$
band) during the fitting process. These are the values derived for the
FDF in \citet{fdflf1} and \citet{fdflf2}, respectively, which are in
good agreement with the MUNICS LFs in the lowest redshift bin. At
higher redshifts, the MUNICS data are not sufficiently deep to
constrain the faint-end slope.

\begin{figure*}

\epsfig{figure=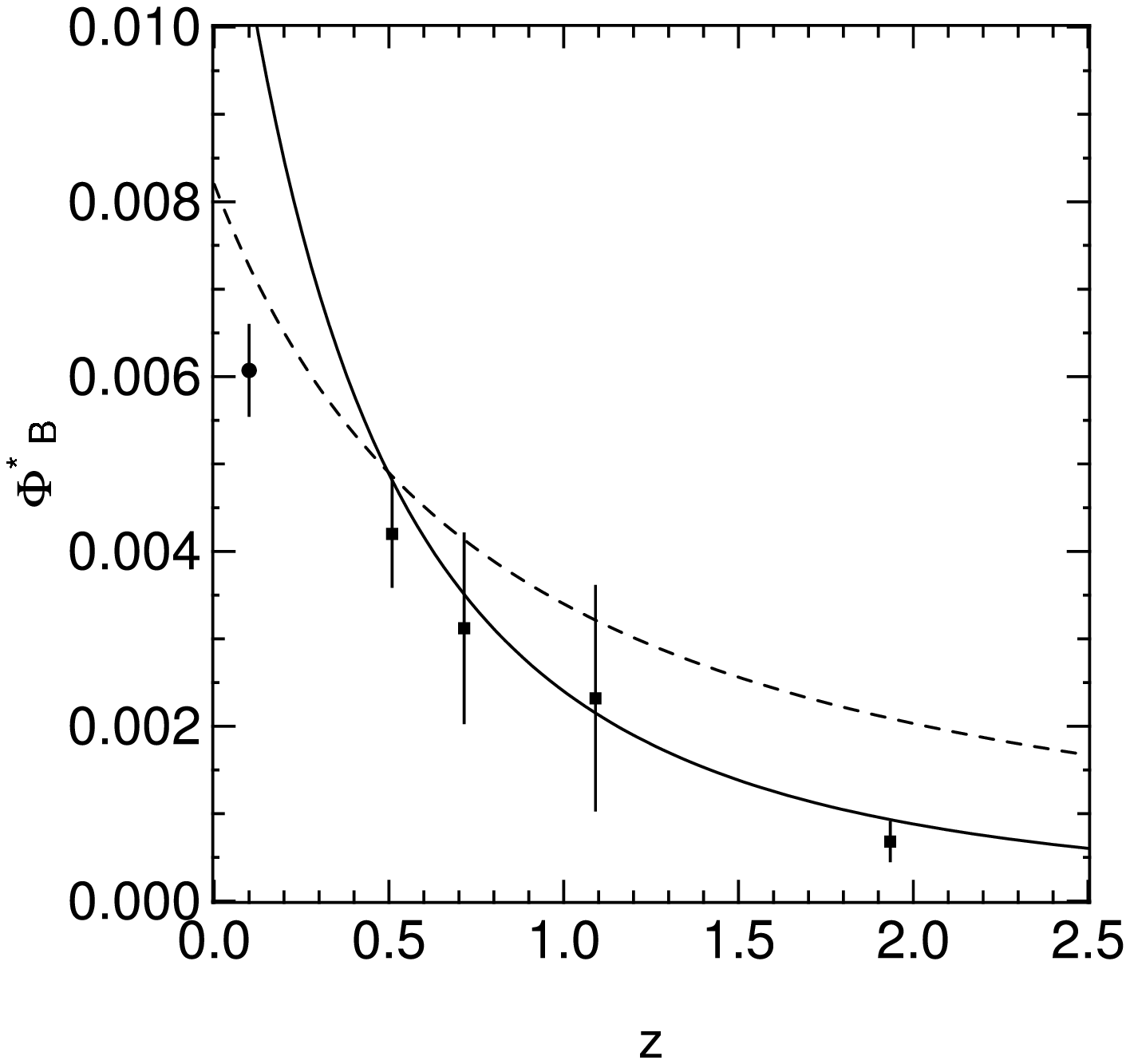,width=0.3\textwidth}
\hspace*{.5cm}
\epsfig{figure=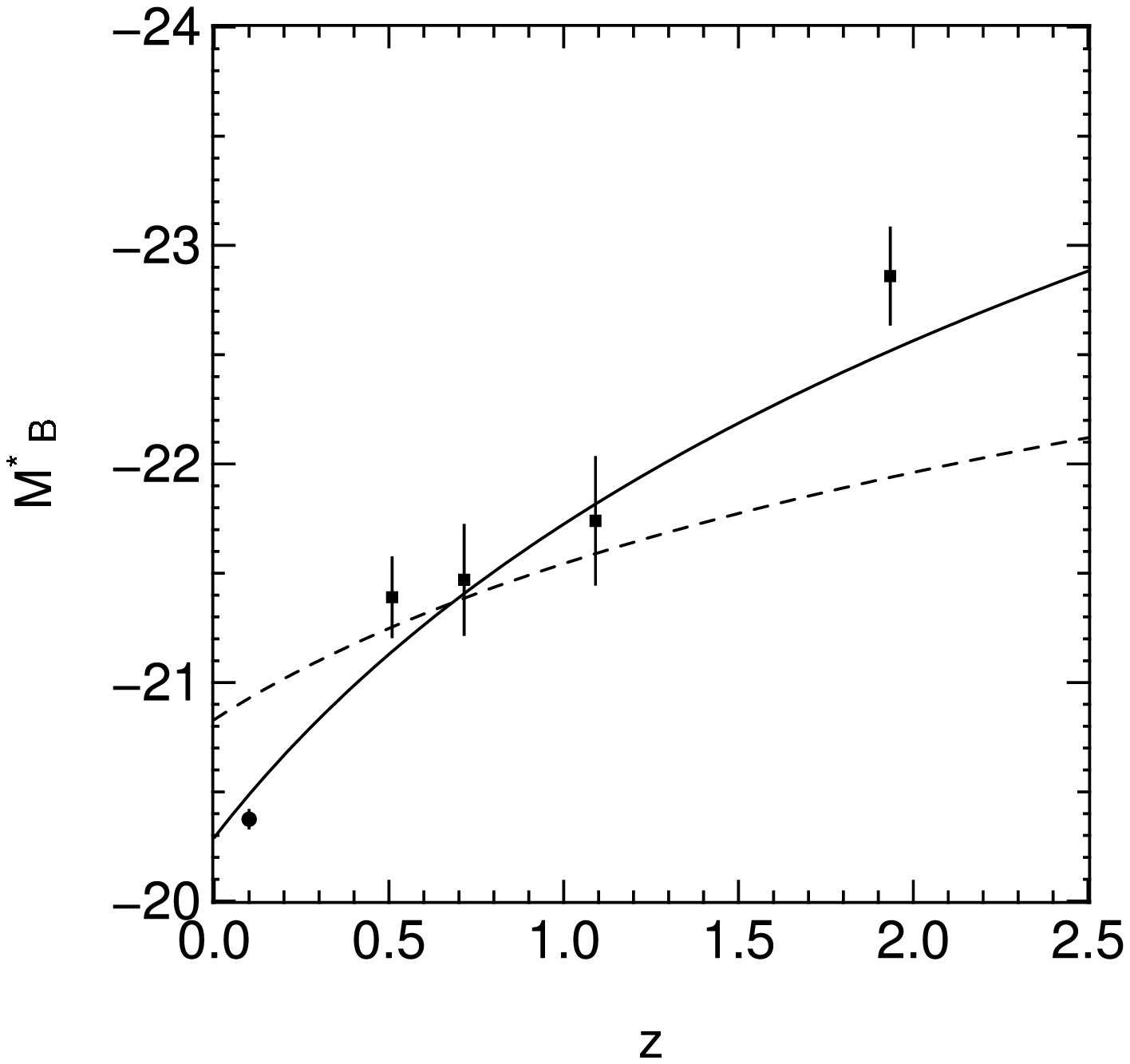,width=0.3\textwidth}

\vspace*{.5cm}

\epsfig{figure=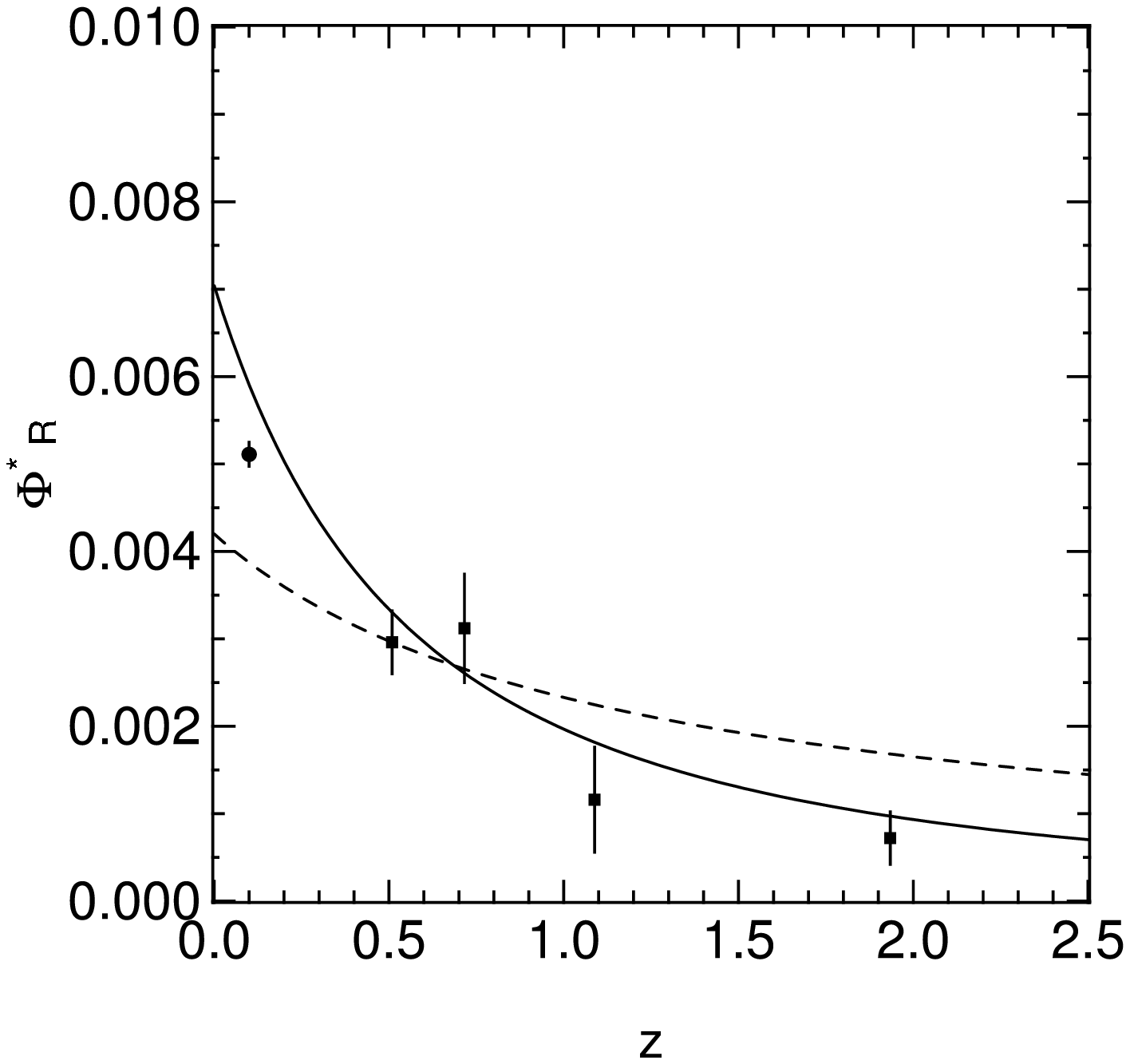,width=0.3\textwidth}
\hspace*{.5cm}
\epsfig{figure=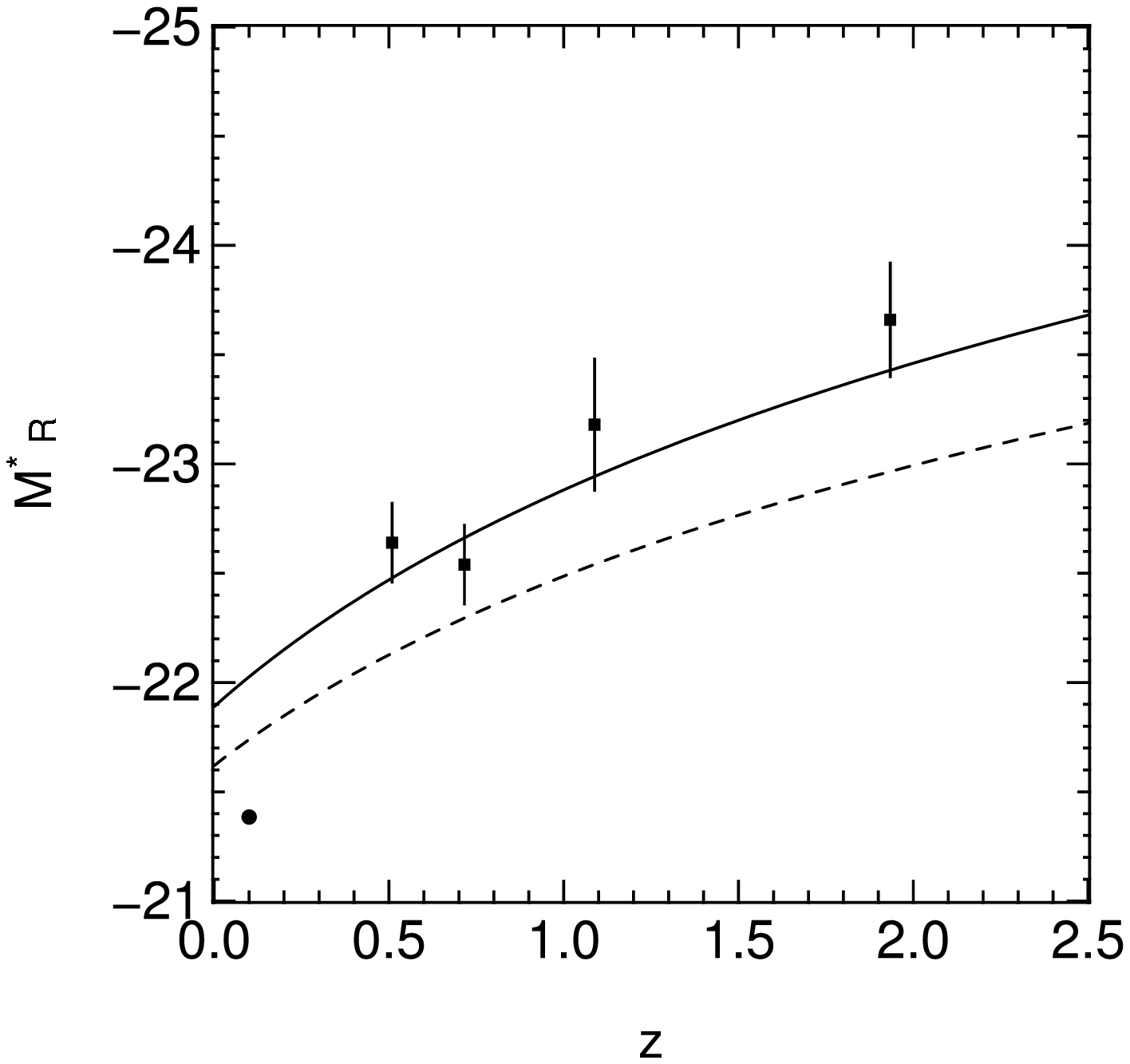,width=0.3\textwidth}

\caption{The redshift evolution of the luminosity function in
  MUNICS\_R for the $B$-band LF (\textit{upper panels}) and the
  $R$-band LF (\textit{lower panels}). In each case we show $\Phi^*$
  versus $z$ (\textit{left-hand panel}) and $M^*$ versus $z$
  (\textit{right-hand panel}) together with the evolutionary model fit
  described in the text (solid line). Note that this is not the fit to
  the Schechter parameters given in Table~\ref{t:plf_r_schechter} and
  represented by the filled squares, but a simultaneous fit to the LF
  values in all magnitude and redshift bins (see text for
  details). The dashed lines denote the evolution as derived from the
  FDF in \citet[][ $B$ band]{fdflf1} and \citet[][ $r'$ band, Case
  3]{fdflf2}.}
\label{f:plf_r_evol}

\end{figure*}

\subsection{Luminosity Function Evolution}

To estimate the rate of evolution of the Schechter parameters $\Phi^*$
and $M^*$ with redshift, we define evolution parameters $a$ and
$b$ as follows:

\begin{eqnarray}
  \nonumber
  M^*    \, (z) & = & M^*    \, (0) \, + \, a \, \mathrm{ln} \left(
  1 \, + \, z \right), \\ 
  \Phi^* \, (z) & = & \Phi^* \, (0) \: \left( 1 \, + \, z \right)^b
   , \; \mathrm{and}
 \label{e:evolnew}\\
  \nonumber
  \alpha \, (z) & = & \alpha \, (0) \;\; \equiv \;\; \alpha .
\end{eqnarray}

Our parametrisation is identical to the one chosen in \citet{fdflf1,
fdflf2} and equivalent to the form $\Phi \propto (1+z)^P$, $L \propto
(1+z)^Q$ sometimes found in the literature (especially in the context
of radio-source evolution). The parameters translate as follows:

\begin{equation}
  a \; = \; - \frac{2.5 \, Q}{\mathrm{ln} 10} \:\:\:\: \mathrm{and} \:\:\:\:
  b \; = P .
\end{equation}

Note that this evolutionary model is different from the one used in
\citet{munics5} and \citet{munics2}, where we used a parametrisation
linear in redshift, $\Phi^* (z) = \Phi^* (0) \left( 1 + c z \right)$
and $M^* (z) = M^* (0) + d z$ with the evolutionary parameters $c$ and
$d$. In this case the parameters translate in the following way: $a =
d z / \ln (1+z) \approx d \; (z \ll 1)$ and $b = \ln (1+cz)/\ln (1+z)
\approx c \; (z \ll 1)$. The advantage of the model used in this paper
is that it is a good representation of the redshift evolution of
$\Phi^*$ and $M^*$ even at higher redshift \citep{fdflf1}. At low
redshifts, however, the evolutionary models are equivalent.

Since there is still some debate about the local galaxy LF (see
e.g.\ the discussion in \citealt{Driver2005}), we decided to derive
the evolutionary parameters from our MUNICS\_R measurements
alone. This means that we have to obtain the best-fitting values for
$a$, $b$, $\Phi^*(0)$, and $M^*(0)$ by minimising the
four-dimensional $\chi^2$ distribution.

\begin{figure}
\centerline{\epsfig{figure=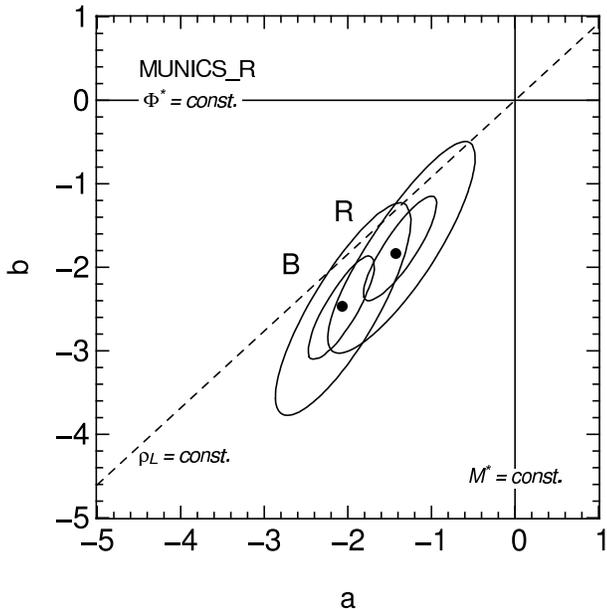,width=0.45\textwidth}}

\caption[MUNICS\_R: Evolution of LF in $B$ and $R$]{Estimates for
  the evolutionary parameters $b$ and $a$ ($1\sigma$ and $2\sigma$
  contours) for the $B$- and $R$-band LFs computed on
  MUNICS\_R. The solid lines show constant $\Phi^*$ and $M^*$,
  respectively, the dashed line indicates constant luminosity density,
  i.e.\ $b = \frac{\mathrm{ln} 10}{2.5}a \simeq 0.921 a$.}
\label{f:plf_evol_r}

\end{figure}

The resulting values for the evolutionary parameters $a$, $b$,
$\Phi^*(0)$ and $M^*(0)$ can be found in Table~\ref{t:plf_evol_r}. We
show the error contours of the evolutionary parameters $a$ and $b$
for the $R$-selected MUNICS catalogue and the LF in the rest-frame $B$
and $R$ filter in Figure~\ref{f:plf_evol_r}. These contours were
derived by projecting the four-dimensional $\chi^2$ distribution to
the $a$-$b$ plane, i.e. for given $a$ and $b$ we used the
values of $M^*(0)$ and $\Phi^*(0)$, which minimise $\chi^2 (a,
b)$.

We find $2\sigma$ evidence for evolution in $\Phi^*$ and $M^*$ for
both filters. Furthermore, the evolution is stronger in $B$ than in
$R$, in agreement with the findings of \citet{fdflf1, fdflf2}. In
Figure~\ref{f:plf_evol_r} we also show the line of constant luminosity
density $\rho_L$. The evolution of the LF parameters seems to follow
this relation rather closely, with $1\sigma$ evidence for a slight
decrease of the luminosity density with redshift..

\begin{table*}
\caption{Parameters $a$, $b$, $M^* (0)$, and $\Phi^* (0)$ for the
  evolution of the LFs in the $B$ and $R$ band derived from
  MUNICS\_R.}
\label{t:plf_evol_r}
\begin{center}
\begin{tabular}{ccccccc}
\hline
Filter & $b$ & $a$ & $\Phi^* (0)$ [$10^{-3}$ Mpc$^{-3}$] & $M^* (0)$ [mag] & $\alpha$ & $\min ( \chi^2_\mathrm{red} )$ \\
\hline
$B$ & $-2.47 \pm 0.58$ & $-2.07 \pm 0.38$ & $13.3 \pm 4.2$ & $-20.29 \pm 0.27$ & $-1.25$ & 0.37 \\
$R$ & $-1.84 \pm 0.61$ & $-1.43 \pm 0.43$ & $7.0 \pm 2.1$ & $-21.89 \pm 0.27$ & $-1.33$ & 0.33 \\
\hline
\end{tabular}
\end{center}
\end{table*}

The same evolutionary trend for the LF can be seen out to higher
redshifts: \citet{fdflf1, fdflf2} use the deep $I$-band
selected FORS Deep Field (FDF; \citealt{FDF1_short}) to trace the evolution
of the LF from the ultraviolet to the near-infrared out to redshifts
$z \; \sim \; 5$ finding similar results to the ones presented
here. That dataset is complimentary to our catalogue, since the FDF
lacks the large local volume of MUNICS which allows us to study the
lower redshift part of the evolution with high statistical accuracy.

In Figure~\ref{f:plf_r_evol} we show the resulting evolutionary model
for $\Phi^*$ and $M^*$ as a function of redshift, together with the
Schechter parameters derived for the LF in the four redshift bins,
local comparison values from the Millennium Galaxy Catalogue
\citep{Driver2005} for the $B$ band and from the Sloan Digital Sky
Survey (SDSS) presented in \citet{Blanton2003s} for the $R$ band, and
the evolution as derived from the FDF \citet{fdflf1, fdflf2}.

%
% SUMMARY
%
\section{Summary and Conclusions}
\label{s:summ}

In this paper we presented results on the evolution of field galaxies
drawn from galaxy catalogues selected in the $B$, $R$, $I$, and $K$
filters in the context of the Munich Near-Infrared Cluster Survey
(MUNICS). The first part of this paper is dedicated to a discussion of
the construction and properties of the $B$, $R$, and $I$-selected
MUNICS catalogues containing $\sim 9000$, $\sim 9000$, and $\sim 6000$
galaxies, respectively. The catalogues reach 50\% completeness limits
for point sources of $B \simeq 24.5$~mag, $R \simeq 23.5$~mag, and $I
\simeq 22.5$~mag and cover an area of about 0.3 square
degrees..  Photometric redshifts are derived for all galaxies
with an accuracy of $\delta z / (1+z) \simeq 0.057$, as demonstrated
by comparing them with a sample of $\sim 600$ spectroscopic redshifts
available for MUNICS galaxies. Star-galaxy classification is performed
by comparing the $\chi^2$ values of the spectral-energy-distribution
fitting within the photometric redshift code. A comparison with
spectroscopy clearly shows the reliability of this approach.

A study of rest-frame $B-K$ colour distributions for the four
different selection bands and different redshifts allows important
conclusions about selection effects and the evolution of galaxy
populations. First, we could show that the selection band influences
the colour distributions only for objects with low $K$-band
luminosities (comparatively low stellar masses), while the
distributions look remarkably similar for high-mass
galaxies. Secondly, the colour distributions get wider and bluer with
decreasing $K$ luminosity, indicating a higher average star-formation
rate and a wider distribution of star-formation rates for less massive
galaxies. Thirdly, there is strong colour evolution with redshift for
the redshift interval $0 \; \lsim \; z \; \lsim \; 1$ in the sense
that with increasing redshift galaxies become bluer (i.e.\ have larger
star-formation activity). This can hardly be seen for the most massive
galaxies, but becomes more and more so for the lowest mass
galaxies. Also, the trend becomes stronger going from $K$-band
selection to $B$-band selection. This means that the increase in star
formation rate from redshift zero to one is largely driven by lower
mass galaxies, and that the most massive galaxies have assembled the
bulk of their stellar mass before redshift unity, in agreement with
our results on the specific star formation rate (the star formation
rate per unit stellar mass; see, e.g., \citealt{Bauer2005};
\citealt{fdfssfr}a,b; \citealt{Juneau2005}). \nocite{munics7}

We investigate the influence of selection band and environment on the
specific star formation rate (SSFR). We find that $K$-band selection
indeed comes close to selection in stellar mass, while $B$-band
selection purely selected galaxies in star formation rate. As a second
order effect, the depth of the other filters of the survey influence
the selection boundary through the model fitting to the objects'
photometry involved in deriving stellar masses. We use a galaxy group
catalogue constructed on the $K$-band selected MUNICS sample to study
possible differences of the SSFR between the field and the group
environment, finding a marginally lower average SSFR in groups as
compared to the field, especially at lower redshifts.

The field-galaxy luminosity function in the $B$ and $R$ band as
derived from the $R$-selected MUNICS catalogue changes with increasing
redshift in the sense that the characteristic luminosity increases but
the number density decreases. This effect is smaller at longer
rest-frame wavelengths and gets more pronounced at shorter
wavelengths. This evolutionary trend continues to higher redshifts
\citep{fdflf1, fdflf2}. Parametrising the redshift evolution of the
Schechter parameters as $M^* (z) = M^* (0) + a \, \ln ( 1 + z )$ and
$\Phi^* (z) = \Phi^* (0) ( 1 + \, z )^b$ we find evolutionary
parameters $a \simeq -2.1$ and $b \simeq -2.5$ for the $B$ band, and
$a \simeq -1.4$ and $b \simeq -1.8$ for the $R$ band.

\section*{Acknowledgments}

The authors would like to thank the staff at Calar Alto Observatory,
ESO, and HET for their support during MUNICS observing runs, as well
as Jan Snigula for help with the catalogues. GF is indebted to John
Schellnhuber for invaluable support during the final stages of this
project. The authors thank Nigel Metcalfe for making number count data
available in electronic form. We thank the anonymous referee for the
careful reading of the manuscript and the comments which helped to
improve the presentation and discussion of our results.  We
acknowledge funding by the Deutsche Forschungsgemeinschaft,
Sonderforschungsbereich 375. This research has made use of NASA's
Astrophysics Data System (ADS) Abstract Service.

\bibliographystyle{mn2e}
\bibliography{mnrasmnemonic,literature,nc_lit_K}

\appendix

\begin{figure*}

\epsfig{figure=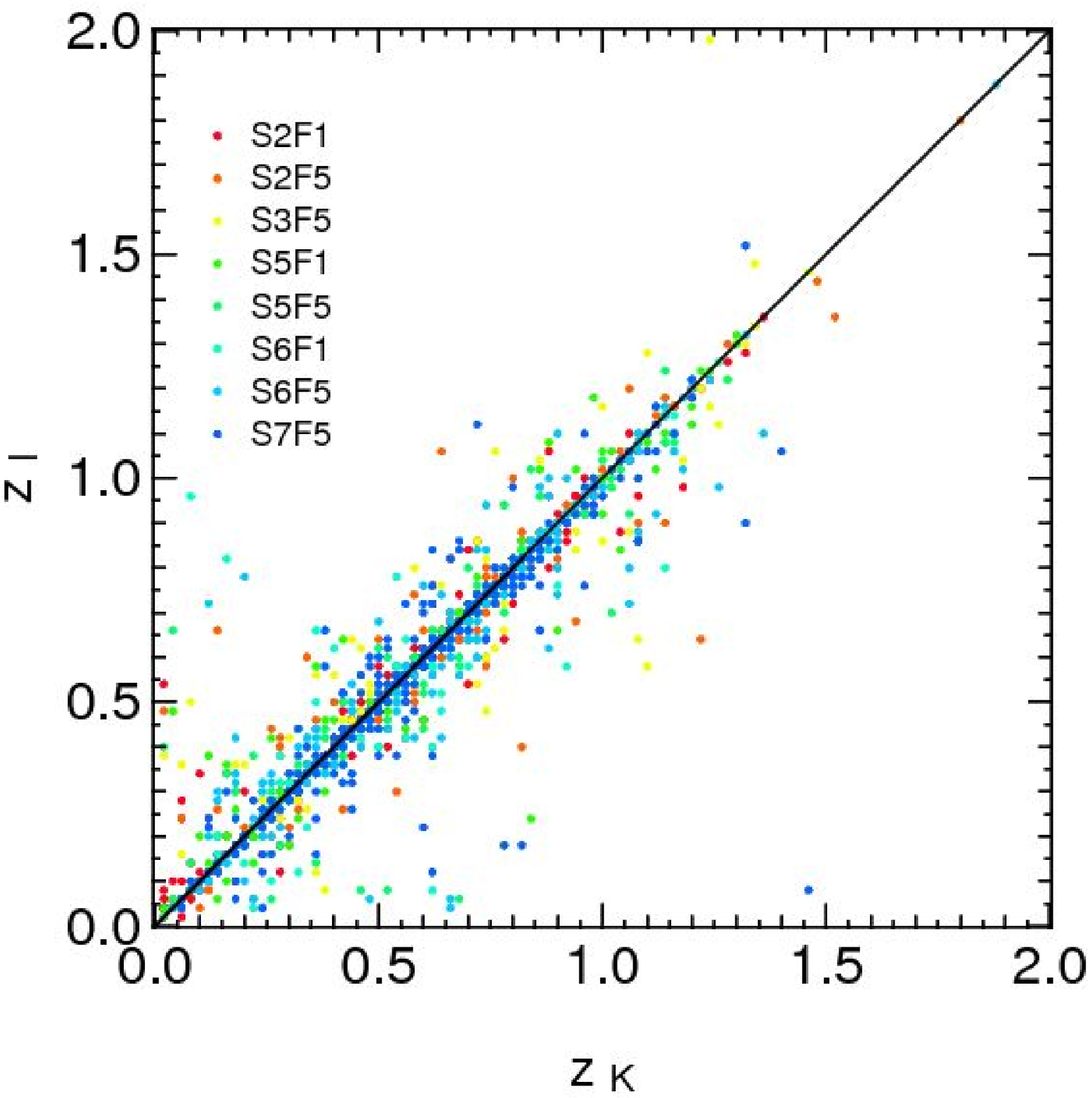,width=0.307\textwidth}
\hspace*{.5cm}
\epsfig{figure=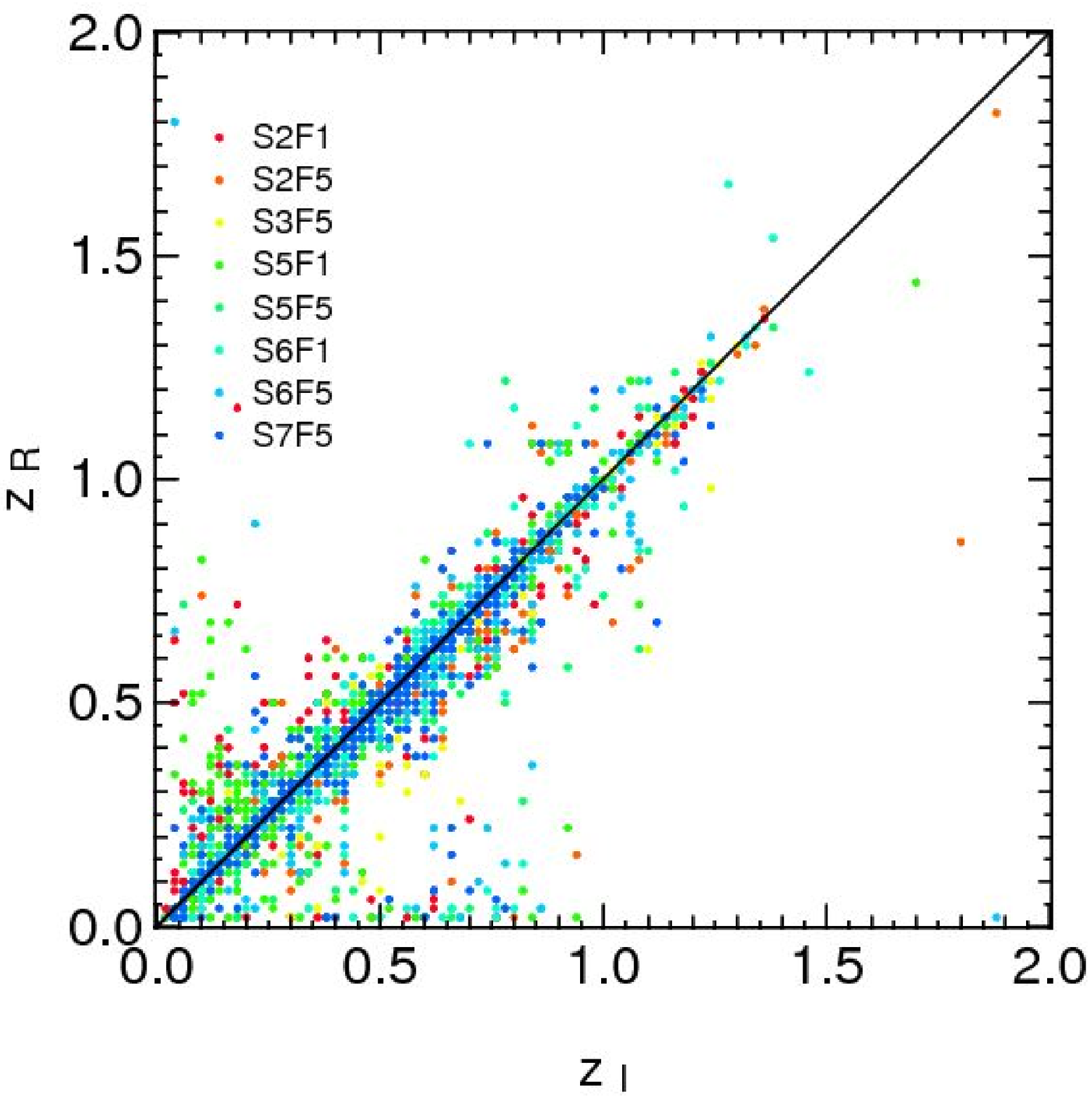,width=0.307\textwidth}
\hspace*{.5cm}
\epsfig{figure=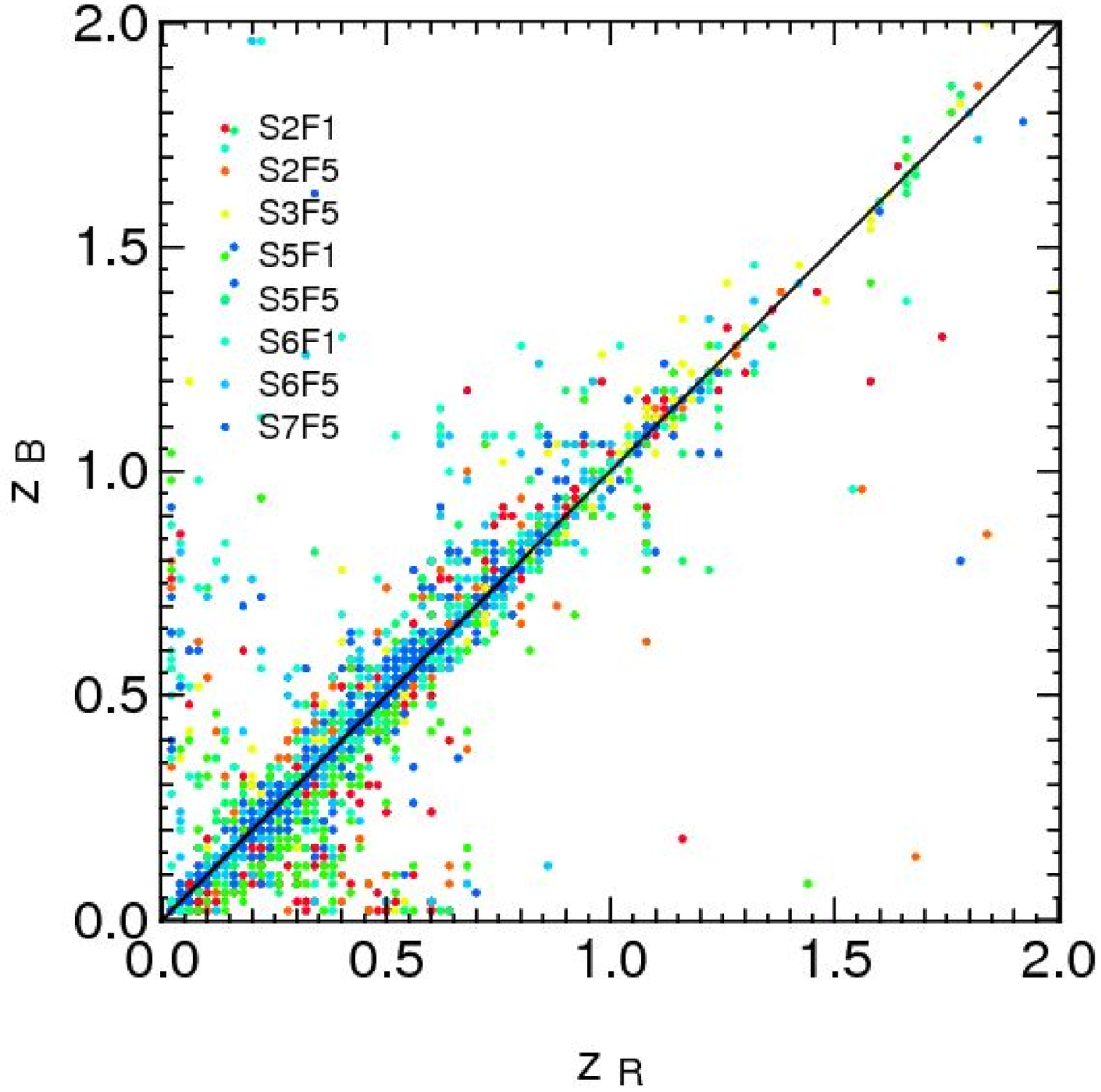,width=0.307\textwidth}

\vspace{.5cm}

\epsfig{figure=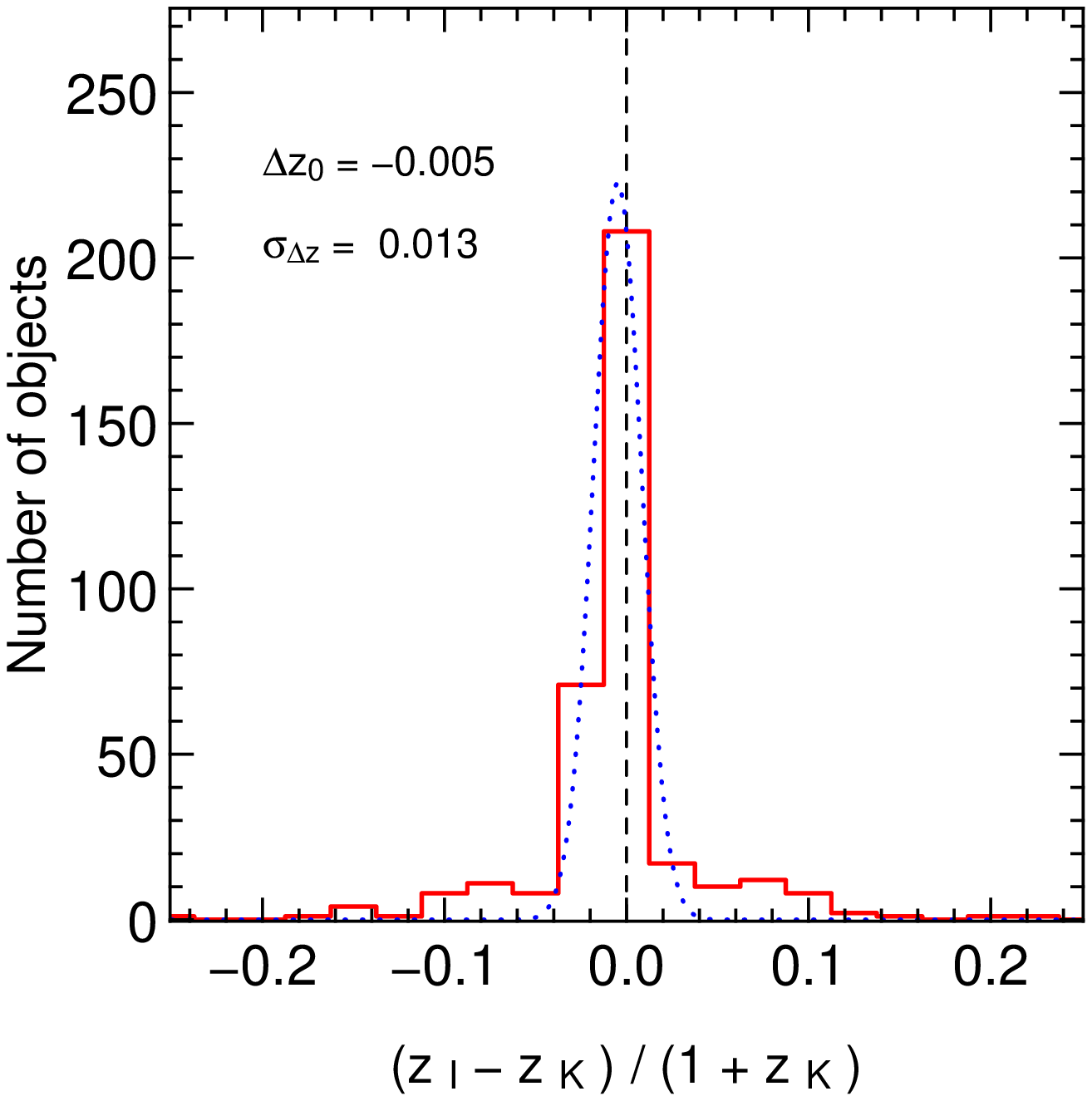,width=0.3\textwidth}
\hspace*{.5cm}
\epsfig{figure=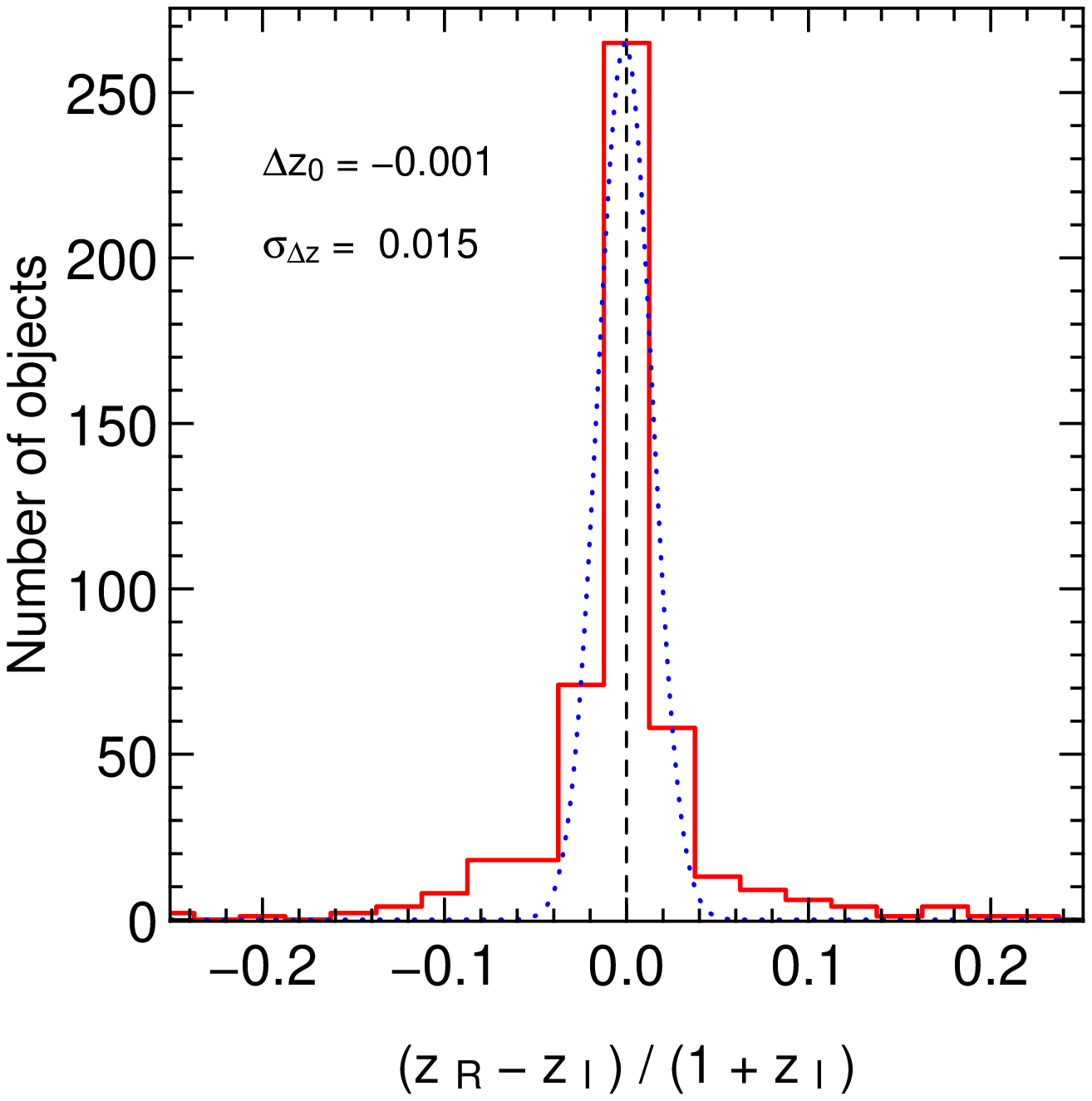,width=0.3\textwidth}
\hspace*{.5cm}
\epsfig{figure=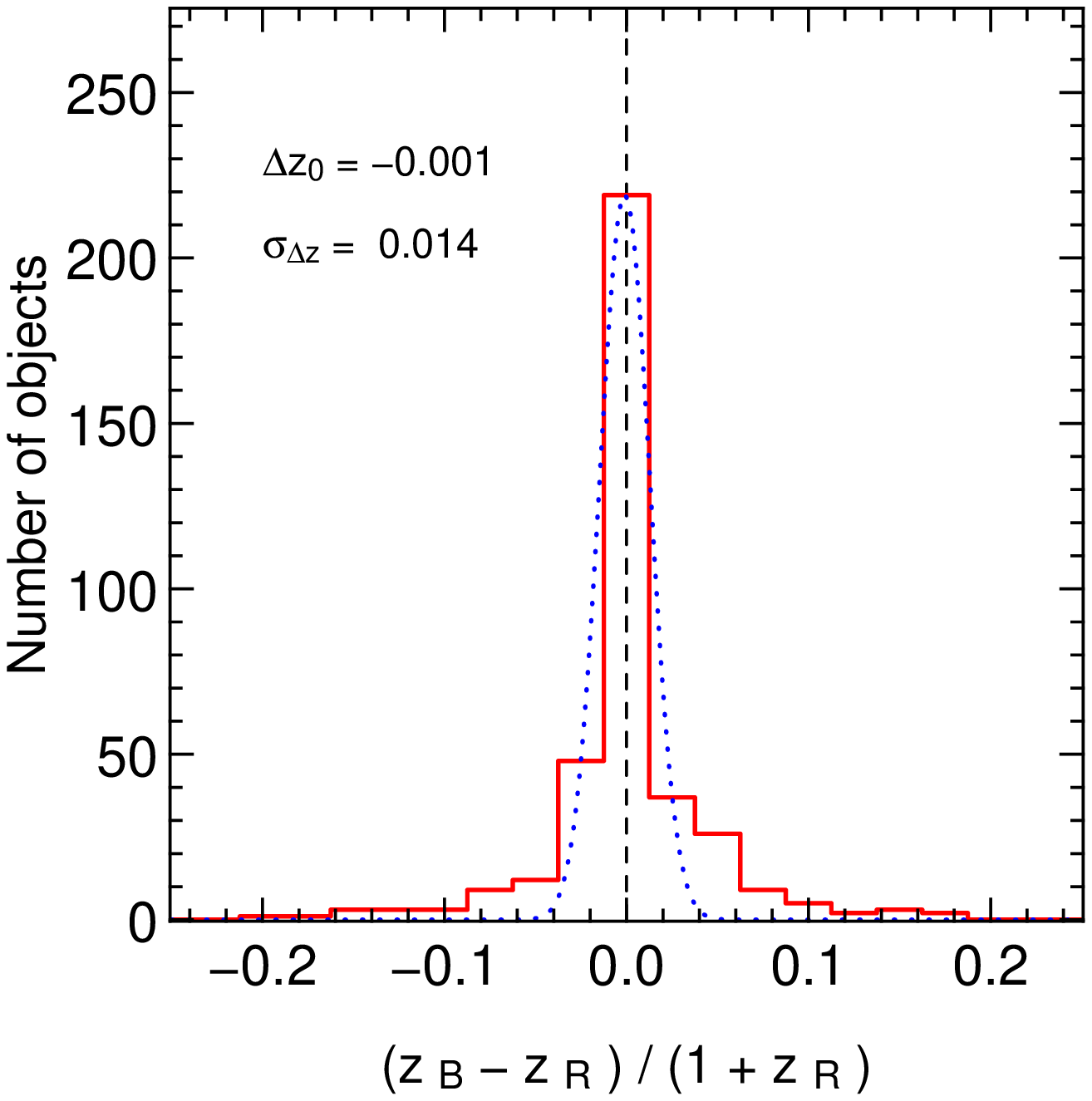,width=0.3\textwidth}

\caption{\textit{Left-hand panels:} Comparison of photometric
  redshifts from the $I$-selected MUNICS catalogue to those from the
  $K$-selected MUNICS sample. \textit{Middle panels:} The same for the
  $R$ and the $I$-selected catalogue. \textit{Right-hand panels:} The
  same for the $B$ and the $R$-selected catalogue. The lower panels
  show histograms of the redshift differences together with a Gaussian
  fit showing the quality of the photometric redshifts.}

\label{f:pz_pz}

\end{figure*}

\section{Photometric Redshifts in Different Samples}
\label{s:pzpz}

It is an important consistency check to compare the photometric
redshifts of an object present in two different selection catalogues.
Note that the two redshifts do not have to be the same in the case of
MUNICS, since small differences in object centring and differences in
the object's shape in the different detection images lead to slightly
different photometry. Direct comparison of magnitudes in the same
filter, but derived in the different selection catalogues show
excellent agreement \citep[comparison plots are published
  in][]{feulnerphd}.

Due to this small differences in the objects' photometry, the result
for a comparison of the photometric redshifts in different catalogues
is a scatter around the one-to-one relation, as shown in
Figure~\ref{f:pz_pz}. However, the general agreement is extremely
good, as can be seen from the histograms of redshift differences also
shown in the Figure. The deviation from a zero redshift difference is
smaller than 0.01 in $z$, and the $\sigma$ of the best-fitting
Gaussian is smaller than 0.02, which is the redshift binning size of
our photometric redshift algorithm.

\label{lastpage}

\end{document}